\documentclass{article}

\usepackage{jheppub}
\usepackage{multirow}
\usepackage{float,color}
\usepackage{mathtools}

\usepackage{etoolbox}
\makeatletter
\patchcmd{\maketitle}{\@fpheader}{}{}{}
\makeatother

\newcommand{\fant}[1]{\phantom{#1}}
\newcommand{\be}{\begin{equation}}
\newcommand{\ee}{\end{equation}}
\newcommand{\wdg}{\wedge}
\newcommand{\ot}{\otimes}
\newcommand{\bigzero}{\mbox{\normalfont\Large\bfseries 0}}
\newcommand{\rvline}{\hspace*{-\arraycolsep}\vline\hspace*{-\arraycolsep}}

\subheader{Review Article \hspace{\fill}\today}

\title{pp-waves in modified gravity}

\abstract{The family of metrics corresponding to the  plane-fronted gravitational waves with parallel propagation, commonly referred to as the family of $pp$-wave metrics,  is studied in the context of various modified gravitational models  in a self-contained and  coherent manner by using a variant of the  null coframe formulation of Newman and Penrose and the exterior algebra of differential forms on pseudo-Riemannian manifolds.}

\keywords{Classical general relativity; Einstein-Maxwell spacetimes, spacetimes with fluids, radiation or classical fields; Wave propagation and interactions; Modified theories of gravity; Scalar-tensor theories of gravity; Chern-Simons gauge theory; Nonminimal electromagnetic coupling;  Exact solutions; Exterior algebra; Newman-Penrose formalism.
\begin{tabbing}
\hspace*{3.4cm}\=\hspace*{1.8cm}\= \kill
\textsc{Pacs Numbers}: \>04.20.-q \>	Classical general relativity\\
              \>04.20.Jb 	\>Exact solutions\\
              \>04.30.Nk 	\>Wave propagation and interactions\\
							\>04.50.Kd \>	Modified theories of gravity\\
              \>11.15.Yc   \>Chern-Simons gauge theory\\
              \>04.40.Nr 	\>Einstein-Maxwell spacetimes, spacetimes \\ 
							\>\>with  fluids, radiation or classical fields
\end{tabbing}
}

\author{Ahmet BAYKAL}
\emailAdd{abaykal@ohu.edu.tr,  failure2communicate@gmail.com}

\affiliation{Department of Physics, Faculty of Arts and Sciences, \"Omer Halisdemir University,  Merkez Yerle\c{s}ke, Bor yolu \"uzeri, 51240   Ni\u gde, TURKEY}

\begin{document}
\maketitle

\section{Introduction}

In  a fortunate turn of events, the Laser Interferometer Gravitational-wave Observatory (LIGO) Scientific Collaboration \cite{LIGO-website,ligo-prl,ligo-prl2} announced  the  detection of gravitational waves emitted by coalescing of two stellar-mass black holes after the original version of the present article had been  posted to the arXiv. The gravitational wave signal GW150914 recorded  by both of the twin LIGO detectors in the first scientific run with an upgraded sensitivity is yet another confirmation the general theory of relativity after a hundred years after Einstein introduced it. The landmark observation heralds the beginning of the gravitational wave astronomy. The subject matter of the gravitational waves, on both experimental and theoretical side, will certainly  attract  the attention  of more researchers and it seems likely that LIGO Scientific Collaboration will keep us busy with further observations of gravitational waves produced by other astrophysical events.

The purpose of the present work is to present a brief and a self-contained discussion of the $pp$-wave solutions to the modified gravitational models in connection with the corresponding solutions in General theory of relativity (GR). In particular, the previous results reported   in \cite{tupper,robinson,mohseni,dereli-sert-pp-wave,dereli-tucker-pp-waves-energy-in-tensor-tensor-model,grumiller-yunes} in connection with modified gravitational models and some other results long-known   in the literature are presented in a unified manner by  making use of the algebra exterior forms and the complex  null tetrad formalism \cite{NP1,NP2} of Newman and Penrose (NP). Perhaps a secondary purpose of the present work is to show how the NP formalism can be applied to a given modified gravitational model 
by using the exterior algebra of differential forms which often turns out to be technically superior compared to the tensorial methods 
\cite{debney-kerr-schild,kerr}. 

The layout of the paper is as follows. The paper is roughly divided into two main parts which are almost equal in length. In the first part, the geometrical techniques of null coframe formalism in connection with a variant of the spin coefficient formalism of Newman and Penrose is developed in  Sect. 2. The mathematical properties of the gravitational plane-fronted waves with parallel propagation, usually referred to as $pp$-waves following the abbreviation originally introduced by Ehlers and Kund,  are briefly reviewed in the notation developed. In most of  the presentation, the Kerr-Schild (or Brinkmann) form of the $pp$-wave metrics is used and the Rosen form  of the metric is discussed in relation to the Kerr-Schild form in the particular case of the plane wave metrics in Sect. 3.    In Sect. 4, gravitational wave solutions of Petrov type N in various  backgrounds are discussed. The discussion of type N solutions belonging to the  Robinson-Trautmann class, which have expanding null geodesic congruence, are excluded from the current version as well. 

The second part is consists of the application of the geometrical techniques to various metric theories of modified gravity that allow a unified treatment presented in Sect. 5. In particular,  the derivation of the field equations  for the Brans-Dicke (BD) theory and the Chern-Simons modified gravity are discussed in a relatively more detailed manner compared to the  discussions of the other models such as the metric $f(R)$ gravity and the gravity model  involving a nonminimally coupled  Maxwell field and a  tensor-tensor gravity theory with  torsion. The $pp$-waves solutions to the general quadratic curvature gravity in four dimensions are also discussed in some detail as well. 

For the current version of the paper, it is fair to say that the  material covered in the first part  as the foundational material as a whole is not fully exploited and applied in all aspects to the second part. Moreover, the choice of modified gravitational models in the second part is motivated merely  by technical ease of the presentation in a unified way. Some solutions in the second part are presented in a slightly more general then their original form.

Finally, the references related to NP formalism and its applications are not, by any means, complete and the reader is referred to the authoritative books 
\cite{exact-sol-SKMHH,exact-sol-griffiths-podolsky,zakharov} which also introduce the NP technique  in full detail and apply it extensively in the context of GR.

\section{Geometrical preliminaries}

In this preliminary section, the essentials of the null tetrad formalism that are required to derive almost any geometrical quantity from the scratch are presented. However, because the family of $pp$-wave metrics has a relatively simple form to deal with, a thorough presentation of the NP formalism, that will be excessive to the purposes in what follows, is avoided.

The use of exterior algebra of differential forms on pseudo-Riemannian manifolds is made use of throughout the paper, and it proves to be practical and powerful in calculations relative to an orthonormal or a null  coframe and a coordinate coframe as well. The expressions for the geometrical quantities relative to a coordinate coframe are only discussed briefly  in connection with the corresponding expressions relative to a null coframe.

\subsection{The definitions of connection and curvature forms relative to a null coframe}

All the calculations in the following will be carried out relative to a set   of orthonormal and/or complex null  basis coframe 1-forms, denoted by $\{\theta^a\}$ for which the metric reads $g=\eta_{ab}\theta^a\ot \theta^b$ with the metric components $\eta_{ab}$ are constants. The mathematical conventions closely follow those of the ``Exact Solutions"  books \cite{exact-sol-SKMHH,exact-sol-griffiths-podolsky}. The signature of the metric is assumed to be mostly plus. The set of basis frame fields is $\{e_a\}$ and the abbreviation $i_{e_a}\equiv i_a$ is used for the contraction operator with respect to the basis frame field $e_a$. $*$ denotes the Hodge dual operator acting on basis forms and $*1=\theta^{0}\wdg \theta^1\wdg \theta^2\wdg \theta^3$ is the oriented volume element. When the Einstein summation convention is used, the exterior products of the basis 1-forms are also abbreviated as $\theta^{a}\wdg \theta^b\wdg \cdots \wdg \theta_c\wdg \theta_d\cdots\equiv \theta^{ab\cdots}_{\fant{aaaa}cd\cdots}$ for the sake of the notational simplicity.
The complex null coframe basis will also be denoted by $\{\theta^a\}=\{k, l, m, \bar{m}\}$ with $a=0, 1, 2, 3$ and an overbar denotes a complex conjugation.
The associated complex basis frame fields will be denoted by $\{e_a\}$.
In terms of the NP-type null coframe basis, the invariant volume element explicitly reads
\be\label{null-vol-form}
*1
=
\frac{1}{4!}\epsilon_{abcd}\theta^{abcd}
=
+ik\wdg l\wdg m\wdg \bar{m}
\ee
where the completely antisymmetric permutation symbol admits the numerical values values $0, \mp i$ with $\epsilon_{0123}=+i$ relative to a null coframe and in this case the indices are raised and lowered by the metric having non-diagonal and constant elements $\eta_{01}=\eta_{10}=-\eta_{23}=-\eta_{32}=1$. For example, a numerical indice 1  goes to 0 accompanied by a sign change whereas a numerical  indice 2 goes to 3 retaining the sign.

In particular, it follows from these definitions that  the self-dual 2-forms that diagonalize the Hodge dual operator  are
\be\label{self-dual-2form-def}
\begin{split}
*(k\wdg m)
&=
ik\wdg m
\\
*(l\wdg \bar{m})
&=
il\wdg \bar{m}
\\
*(k\wdg l-m\wdg\bar{m})
&=
i(k\wdg l-m\wdg\bar{m}).
\end{split}
\ee
The anti-selfdual 2-forms follow from the complex conjugation of the Hodge duality relations given in (\ref{self-dual-2form-def}) above.

The first structure equations of Cartan with vanishing torsion read
\be\label{cse1}
\Theta^a
=
D\theta^a
=
d\theta^a+\omega^{a}_{\fant{a}b}\wdg \theta^b=0
\ee
with $\Theta^a=\frac{1}{2}T^{a}_{\fant{a}bc}\theta^{b}\wdg \theta^{c}$ where $T^{a}_{\fant{a}bc}$ represent the components the torsion tensor.
$D$ is the covariant exterior derivative acting on tensor-valued forms. A suitable definition and its relation to covariant derivative can be found, for example, in \cite{straumann,benn-tucker}.
In terms of the complex connection 1-forms and the null coframe,  the first structure equations of Cartan explicitly read
\be\label{first-SE}
\begin{split}
&dk
+
\omega^{0}_{\fant{0}0}\wdg k
+
\bar{\omega}^{0}_{\fant{0}3}\wdg m
+
\omega^{0}_{\fant{0}3}\wdg \bar{m}
=
0,
\\
&
dl
-
\omega^{0}_{\fant{0}0}\wdg l
+
\omega^{1}_{\fant{0}2}\wdg m
+
\bar{\omega}^{1}_{\fant{0}2}\wdg \bar{m}
=0,
\\
&
dm
+
\bar{\omega}^{1}_{\fant{0}2}\wdg k
+
\omega^{0}_{\fant{0}3}\wdg l
-
\omega^{3}_{\fant{0}3}\wdg {m}
=0,
\end{split}
\ee
where the complex conjugate of the last equation has been omitted for convenience.
Because  the Levi-Civita  connection is metric compatible, one has $D\eta_{ab}=d\eta_{ab}-\eta_{ac}\omega^{c}_{\fant{a}b}-\eta_{bc}\omega^{c}_{\fant{a}a}$ which implies the antisymmetry $\omega_{ab}+\omega_{ab}=0$ for a null (and also for an orthonormal) coframe. Consequently, $\omega^{1}_{\fant{a}0}=\omega^{0}_{\fant{a}1}=\omega^{2}_{\fant{a}3}=\omega^{3}_{\fant{a}2}=0$
and there are three complex connection 1-forms
\be\label{3-connection-forms}
\omega^{0}_{\fant{a}3},\quad \omega^{1}_{\fant{a}2},\quad  \tfrac{1}{2}(\omega^{0}_{\fant{a}0}-\omega^{3}_{\fant{a}3})
\ee
related to the other connection 1-forms by complex conjugation. Explicitly, one has the conjugation relations 
$\bar{\omega}^{0}_{\fant{a}2}={\omega}^{0}_{\fant{a}3}$, $\bar{\omega}^{1}_{\fant{a}2}={\omega}^{1}_{\fant{a}3}$, 
$\bar{\omega}^{0}_{\fant{a}0}={\omega}^{0}_{\fant{a}0}$ and $\bar{\omega}^{3}_{\fant{a}3}=\bar{\omega}^{2}_{\fant{a}2}=-\bar{\omega}^{3}_{\fant{a}3}$,  The charge conjugation amounts to the interchange $2\leftrightarrow 3$ of the
null coframe indices.  The twelve NP spin coefficients can be identified as the components of the above complex connection 1-forms \cite{exact-sol-SKMHH}.
In contrast to the structure equations expressed in terms of six real connection 1-forms relative to an orthonormal coframe, it is possible to write the structure equations using only three complex connection 1-forms displayed in (\ref{3-connection-forms}) as in (\ref{first-SE}).

The curvature 2-form $\Omega^{a}_{\fant{a}b}$ with
$\Omega^{a}_{\fant{a}b}=\frac{1}{2}R^{a}_{\fant{a}bcd}\theta^{c}\wdg \theta^{d}$ satisfies the second structure of equations of  Cartan
\be
\Omega^{a}_{\fant{a}b}
=
d\omega^{a}_{\fant{a}b}
+
\omega^{a}_{\fant{a}c}\wdg \omega^{c}_{\fant{a}b}
\ee
and in terms of the complex connection 1-forms defined above and  the complex curvature 2-forms defined accordingly, the complex structure equations
can be written in the form
\be\label{SE-2}
\begin{split}
\Omega^{0}_{\fant{0}3}
&=
d{\omega}^{0}_{\fant{0}3}
-
{\omega}^{0}_{\fant{0}3}\wedge\left(\omega^{0}_{\fant{0}0}
-
\omega^{3}_{\fant{0}3}\right),
\\
\Omega^{1}_{\fant{0}2}
&=
d{\omega}^{1}_{\fant{0}2}
+
{\omega}^{1}_{\fant{0}2}\wdg\left(\omega^{0}_{\fant{0}0}
-
\omega^{3}_{\fant{0}3}\right),
\\
\Omega^{0}_{\fant{0}0}-\Omega^{3}_{\fant{0}3}
&=
d(\omega^{0}_{\fant{0}0}-\omega^{3}_{\fant{0}3})
+
2{\omega}^{0}_{\fant{0}3}\wedge{\omega}^{1}_{\fant{0}2},
\end{split}
\ee
where the use of complex differential forms reduces the number of independent tensor-valued  2-forms by half compared to the number of  corresponding
components defined relative to an orthonormal coframe.

In four dimensions, the curvature 2-form can be decomposed into the  irreducible parts in the form \cite{exact-sol-SKMHH}
\be\label{curvature-general-expansion}
\Omega^{a}_{\fant{a}b}
=
C^{a}_{\fant{a}b}+\tfrac{1}{2}\left(\theta^a\wdg S_b-\theta_b\wdg S^a\right)+\tfrac{1}{12}R\theta^a\wdg \theta_b
\ee
where $C^{a}_{\fant{a}b}$ is the traceless fourth-rank Weyl tensor, and the second term on the right hand side is expressed in terms of second-rank
traceless Ricci 1-form $S^a\equiv R^a-\tfrac{1}{4}R\theta^a$ and the last term is the scalar trace. It is possible to show (see, for example Ref. \cite{obukhov-PGT-rev}) that each term has definite self-duality/self anti-self-duality property by using the defining  relation (\ref{curvature-general-expansion}). Namely, the first and the third terms on the right hand side are self-dual, whereas the traceless Ricci part constitutes the anti-self-dual.
Now, using the above definitions, it is possible to express the decomposition of the curvature 2-form (\ref{curvature-general-expansion}) relative to a complex null coframe in terms of the original NP curvature scalars  as
\be\label{null-coframe-curvature-scalars}
\begin{split}
{\Omega}^{0}_{\fant{0}3}
&=
-
\Psi_{0}l\wdg \bar{m}
-
\Psi_{1}(k\wdg l-m\wdg \bar{m})
+
\Psi_{2}k\wdg m
\\
&\phantom{=}
-\Phi_{00}l\wdg m-\Phi_{01}(k\wdg  l+m\wdg \bar{m})+\Phi_{02}k\wdg \bar{m}
+
\tfrac{1}{12} Rk\wdg m,
\\
\tfrac{1}{2}({\Omega}^{0}_{\fant{0}0}-{\Omega}^{3}_{\fant{0}3})
&=
+
\Psi_{1}l\wdg \bar{m}
+
\Psi_{2}(k\wdg l-m\wdg \bar{m})
-
\Psi_{3}k\wdg m
\\
&\phantom{=}
+
\Phi_{10}l\wdg m
+
\Phi_{11}(k\wdg l+m\wdg\bar{m})
-
\Phi_{12}k\wdg \bar{m}
-
\tfrac{1}{24}R(k\wdg l-m\wdg\bar{m}),
\\
{\Omega}^{1}_{\fant{0}2}
&=
+
\Psi_{2}l\wdg \bar{m}
+
\Psi_{3}(k\wdg l-m\wdg \bar{m})
-
\Psi_{4}k\wdg m
\\
&\phantom{=}
+
\Phi_{20}n\wdg m
+
\Phi_{21}(k\wdg l+m\wdg\bar{m})
-
\Phi_{22}k\wdg \bar{m}
+
\tfrac{1}{12}Rl\wdg \bar{m}.
\end{split}
\ee
The scalar NP field equations, i.e., the Ricci identities in   component form,  can be reproduced form Eqs. (\ref{SE-2}) and (\ref{null-coframe-curvature-scalars}) by also taking the original definitions of the NP spin coefficients into account.
In four spacetime dimensions, there are six number number of real basis 2-forms and in the complex  NP forms there are three
independent basis 2-forms which are self-dual by definition while their complex conjugates are anti-self-dual by definition.
The self dual and anti-self dual components of the curvature 2-form are displayed explicitly in Eqs.  (\ref{null-coframe-curvature-scalars}) above.

The $pp$-waves  on flat backgrounds are in general of Petrov type N, and it is possible to find a NP null coframe with $\Psi_0=\Psi_1=\Psi_2=\Psi_3=0$
and therefore the only nonvanishing Weyl curvature spinor is $\Psi_4$ by definition. (See also the formula given in Appendix 1) As will explicitly be studied
in Sect. 3, a $pp$-wave-type gravitational wave can also be defined on backgrounds having, for example,   Petrov type  D or O.
However, evidently it is more difficult to extend such gravitational  wave solutions constructed on background spacetimes with a particular Petrov type to a modified gravitational model because the background spacetimes  may not be a solution the field equations for the modified gravity and  in general, such solutions  also require modification of the background spacetimes as well.

The above exterior algebra equations and the definitions belonging to  the null coframe formalism closely follow the variant  of the NP formalism presented in \cite{exact-sol-SKMHH}. The mathematical formula introduced above is sufficient for the description of gravitational wave metrics in the context of modified gravity models starting from scratch provided that the field equations are formulated accordingly by using the algebra of differential forms. Thus, the above geometrical formulas are sufficient for the  formulation of the field equations in a form suitable for the discussion below. Except for  Sect. 3.6 in which a particular tensor-tensor model of gravity allowing a torsion that can consistently be set to zero is studied, the discussion on the modified gravity models is confined to the pseudo-Riemannian case.

\subsection{The geometrical description of \texorpdfstring{$pp$}{}-wave metrics}

The $pp$-wave metrics has been introduced quite a long time ago by Brinkmann  \cite{brinkmann} and shortly after that by Jeffrey and Baldwin  \cite{jeffrey-baldwin}. Subsequently, they are interpreted as the metrics representing the gravitational waves \cite{peres-1959} by Peres.  From an idealized point of view, the $pp$-waves metrics can be regarded as a far-field  description of an isolated astrophysical source radiating  gravitational waves.

It is well-known that the family of the $pp$-wave  metric can conveniently be defined as a subclass of Kundt metrics by introducing a covariantly constant geodesic null congruence  \cite{kundt1,kundt2,ehlers-kundt} with all the optical scalars corresponding to shear, divergence and twist  vanishing.
The Killing symmetries of the $pp$-wave metrics have studied by Sippel and Goenner, and by Bondi et al.  \cite{sippel-goenner,bondi-pirani-robinson} for variety of profile functions in both pseudo-Riemannian  and Riemann-Cartan geometry settings (see, also Chapter 23 \cite{exact-sol-SKMHH}).  Noether gauge symmetries of the $pp$-wave metric have recently been discussed  
in \cite{camci}.

Another well-known peculiar property of the $pp$-wave metric is that all the polynomial scalar invariants vanish \cite{peres-1960,schmidt,pravda}. It is of Petrov type N with only one non-vanishing complex Weyl curvature spinor \cite{exact-sol-SKMHH}. The classical gravitational plane waves are shown \cite{deser-polarization} to be unaffected by the vacuum  polarization effects to all loop orders.

In terms of the global null coordinates $\{x^\alpha\}=\{u, v, \zeta, \bar{\zeta}\}$ for $\alpha=0, 1, 2, 3$ respectively, the $pp$-wave ansatz in so-called Kerr-Schild form can be expressed as
\be\label{pp-wave-ansatz}
g=-du\ot dv-dv\ot du-2H du\ot du+d\zeta\ot d\bar{\zeta}+d\bar{\zeta}\ot d\zeta
\ee
with a real profile function $H=H(u,\zeta, \bar{\zeta})$. For $H=0$, the metric becomes the Minkowski background.
The real null vector $\partial_v$, the four-fold repeated principle null direction of the Weyl tensor, defines the direction of propagation and that $u=$ constant surfaces are the flat transverse planes.

The particular form of the $pp$-wave metric given in (\ref{pp-electro-ansatz}) is known to be the  Brinkmann form and that there is yet another useful form which is known as the Rosen form \cite{rosen}. By a suitable coordinate transformation, the metric in the Brinkmann form in (\ref{pp-electro-ansatz}) can be related to the corresponding Rosen form which explicitly illustrates the transverse character of the $pp$-wave metrics. For the particular case of the plane wave metrics, the explicit coordinate transformations  relating the two forms can be found in \cite{exact-sol-griffiths-podolsky}. In most of the discussion in what follows, the Brinkmann form for  the family of the $pp$-wave metrics will be used.

To begin with, the metric (\ref{pp-electro-ansatz}) is to be cast into the following familiar complex null form
\be\label{NP-general-form-metric}
g=
-k\ot l-l\ot k+m\ot\bar{m}+\bar{m}\ot m
\ee
in terms of a null basis coframe 1-forms $k, l ,m,\bar{m}$. In practical calculations, such a null coframe can be constructed with the help of a set of an orthonormal basis 1-forms as well. Although there are different possible choices for the frames and associated coframes in the literature for the $pp$-wave metric, a natural choice for the set of basis coframe 1-forms is
\be\label{pp-coframe-def}
\theta^0=k=du,\qquad  \theta^1=l=Hdu+dv,\qquad  \qquad \theta^2=m=d\zeta
\ee
in terms of  the complex null coordinates and that $\theta^3=\bar{\theta^2}$.
The other choices of the basis coframe 1-forms can be  related to (\ref{pp-coframe-def}) by, for example, the interchanges $k\leftrightarrow l$ and $m\leftrightarrow \bar{m}$ (See Appendix 1).
The corresponding volume 4-form defined up to an orientation is explicitly given by
\be
*1=idu\wdg dv\wdg d\zeta\wdg d\bar{\zeta}
\ee
that is identical to that of Minkowski volume 4-form. The set of orthonormal  basis frame fields associated with the above coframe can be written as
\be\label{associated-frame-fields}
e_0\equiv D=-\partial_v,\qquad  e_1\equiv \Delta=-\partial_u+H\partial_v,\qquad  e_2\equiv\bar{\delta}=\partial_{\bar{\zeta}}, \qquad  e_3\equiv 
\delta=\partial_{{\zeta}}.
\ee
Following to the original notation of the NP formalism, the null basis frame fields are denoted by $D, \Delta, \bar{\delta}, \delta$.
The definition of the frame fields (\ref{associated-frame-fields}) are identical to the frame fields adopted in \cite{exact-sol-SKMHH}.
The set of frame fields are useful in the calculations making use of the tensorial methods and when this is the case, both of the minus signs in the defining relations of $e_0$ and $e_1$ are usually  transferred to the   coframe fields. The geometrical quantities for the $pp$-wave metric then can be calculated readily by using the commonplace techniques of the exterior algebra of differential forms.

By definition, the only non-vanishing exterior derivative of basis coframe 1-forms can be expressed in the form
\be\label{exterior-l}
dl
=-H_\zeta k\wdg m-H_{\bar{\zeta}} k\wdg \bar{m}.
\ee
In Eq. (\ref{exterior-l}) and in the expressions in what follows, a coordinate subscript to a function denotes the partial derivative with respect to the coordinate. Consequently, by making use of the derivative expression in connection with the first structure equations, one readily finds that the only non-vanishing connection 1-form is
\be\label{pp-spin-connection-coefficient}
\omega^{1}_{\fant{a}2}=H_{\zeta}k.
\ee
As a result, there is only one non-vanishing spin coefficient for the $pp$-wave metric ansatz.
Now using (\ref{pp-spin-connection-coefficient}), it is straightforward to show that the vector field $k^ae_a$, associated to the basis 1-form $\theta^0=k$, is covariantly constant and  real.

Turning now to the second structure equations (\ref{SE-2}), it is easy to see  that relative to the NP-type complex null coframe (\ref{pp-coframe-def}), there are only two non-vanishing curvature components of the curvature 2-form $\Omega^{1}_{\fant{a}2}$. The $pp$-wave metric ansatz is known to linearize the curvature 2-forms  and accordingly, the only non-vanishing curvature 2-form components then take the form
\be\label{pp-curavture-form}
\Omega^{1}_{\fant{a}2}
=
d\omega^{1}_{\fant{a}2}
=
-H_{\zeta\zeta} k\wdg m-H_{\zeta\bar{\zeta}} k\wdg \bar{m}
\ee
with $\Omega^{0}_{\fant{a}3}=\Omega^{0}_{\fant{a}0}=\Omega^{3}_{\fant{a}3}=0$.
The non-vanishing components of the Riemann tensor are $R^{13}_{\fant{aa}02}$ and $R^{13}_{\fant{aa}03}$ relative to the null coframe. Consequently, one  finds  $R=0$. The set of basis 2-form $k\wdg m$, $l\wdg\bar{m}$ and $\tfrac{1}{2}(k\wdg l-m\wdg\bar{m})$ defined by (\ref{self-dual-2form-def}) are self-dual whereas their complex conjugates define the set of anti-self-dual 2-forms relative to the null coframe and the volume element defined in Eq. (\ref{null-vol-form}). The set of all self-dual and anti-self-dual 2-forms form a convenient basis for the 2-forms.

Thus, for example, the $k\wdg m$ component of the curvature 2-form (\ref{pp-curavture-form}) corresponds to the component of the complex Weyl 2-form.
The curvature spinors  can be obtained by comparing the general expression (\ref{curvature-general-expansion}) with the curvature expression in Eq. (\ref{pp-curavture-form}). One finds
\be
C^{1}_{\fant{a}2}
=
-H_{\zeta\zeta} k\wdg m, \quad\qquad R^1=-2H_{\zeta\bar{\zeta}} k
\ee
for the $pp$-wave metric ansatz (\ref{pp-wave-ansatz}). Accordingly, the non-vanishing curvature scalars are given by
 \be
 \Psi_4=H_{\zeta\zeta},\qquad   \Phi_{22}=H_{\zeta\bar{\zeta}}.
 \ee
The solution to the homogeneous vacuum equations $\Phi_{22}=0$ can be written in terms of an arbitrary  function $h=h(u,\zeta)$ analytic in $\zeta$ as
\be
H(u,\zeta, \bar{\zeta})
=
h(u,\zeta)+\bar{h}(u, \bar{\zeta})
\ee
and consequently, the surviving curvature component $\Psi_4$ then takes the form 
\be
\Psi_4(u, \zeta)
=
h_{\zeta\zeta}.
\ee

The effect of gravitational wave, which is  contained in the curvature component $\Psi_4$,  can be observed as the vibrations of test particle in the plane transverse to the propagation vector $k$.
In vacuum, the amplitude of a $pp$-waves  is then determined by the non-vanishing component of the Weyl tensor $|\Psi_4|$ whereas the corresponding   polarization modes are determined by the angle $\varphi$ in $\Psi_4=|\Psi_4|e^{i\varphi}$.  
By introducing a suitable frame, it is possible to show the real and imaginary parts can be identified with the usual ``+" and ``$\times$" transverse polarization modes obtained by linearizing the Einstein field equations around the Minkowski spacetime. In general, the explicit form of the Weyl spinor $\psi_4$ can be found  after the profile function is determined from the Einstein field equations.

  As a side remark note that in general  the Einstein field equations $G_{ab}=\kappa^2T_{ab}$ can conveniently be implemented directly into the curvature expression as the anti-self-dual part of the curvature expansion (\ref{curvature-general-expansion}), e.g.,  $\Phi_{ik}$ components of the curvature 2-forms in (\ref{null-coframe-curvature-scalars}), in the NP formalism.  In addition, the use of exterior algebra also  offers  some alternate means to calculate the Ricci spinors as will be exemplified below in the case of the $pp$-wave metrics.

To facilitate the comparison with the coordinate expressions in the literature, it is possible to relate the curvature 2-form expression relative to the orthonormal coframe easily in the particular case of the $pp$-wave metric ansatz. For this purpose, first note that the curvature expression $\Omega^{1}_{\fant{a}2}=d\omega^{1}_{\fant{a}2}$ can be written in the form $\Omega^{1}_{\fant{a}2}=dH_\zeta\wdg k$.
Considering the coordinate expressions of the tensorial quantities labeled by the Greek indices, one arrives at the relation
\be\label{second-pair-expanded}
\Omega^{1}_{\fant{a}2}=\partial_{\zeta} \partial_{[\alpha}  Hk_{\beta]}dx^\alpha\wdg dx^\beta
\ee
where $\partial_{\alpha}\equiv {\partial}/{\partial x^{\alpha}}$ and the expansion $k=k_\beta dx^\beta$ of the basis 1-form $k$ are used. The square brackets implies  the antisymmetrized indices. Furthermore, it is now convenient to rewrite the partial derivative of the profile function  with respect to the complex coordinate $\zeta$ in the form
\be
\Omega^{1}_{\fant{a}2}=\bar{m}^\nu(\partial_\nu\partial_{[\alpha}  H) k_{\beta]}dx^\alpha\wdg dx^\beta
\ee
using the definition of the basis frame vectors.
As the curvature 2-form $\Omega^{1}_{\fant{a}2}$ can be expanded into  the coordinate basis with respect to the last pair of indices as  in (\ref{second-pair-expanded}), the first pair of indices can also be expressed in terms of the  contractions of the Riemann tensor with the coordinate basis frame fields. Consequently, it follows from the definition  of the curvature 2-form that
\be
\Omega^{1}_{\fant{a}2}=\tfrac{1}{2}l^\mu \bar{m}^\nu R_{\mu\nu\alpha\beta}dx^\alpha\wdg dx^{\beta}.
\ee
Finally, by making use of the fact that $g(k,l)=g_{\mu\nu}k^\mu l^\nu=-1$, one readily finds that the coordinate components of the Riemann tensor are given by the relation
\be\label{coord-curvature-exp}
R_{\mu\nu\alpha\beta}
=
-4k_{[\mu}{(}\partial_{\nu]}\partial_{[\alpha} H{)} k_{\beta]}.
\ee
The corresponding expression for the Ricci tensor components can be found, by  the contraction $R_{\mu\nu}=g^{\alpha\beta}R_{\mu\alpha\nu\beta}$,
as
\be\label{coord-Ricci-exp}
R_{\mu\nu}
=
k_{\mu}k_\nu g^{\alpha\beta}\partial_\alpha\partial_\beta  H
=
2k_{\mu}k_\nu\partial_\zeta\partial_{\bar{\zeta}}H
\ee
by using the fact that $k^\mu\partial_\mu H=0$ by definition.
Consequently,  the scalar curvature defined by $R=g^{\mu\nu}R_{\mu\nu}$ vanishes identically as a consequence of  the fact that $k^\mu\partial_\mu $ is a null vector. It is now straightforward  to see that any contraction  of the vector $k^\mu\partial_\mu$ with the curvature tensor (\ref{coord-curvature-exp} )vanishes identically. Explicitly, the curvature tensor expression imply the relations $k^\mu R_{\mu\nu\alpha\beta}=0$ and $k^\mu R_{\mu\nu}=0$.

Alternatively, Eq. (\ref{coord-curvature-exp}) can also be obtained by using the Christoffel symbols $\Gamma^{\alpha}_{\mu\nu}$.
Although the  orthonormal  coframe expression (\ref{pp-curavture-form}) for the curvature tensor  is not particularly less convenient then the corresponding expression relative to  the coordinate  basis from a merely technical point of view, the coordinate expression (\ref{coord-curvature-exp}) is more frequently used in the literature. 

Now returning  to the calculation of the Einstein form relative to the orthonormal coframe,
the Einstein 3-form can be calculated from the general formula
\be
*G^a
=
-
\tfrac{1}{2}\Omega_{bc}\wdg *\theta^{abc}
\ee
where $G_a=G_{ab}\theta^b$ is a covector-valued 1-form which can be expressed in terms of Ricci tensor and  curvature scalar as $G_a\equiv (R_{ab}-\tfrac{1}{2}\eta_{ab}R)\theta^b$. Explicitly, by specializing the indices to the null coframe introduced above and
with the help of Hodge duality relations, one arrives at
\be
*G^1
=
\bar{\Omega}^{1}_{\fant{a}2}\wdg i\bar{m}-\Omega^{1}_{\fant{a}2}\wdg i{m}
=
-2H_{\zeta\bar{\zeta}}*k.
\ee
As it is noted previously, because the scalar curvature vanishes identically, one  has $G^1=R^1$ and  thus the Ricci tensor is of the form $R_{ab}\theta^a\ot \theta^b=2H_{\zeta\bar{\zeta}}k\ot k$ or equivalently, in components $R_{\mu\nu}=2H_{\zeta\bar{\zeta}}k_\mu k_\nu$  after expanding the basis 1-form $k$ to its coordinate components. The expression for the Ricci tensor components is consistent  with the previous coordinate expression (\ref{coord-Ricci-exp}).
In addition, by using the Ricci spinor  definitions \cite{exact-sol-griffiths-podolsky}, one finds $\Phi_{22}\equiv\tfrac{1}{2}l^\mu l^\nu R_{\mu\nu}=H_{\zeta\bar{\zeta}}$.

It is also worth noting that the contracted second Bianchi identity reduces to $D*G^1=d*G^1=0$ and, consequently,  the Einstein 3-form can be rewritten as an exact form \cite{frauendiener-thirring-form-null-coframe}. For this purpose, one first notes that
\be
*G^1
=
id (\bar{\omega}^{1}_{\fant{a}2}\wdg m-\omega^{1}_{\fant{a}2}\wdg \bar{m})
\ee
by making use of  $dm=dd\zeta\equiv0$. This alternate expression can be simplified further.
It is possible to verify by direct calculation that for the $pp$-wave ansatz (\ref{pp-wave-ansatz}), the real null basis 1-forms $k, l$ satisfy the following relation
\be
d*dl
=
-2H_{\zeta\bar{\zeta}}*k
\ee
by making use of (\ref{exterior-l}). Note also that $d*l=0$ by definition. Consequently, one ends up with the relation
\be\label{simplified-einstein-form-for-pp}
*G^1=d*dl
\ee
illustrating the linearization of the components of  the metric ansatz in an explicit and  compact form.

To summarize the curvature calculations for the pp-wave metric,
the vacuum equations for the $pp$-wave metric then reduces to the Laplaces's equation $H_{\zeta\bar{\zeta}}=0$ on the transverse plane spanned by the complex coordinate $\zeta$. This equation has the  general solution of the form
$
H(u, \zeta, \bar{\zeta})
=
h(u, \zeta)+\bar{h}(u, \bar{\zeta})
$
with  $h(u, \zeta)$ being an arbitrary complex function of the coordinates $u, \zeta$ and it is analytic in $\zeta$. Because the field equations do not determine the profile function fully, it is possible to construct metrics various metrics, for example, involving distribution functions of the form $\delta(u)$ with null point particle sources.

The $pp$-wave ansatz in the form above is restrictive in admitting a matter source. However, the discussion can be generalized to include a null electromagnetic field in a straightforward way as follows. The  self-dual Faraday 2-form can be defined as
\be\label{maxwell-spinor-def}
\mathcal{F}
=
\tfrac{1}{2}(F-i*F)
=
-\Phi_{0}l\wdg \bar{m}-\Phi_{1}(k\wdg l-m\wdg \bar{m})+\Phi_{2}k\wdg m
\ee
in terms of  the complex Maxwell spinor scalars $\Phi_{k}$. Using equation (\ref{maxwell-spinor-def}), the Maxwell spinors $\Phi_k$ can be expressed in terms of the contractions with the frame fields as
\begin{align}
\Phi_{0}
&=
 F_{\alpha\beta}k^\alpha m^\beta,
\\
\Phi_{1}
&=
 \frac{1}{2}F_{\alpha\beta}\left(k^\alpha l^\beta+\bar{m}^\alpha m^\beta\right),
\\
\Phi_{2}
&=
 F_{\alpha\beta}\bar{m}^\alpha l^\beta.
\end{align}

The Maxwell's  equations can be written as equations for 3-forms in the form $dF=0$ and $d*F=0$.
These equations can also be rewritten  in the complex form as $d\mathcal{F}=0$ in terms of self-dual Maxwell 2-form  
defined by $\mathcal{F}\equiv \frac{1}{2}(F-i*F)$. In terms of  ${F}$, the components of the energy-momentum 3-forms for the Faraday 2-form field  read
\be\label{maxwell-en-mom-3-form}
*T^a[F]
=
\tfrac{1}{2}(i^a F\wdg *F-F\wdg i^a*F),
\ee
and note that these expressions are valid relative to an orthonormal frame and to a null coframe. $*T^a[F]$ can also be expressed in terms of self-dual 2-form $\mathcal{F}$ as well by making use of (\ref{maxwell-spinor-def}).

For the $pp$-wave metric, it is straightforward to see that a 2-form $\mathcal{F}$ compatible with the metric ansatz (\ref{pp-electro-ansatz}) can have  only one non-vanishing component. For $\mathcal{F}=\Phi_{2}k\wdg {m}$ with the non-vanishing Maxwell spinor $\Phi_2=\Phi_2(u, \zeta, \bar{\zeta})$. The Maxwell's equations in the complex form
\be
d\mathcal{F}
=
\partial_{\bar{\zeta}}\Phi_{2}k\wdg m\wdg \bar{m}=0
\ee
implies that $\partial_{\bar{\zeta}}\Phi_{2}=0$ and therefore one can define a convenient four-potential 1-form. Explicitly, it is possible to define  $\Phi_2=\Phi_2(u, \bar{\zeta})\equiv\partial_{{\zeta}}{f}(u, {\zeta})$ with $f(u, \zeta )$ being an arbitrary function analytic  in the variable $\zeta$. Consequently, the Faraday 2-form $F$ can be derived from the gauge potential $A=A_a\theta^a$ that has the expression
\be\label{pp-electro-ansatz}
A
=
[f(u, {\zeta})+\bar{f}(u, \bar{\zeta})]du.
\ee
The general solution to the homogeneous Maxwell's equation $d\mathcal{F}=0$ then can be expressed in the form of an expansion 
in $\zeta$ as
\be
f(u,\zeta)
=
\sum^{\infty}_{-\infty}f_n(u)\zeta^n.
\ee

In terms of the self-dual Faraday 2-form $\mathcal{F}$, and relative to the complex null coframe, the energy-momentum form
has the only non-vanishing component
\be
*T^1[F]
=
-2\Phi_{2}\bar{\Phi}_{2}*k
=
-2\partial_{{\zeta}}f(u, {\zeta})\partial_{\bar{\zeta}}\bar{f}(u, \bar{\zeta})*k.
\ee
For the metric ansazt (\ref{pp-wave-ansatz}), the electrovacuum field equations  $*G^a=\kappa^2*T^a[F]$ reduce to
$*G^1=\kappa^2*T^1[F]$, or equivalently, $\Phi_{22}=\kappa^2|\Phi_{2}|^2$, explicitly reads
\be\label{electrovac-gr-eqn}
H_{\zeta\bar{\zeta}}
=
\kappa^2\partial_{{\zeta}}f(u, {\zeta})\partial_{\bar{\zeta}}\bar{f}(u, \bar{\zeta}).
\ee
$\kappa^2\equiv8\pi G/c^4 $ with $G$ and $c$ stand for the Newton's gravitational constant and the speed of light, respectively.
Note that the field equations can also be written as an equation  for 3-forms as
\be\label{3-form-eqn-gr}
d*dl=-\kappa^2f_\zeta\bar{f}_{\bar{\zeta}}*k.
\ee

In general, one can show that the Einstein-Maxwell field equations can be expressed in the form $\Phi_{ik}=\kappa^2\Phi_{i}\bar{\Phi}_{k}$ in terms of the
Ricci and the Maxwell spinors with $i,k=0,1,2$ provided that the complex components of the curvature 2-forms (\ref{SE-2}) are identified in terms of the Ricci spinors $\Phi_{ik}$ with $i,k=0, 1, 2$ given  in (\ref{null-coframe-curvature-scalars}).

The general solution to  the inhomogeneous field equation (\ref{electrovac-gr-eqn}) can be written in the form
\be\label{electrovac-gr-sol}
H(u, \zeta, \bar{\zeta})
=
h(u, \zeta)+\bar{h}(u, \bar{\zeta})
+
\kappa^2f(u, \zeta)\bar{f}(u, \bar{\zeta})
\ee
with $h$ and $f$ being arbitrary functions of the coordinates $u$ and $\zeta$ and analytic  in $\zeta$.
As has been stated above, the field equations do  not determine the $u$-dependence of the profile function and that the complex function $h$ and because the superposition principle holds as an exception, $h$ can be expanded as
\be\label{h-expansion}
h(u,\zeta)
=
\sum^{\infty}_{n=2}a_n(u)\zeta^n+\mu\ln\zeta+\sum^{\infty}_{n=1}b_n(u)\zeta^{-n}.
\ee
The superposition principle also implies that the $pp$-waves propagating in the same direction do not interact.
$n=0$ and $n=1$ terms in the first sum are omitted because these terms can be eliminated by the coordinate transformation defined by
\be
\zeta=\zeta'+a(u), \qquad v'= v+b(u)+\dot{a}\bar{\zeta}+\dot{\bar{a}}\zeta,\qquad u'=u
\ee
that leave the form of the metric (\ref{pp-electro-ansatz}) invariant provided that the new profile function  $H'$ is identified as $H'=H+\ddot{a}\bar{\zeta}+\ddot{\bar{a}}\zeta-\dot{a}\dot{\bar{a}}+\dot{b}$ where  $\dot{}\equiv d/du$.
Therefore, it is possible to choose the complex functions $a(u)$ and $b(u)$ such that the profile function has  the form (\ref{h-expansion})
without any loss of generality.

In (\ref{h-expansion}), the term of the form  $\mu\ln \zeta$ requires a null particle source with $T_{00}\sim \delta(u)$.  This impulsive $pp$-wave solution is known as the Aichelburg-Sexl solution \cite{exact-sol-SKMHH}. It is originally  obtained \cite{aichelburg-sexl} by boosting the Schwarzchild solution to the speed of light in the limit  the Schwarzchild mass reducing to  zero. 

The impulsive $pp$-waves solutions can also be constructed by the geometrical method \cite{exact-sol-griffiths-podolsky,penrose-cut-paste} of Penrose, by cutting the Minkowski background along a null hypersurface and then reattaching the two parts with a warp. The `Cut and Paste' method of Penrose was used to construct a spherical impulsive gravitational wave solution \cite{nutku-penrose,aliev-nutku} as well. Dray and 't Hooft introduced the ``coordinate shift method'' \cite{dray-thooft} to construct non-expanding impulsive waves in non-flat backgrounds.

One can construct the $pp$-wave solutions with a profile function  involving some other dependence on the real null coordinate $u$, instead of a Dirac delta function. For example, a sandwich $pp$-wave metric \cite{exact-sol-griffiths-podolsky} which has a discontinuity that can be expressed in terms of a a particular function having nonzero values only over  a finite interval $u_1\leq u\leq u_2$ can be constructed.

The particular vacuum solution with $h=a_2(u)\zeta^2$ corresponds to gravitational  plane waves (or homogeneous $pp$-waves) having a constant wave amplitude. The solutions with profile function of the form $b_n(u)\zeta^{-n}$ also require null particles with multipole structure 
\cite{null-particles-with-multipole-structure1,null-particles-with-multipole-structure2}. The solutions with the terms of the form $a_{n}\zeta^n$ has recently been studied in \cite{podolsky-vesely1,podolsky-vesely2,podolsky-vesely3,podolsky-vesely4,sakalli-halilsoy,sakalli-halilsoy-2} and shown that the geodesic structure of these $pp$-wave spacetimes leads to chaotic motion of the test particles.

It is also possible to construct solutions corresponding to a linear superposition of two distinct null  electromagnetic field.
Let us consider a four potential $A$ in Eq. (\ref{pp-electro-ansatz})  expressed as a superposition of the form
\be
f(u, {\zeta})
=
f_1(u, {\zeta})+f_2(u, {\zeta})
\ee
where the two independent functions $f_1$ and $f_2$ are analytic in $\zeta$. Then, for the superposed electrovacuum metric, the profile function is given by
\be
H(u, \zeta)
=
h(u, \zeta)+\bar{h}(u, \bar{\zeta})
+
\kappa^2
\left\{
|f_1|^2
+
|f_2|^2
+
f_1\bar{f}_2
+
f_2\bar{f}_1
\right\}.
\ee
Although the $pp$-wave metric ansatz linearizes the Einstein tensor, a nonlinearity arise from the electromagnetic energy-momentum tensor \cite{nutku1, nutku2}. On the other hand, the superposition is allowed for the profile function, as well as it is valid at the level of the corresponding curvature tensor as can be observed from Eqs. (\ref{pp-spin-connection-coefficient}) and (\ref{pp-curavture-form}).

The profile function of an electrovacuum solution can be combined with that of a vacuum solution defined at different regions of the transverse plane \cite{bonnor-beam,exact-sol-griffiths-podolsky} to have a new profile function  of the form
\be\label{bonnor-beam-profile-funct}
H(u, \zeta, \bar{\zeta})
=
\left\{
\begin{aligned}
&\alpha^2(u)(|\zeta-\zeta_0|^2-r^2),\hspace{12.3mm} |\zeta-\zeta_0|\leq r\\
&\alpha^2(u)r^2\ln (|\zeta-\zeta_0|^2/r^2),\qquad |\zeta-\zeta_0|> r
\end{aligned}
\right.
\ee
The composite function $H$  is defined to be continuous across the  boundary $|\zeta-\zeta_0|=r$.
Note also that the logarithmic part of the profile function  is in accordance with the cylindrically symmetric  metric exterior to an infinite line source that can be obtained in the linear approximation \cite{bonnor-beam} and that the logarithmic singularity $|\zeta-\zeta_0|\mapsto\infty$ is a coordinate singularity as one can show by studying the geodesics of  test particles. The function (\ref{bonnor-beam-profile-funct}) requires an electromagnetic ansatz of the form  $f(u, \zeta)=\alpha(u)(\zeta-\zeta_0)$ for the region $|\zeta-\zeta_0|\leq r$ on the transverse planes and that the corresponding metric represents the gravitational field of  an infinite  uniform beam of light with circular cross-section centered at $\zeta_0$ and having a radius a finite radius $r$. It is conformally flat in the region $|\zeta-\zeta_0|\leq r$ whereas  $\Psi_4=-(\alpha r)^2 (\zeta-\zeta_0)^{-2}$ outside the beam. One can superpose \cite{bonnor-beam,exact-sol-griffiths-podolsky} any number of profile functions of the form (\ref{bonnor-beam-profile-funct}) for the non-intersecting parallel light beams having distinct centers of symmetry.

\subsection{Linear polarization modes for a plane gravitational wave}\label{polarization-sect}

The polarization modes of a gravitational wave is determined by the geodesic deviation equation for a congruence of timelike geodesic vector fields \cite{straumann}.
The polarization modes of the plane gravitational waves (and their superposition) have quite recently been discussed in \cite{cropp-visser} by using both the Brinkman and the Rosen forms of the metric.  This brief subsection is devoted to a discussion of the polarization modes for the vacuum plane waves using the Brinkman form.

The vacuum solution with the homogeneous function of the form
\be
h(u,\zeta)=h(u)\zeta^2
\ee
corresponds to gravitational waves with constant wave amplitude, the so-called plane waves as a particular subclass of  $pp$-waves. With the benefit of hindsight, it is convenient to write  the function $h(u)$ out explicitly as 
\be 
h(u)=h_+(u)-ih_\times(u)
\ee  
in terms of two real functions $h_+(u)$ and $h_\times(u)$, the corresponding real profile function $H(u, \zeta, \bar{\zeta})$ then can be expressed in the form
\be
H
=
h_+(u)\left(\zeta^2+\bar{\zeta}^2\right)-ih_\times(u)\left(\zeta^2-\bar{\zeta}^2\right).
\ee
Thus, the profile function for the general vacuum solution is composed of the superposition of two linear polarization modes. The real part $h_+(u)$ gives ``+" polarization mode whereas the imaginary part $h_{\times}(u)$ becomes  the amplitude for the  ``$\times$" polarization mode.
Consequently, the real and imaginary parts of the function $h$ correspond to the real and the imaginary parts of the  Weyl curvature spinor respectively. 
The corresponding Weyl spinor has the explicit form
\be
\Psi_4
=
h_+(u)-ih_\times (u)
\ee
in this particular case.

By  definition, the linear   polarization modes of a plane wave depend on the coordinate system at hand, and thus one has the freedom to rotate the coordinates in the transverse plane  to have a definite polarization mode. In order to obtain the general transformation rules for the polarization modes $h_+$ and $h_\times$,  let us now consider a rotated complex spatial coordinate $\zeta'$ defined by  $\zeta=e^{i\varphi}\zeta'$ for some fixed angle $\varphi$ which leave the $pp$-wave metric ansatz (\ref{pp-wave-ansatz}) intact. Expressed in terms of the primed coordinates, the function $h(u)$ then takes the form
\be
h(u)
=
\left[\left(h_+\cos 2\varphi +h_\times\sin 2\varphi\right)-i\left(h_\times\cos 2\varphi -h_+\sin 2\varphi\right)\right]\zeta'^2.
\ee
One can read off the primed polarization modes as 
\begin{align}
h'_+
&=
+h_+\cos 2\varphi +h_\times\sin 2\varphi,\label{pol-rot1}\\
h'_\times
&=
-h_+\sin 2\varphi+h_\times\cos 2\varphi,\label{pol-rot2}
\end{align}
which illustrate the helicity property of a massless spin-2 field obtained by linearizing the  Einstein's field equations around a flat
Minkowski background. 

As a consequence of the  rotation formulas (\ref{pol-rot1}) and (\ref{pol-rot2}), note that  the linear polarization modes $h_+$ and $h_\times$ are interchanged by a rotation of the transverse coordinates  by an amount $\pi/4$. 

In extended theories of gravity, such as Brans-Dicke sclar-tensor theory, usually gravitational waves appear  to have more than two polarization modes which are recently discussed in \cite{alves-miranda-araujo} see,  also \cite{eardley-lee-lightman,eardley-lee-lightman2}. 

Halilsoy \cite{halilsoy-cyl-gr-waves} discussed the linear polarization modes of interacting, cylindrical gravitational waves using a form 
of a metric ansatz slightly more general than the one  originally introduced in \cite{einstein-rosen} by Einstein and Rosen.

\subsection{Canonical coframe}

It is also possible to find a coframe where $\Psi_4$ is  normalized to unity. Let us consider two coframes related by a Lorentz boost of the form
\be\label{null-coframe-tr}
\tilde{k}=a k,\qquad \tilde{l}=a^{-1}l,\qquad \tilde{m}=m
\ee
with $a=a(u,\zeta,\bar{\zeta})$. The corresponding connection 1-forms are then related by
\be
\begin{split}
&\tilde{\omega}^{0}_{\fant{a}3}=a\omega^{0}_{\fant{a}3},
\qquad
\tilde{\omega}^{1}_{\fant{q}2}=a^{-1}\omega^{1}_{\fant{a}2},
\\
&\tilde{\omega}^{0}_{\fant{a}0}-\tilde{\omega}^{3}_{\fant{a}3}=\omega^{0}_{\fant{a}0}-\omega^{3}_{\fant{a}3}+a^{-1}da,
\end{split}
\ee
and, using these results one finds that the tilde curvature 2-forms can be found as 
\be
\begin{split}
&\tilde{\Omega}^{0}_{\fant{a}3}=a\Omega^{0}_{\fant{a}3},
\qquad
\tilde{\Omega}^{1}_{\fant{q}2}=a^{-1}\Omega^{1}_{\fant{a}2},
\\
&\tilde{\Omega}^{0}_{\fant{a}0}-\tilde{\Omega}^{3}_{\fant{a}3}=\Omega^{0}_{\fant{a}0}-\Omega^{3}_{\fant{a}3}.
\end{split}
\ee
For the coframe (\ref{pp-coframe-def}) corresponding to $pp$-wave metric ansatz (\ref{pp-wave-ansatz}) the transformation formulas yield
the following  nonvanishing connection 1-forms: 
\be
\tilde{\omega}^{1}_{\fant{q}2}=a^{-2}H_\zeta \tilde{k},
\qquad \tilde{\omega}^{0}_{\fant{a}0}-\tilde{\omega}^{3}_{\fant{a}3}=a^{-1}da.
\ee
The corresponding curvature 2-forms are given by
\be
\tilde{\Omega}^{1}_{\fant{q}2}
=
-a^{-2}H_{\zeta\zeta} \tilde{k}\wdg \tilde{m}-a^{-2}H_{\zeta\bar{\zeta}} \tilde{k}\wdg \tilde{\bar{m}}.
\ee
Consequently, the normalization requirement $\tilde{\Psi}_4=1$  determines the transformation parameter as
\be
a(u,\zeta, \bar{\zeta})=\left(f_{\zeta\zeta}\right)^{1/2}
\ee
with an arbitrary function  $f$ is analytic in $\zeta$.
Using this result in the connection transformation one finds
\be\label{canonical-coframe-conn-exp}
\tilde{\omega}^{0}_{\fant{a}0}-\tilde{\omega}^{3}_{\fant{a}3}
=
\frac{1}{2}(f_{\zeta\zeta})^{-2}f_{u\zeta\zeta}\tilde{k}
+
\frac{1}{2}(f_{\zeta\zeta})^{-1}f_{\zeta\zeta\zeta}\tilde{m}.
\ee
For the plane waves the function $f$ has the form  $f(u,\zeta)=g(u)\zeta^2$ for some arbitrary complex function $g(u)$ therefore 
the last term in (\ref{canonical-coframe-conn-exp}) drops out for the plane waves and
the remaining component of the connection 1-form in (\ref{canonical-coframe-conn-exp}) corresponds to the NP spin coefficient $\gamma$ \cite{Coley-McNutt-Milson-CQG} in the tilde coframe. 

The coframe in which the Weyl curvature normalized to unity, the tilde coframe above, is called a canonical coframe and the first step in the Karlhede classification  of the Riemannian spacetimes \cite{karlhede,karlhede2} based on the derivatives of the Riemann tensor is to find such a canonical coframe. 
In the case of a plane wave spacetime the invariant classification then  depends on the spin coefficient  $\gamma$ and its derivatives.

\subsection{Plane polar coordinates for the transverse planes}\label{polar-coords}

For some particular $pp$-waves to be discussed below, it is more convenient to adopt the plane polar coordinates for the transverse planes. Thus, it is  
useful to cast the geometrical formulas and the tensorial quantities of the $pp$-waves using the plane polar coordinates.

Let us consider the null coframe 
\be\label{polar-null-coframe}
k=du,\qquad l=dv+Hdu,\qquad m=\frac{1}{\sqrt{2}}\left(d\rho+i\rho d\phi\right)
\ee
defined in terms of plane polar coordinates $\{\rho. \phi\}$ for the $pp$-wave metric ansatz with the profile function $H=H(u,\rho,\phi)$.
The basis null frame fields $\{e_a\}$ associated to the null coframe above is
\be
D=-\partial_v,\qquad \Delta=-(\partial_u-H\partial_v),\qquad
\delta
=
\frac{1}{\sqrt{2}}\left(\partial_\rho+\frac{i}{\rho}\partial_\phi\right).
\ee
In a typical computation relative to the null coframe  (\ref{polar-null-coframe}), it is convenient to express the exterior derivative as
\be
d=
-k\Delta-lD+m\bar{\delta}+\bar{m}\delta.
\ee
The nonvanishing connection 1-forms for the null coframe (\ref{polar-null-coframe}) are
\be
\omega^{1}_{\fant{a}2}=\bar{\delta}H k,\qquad \omega^{3}_{\fant{a}3}=-\frac{1}{\sqrt{2}\rho}(m-\bar{m}).
\ee
One can show that the corresponding nonvanishing curvature 2-form can be written in the form  
\be
\Omega^{1}_{\fant{a}2}
=
-
\rho\bar{\delta}\left(\frac{1}{\rho}\bar{\delta}H\right)k\wdg m
-
\frac{1}{\rho}\delta\left(\rho\bar{\delta}H\right)k\wdg \bar{m}.
\ee
Consequently, the Ricci spinor $\Phi_{22}$ is of the form
\be
\Phi_{22}
=
\frac{1}{2}
\left(
\frac{1}{\rho}\frac{\partial}{\partial \rho}\rho\frac{\partial}{\partial \rho}+\frac{1}{\rho^2}\frac{\partial^2}{\partial \phi^2}
\right)H
\ee
where the expression on the right hand side  is the Laplacian acting on the profile function  on the transverse planes in terms of plane polar coordinates:
\be
\frac{1}{\rho}\delta(\rho\bar{\delta}H)
=
\frac{1}{2}\left(
\frac{1}{\rho}\frac{\partial}{\partial \rho}\rho\frac{\partial}{\partial \rho}+\frac{1}{\rho^2}\frac{\partial^2}{\partial \phi^2}\right)
H
=
2\partial_\zeta\partial_{\bar{\zeta}}H
\ee
where the coordinates are related by
\be
\zeta=\frac{1}{\sqrt{2}}\rho e^{i\phi}
\ee
up to an irrelevant phase angle.
Likewise, the nonvanishing Weyl curvature spinor has the expression
\be
\Psi_4
=
-
\frac{1}{2}
\left(
H_{\rho\rho}-\frac{1}{\rho}H_{\rho}-\frac{1}{\rho^2}H_{\phi\phi}
\right)
+
i
\left(
\frac{1}{\rho}H_{\rho\phi}
-
\frac{1}{\rho^2}H_{\phi}
\right)
\ee
where a coordinate subscript for the profile function  stands for a partial derivative with respect to that coordinate as before.
The vacuum field equations $\Phi_{22}=0$ can be used to simplify the expression for $\Psi_4$. Thus, for the vacuum the Weyl curvature 
reduces to
\be\label{polar-reduced-weyl-curv}
\Psi_4
=
-
\frac{1}{\rho^2}\left(
\rho H_{\rho}
+
H_{\phi\phi}
\right)
+
\frac{i}{\rho^2}
\left(
\rho H_{\rho\phi}
-
H_{\phi}
\right).
\ee

It is instructive to plug the profile function  for a vacuum plane wave with the profile function of the form $H(u,\zeta,\bar{\zeta})=h(u)(\zeta^2+\bar{\zeta}^2)$ into the reduced expression (\ref{polar-reduced-weyl-curv}). For the profile function of a plane wave expressed in terms of the  plane polar coordinates  of the form
\be
H(u, \rho,\phi)
=
h(u)\rho^2\cos 2\phi
\ee
the Weyl curvature component then  takes the expected form
\be\label{polar-superposed-pol}
\Psi_4
=
h(u)e^{-2i\phi}
\ee
which justifies the simplified expression on the right hand side of (\ref{polar-reduced-weyl-curv}). 
 As discussed in the Section \ref{polarization-sect}, for a complex function $h(u)$, the profile function (\ref{polar-superposed-pol}) represents the amplitude of a superposition of $+$ and $\times$ polarization modes and that one can rotate the transverse coordinates by a suitable angle to gauge  a particular polarization mode.

\subsection{\texorpdfstring{$pp$}{}-wave solutions with null particles having multipole structure}

It is possible to construct the impulsive  $pp$-wave solutions with  null particles sources  with arbitrary multipole structure
\cite{null-particles-with-multipole-structure1}. For this purpose, it is convenient to use the plane polar coordinates $\rho,\phi$ for the transverse planes
in the preceding subsection.

One starts with the partial differential equation that the profile function satisfies with the source term of the form $T_{00}=J(\rho,\phi)\delta(u)$
representing a null point particle on the wavefront $u=0$.
Einstein field equations for the profile function $\delta(u)H(\zeta,\bar{\zeta})$ can be written as the following partial differential equation
\be
\left(
\frac{1}{\rho}\frac{\partial}{\partial \rho}\rho\frac{\partial}{\partial \rho}
+
\frac{1}{\rho^2}\frac{\partial^2}{\partial \phi^2}
\right)H(\rho,\phi)
=
\kappa^2 J(\rho,\phi)
\ee
expressed in terms of the Laplacian in the plane polar coordinates.
The angular variable can be separated by considering   $H=h+\bar{h}$ with
\be
h(\rho,\phi)
=
\sum_{m}h_m(\rho)e^{-im(\phi-\phi_m)}
\ee
and also
 \be
  J(\rho,\phi)=\sum^{\infty}_{m=1}j_m(\rho)\cos[m(\phi-\phi_m)]
\ee
with an arbitrary phase angle $\phi_m$ for each $m$ mode accordingly as well. It is convenient to define the complex source $j_m(\rho)e^{-im(\phi-\phi_m)}$.
For each $m$ mode, the function $f$ is of the form
\be
h_m(\rho)=
\begin{cases}
&a_0-b_0\ln\rho,\qquad \qquad m=0,\\
&a_m\rho^m+b_m\rho^{-m},\qquad m\geq1.
\end{cases}
\ee
The terms with $a_m=0$ for all $m$ will be discarded, assuming that $h_m(\rho)\sim\rho^{-m}$. Thus, for the functions $h_m$, one has the following relations
\be\label{h-recurrence}
h_{m+1}=-mh_m',\qquad h_{m+1}=\rho^{-1}h_m.
\ee
The differential equation that $h_m$ satisfy in this case can be used to derive the recurrence relation that $j_m$'s satisfy.
Explicitly, by taking the derivative of the equation
\be
h_m''+\frac{1}{\rho}h_m'-\frac{m^2}{\rho^2}h_m=j_m
\ee
one finds
\be
h_m'''+\frac{1}{\rho}h_m''-\frac{m^2}{\rho^2}h_m'-\frac{1}{\rho^2}h_m'+\frac{2m^2}{\rho^3}h_m=j'_m.
\ee
By making use of the recurrence relations (\ref{h-recurrence}), and by grouping the first three terms and the remaining two terms on the left hand side together, this equation  can be rewritten in the form
\be
-m\left(h_{m+1}''+\frac{1}{\rho}h_{m+1}'-\frac{m^2}{\rho^2}h_{m+1}\right)+\frac{m}{\rho^2}h_{m+1}+\frac{2m^2}{\rho^2}h_{m+1}=j'_m.
\ee
Finally, by identifying all the terms on the hand side left in terms of $j_{m+1}$, one finds the key recurrence relation
\be
j_{m+1}=-\frac{1}{m}j_m
\ee
for $m\geq1$. Consequently, this recurrence relation allows one to generate  $j_m$ from $j_1$ as
\be
j_m
=
\frac{(-1)^m}{(m-1)!}\frac{d^{m-1}}{d\rho^{m-1}}j_1.
\ee
Furthermore, one can make use of  the relation $j_1=-j_0'$ to have
\be
j_m
=
\frac{(-1)^m}{(m-1)!}\frac{d^{m}}{d\rho^{m}}j_0.
\ee

On the wavefronts $u=0$, one has the null particle with the multipole structure of the superposition form as
\be
J
=
-\frac{b_0}{4}\delta(\rho)
-
\sum^{\infty}_{m=1}\frac{b_m}{4}\frac{(-1)^m}{(m-1)!}\cos[m(\phi-\phi_m)]\delta^{(m)}(\rho)
\ee
where $\delta^{(m)}(\rho)$ stands for the $m^{th}$ derivative of the Dirac delta function with respect to $\rho$.
The explicit form of the profile function of the source is then given by
\be
H(\rho,\phi)
=
-{b_0}\ln\rho-\sum^{\infty}_{m=1}b_m\rho^{-m}\cos[m(\phi-\phi_m)]\delta(u).
\ee
The $m$th component describes a null particle with multipole solution of order $m$ located at the wavefront and that a superposition of such solutions
are also allowed to construct  solutions with a desired multipole structure.

The solutions constructed in this subsection constitute generalization of the Aichelburg-Sexl solution \cite{aichelburg-sexl} which can be considered as the
metric with a single null particle located at the origin of the plane wavefront with $h(u, \zeta)\sim \delta(u)\ln\zeta$ component in the general series solution 
(\ref{h-expansion}) for the profile function.

\subsection{Spinning null fluids: Gyratons}

The original metric introduced by Brinkman in 1925 \cite{brinkmann} was in fact of the form
\be\label{brinkmann-original-form}
g=-du\ot dv-dv\ot du-2H(u,x^j)du\ot du+a_i(u,x^j)(dx^i\ot du+du\ot dx^i)+\delta_{ij}dx^i\ot dx^j
\ee
with the metric components $a_i(u,x^j)$ corresponding to cross-terms present.
However, note that it is possible transformed the cross-terms away by an appropriate  coordinate transformation locally.
The coordinates for the flat transverse space are denoted by $\{x^j\}$.
Bonnor \cite{bonnor-spinning-null-fluid} first studied the axially symmetric  $pp$-waves with null fluid sources having an intrinsic
spin.  Such a class of metrics   was also studied by Griffiths \cite{griffiths-neutrino} to describe the gravitational field of  a beam of neutrinos.
Much later, the family of the metrics of the form (\ref{brinkmann-original-form}) describing the gravitational field of
null particles with  localized spin has been reconsidered by Frolov and his collaborators \cite{frolov-fursaev,frolov-israel-zelnikov} who also introduced the
name "gyratons" for such metrics. Around at the same time, Frolov and Lin \cite{frolov-lin} presented the gyratonic gravitational wave solutions in supergravity.  The charged gyratons have also been studied by Frolov and Zelnikov \cite{charged-gyratons}.
The gyratonic  metrics in various backgrounds  have been studied by Kadlecov\'{a} and her collaborators 
\cite{kadlecova-et-al,kadlecova-melvin,kadlecova-phd-thesis}. \v{S}varc and Podolsk\`{y}  proved \cite{svarc-podolsky-absence} that the family of Robinson-Trautman type N metrics with does not admit gyratonic wave solutions recently.

The introductory presentation below closely follows the work \cite{podolsky-steinbauer-svarc} by Podolsk\`y, Steinbauer and \v{S}varc studying  the impulsive limit of the gyratonic $pp$-waves.

It is convenient to adopt plane polar coordinates $\rho,\phi$ for the transverse plane for which the metric takes the form
\be\label{gyraton-metric-polar-form}
g=-du\ot dv-dv\ot du-2H(u,\rho,\phi)du\ot du-J(du\ot d\phi+d\phi\ot du)+d\rho\ot d\rho+\rho^2d\phi\ot d\phi
\ee
where the metric functions are $H=H(u,\rho,\phi)$ and $J=J(u,\rho,\phi)$. The metric function $J$, which is responsible for the frame dragging effects,  is related to the angular momentum of the gyratonic source and cannot be gauged away globally but only locally. The integrability conditions for $J$ requires $J_\rho=0$.
Specifically for the metric (\ref{gyraton-metric-polar-form}), the line integral 
\be
\frac{1}{4\pi}\oint_{C}J(u,\phi)d\phi
\ee
gives the angular momentum of the source up to a constant multiple  \cite{podolsky-steinbauer-svarc}.

A suitable set of coframes that allows  to compare the related formula is of the form
\be\label{gyraton-coframe}
k=du,\qquad l=dv+Hdu+Jd\phi,\qquad m=\frac{1}{\sqrt{2}}\left(d\rho+i\rho d\phi\right)
\ee
whereas
the set of frame fields associated to the coframe take the form
\be
\Delta
=
-(\partial_u-H\partial_v),\qquad
D
=
-\partial_v,
\qquad
{\delta}
=
\frac{1}{\sqrt{2}}
\left[
\partial_\rho
+
\frac{i}{\rho}
\left(
\partial_\phi
-
\sqrt{2}J\partial_v\right)
\right].
\ee
The frame fields are especially useful in writing the expressions for the curvature and the connection forms in a compact manner.
In computations, the expressions for $\delta$ and $\bar{\delta}$ simplifies due to the fact that $\partial_v$ is a  null Killing
vector field. The non-vanishing connection 1-forms corresponding to the above coframe are
\begin{align}
\omega^{1}_{\fant{a}2}
&=
\left(
\bar{\delta}H+\frac{i}{\sqrt{2}\rho}J_u
\right)k
+
\frac{i}{2\rho}J_\rho \bar{m},
\\
\omega^{3}_{\fant{a}3}
&=
\frac{i}{2\rho}J_\rho k -\frac{1}{\sqrt{2}\rho}(m-\bar{m}).
\end{align}
The Cartan's second structure equations the lead to the following expressions  for the components of the curvature 2-forms 
\begin{align}
\Omega^{1}_{\fant{a}2}
&=
\left[
-\rho\bar{\delta}\left(\frac{1}{\rho}
\bar{\delta}H
\right)
+
\frac{i}{\sqrt{2}}\delta\left(\frac{1}{\rho} J_u\right)
-
\frac{i}{2\rho^2}J_{u\phi}
\right]k\wdg m
\nonumber\\
&\fant{=}
-
\left[
\frac{1}{\rho}\delta\left(\rho\bar{\delta}H\right)
-
\frac{1}{2\rho^2}J_{u\phi}
+
\frac{1}{4}\left(\frac{1}{\rho}J_\rho\right)^2
\right]
k\wdg \bar{m}
+
\frac{i}{2}\bar{\delta}\left(\frac{1}{\rho}J_\rho\right)m\wdg\bar{m}
\\
\Omega^{3}_{\fant{a}3}
&=
-
\frac{i}{2}\bar{\delta}\left(\frac{1}{\rho}J_\rho\right)k\wdg m
-
\frac{i}{2}\delta\left(\frac{1}{\rho}J_\rho\right)k\wdg\bar{m}\label{gyraton-curv-forms}
\end{align}
for the connection 1-forms above. As before, the curvature spinors can be read off from the curvature expressions as
\begin{align}
\Phi_{22}
&=
-\frac{1}{2}\left(
\frac{1}{\rho}\frac{\partial}{\partial \rho}\rho\frac{\partial}{\partial \rho}+\frac{1}{\rho^2}\frac{\partial^2}{\partial \phi^2}
\right) H
+\frac{1}{2\rho^2}J_{u\phi}-\frac{1}{4}\left(\frac{1}{\rho}J_\rho\right)^2
\\
\Phi_{21}
&=
-\Psi_3
=
\frac{i}{4}\bar{\delta}\left(\frac{1}{\rho}J_\rho\right)
\\
\Psi_4
&=
-\rho\bar{\delta}\left(\frac{1}{\rho}\bar{\delta}H\right)
+
\frac{i}{\sqrt{2}}\delta\left(\frac{1}{\rho}J_u\right)
-
\frac{i}{2\rho^2}J_u.
\end{align}

As a consequence of the cross term in the metric ansatz (\ref{brinkmann-original-form}), it  is not of Petrov type N in general unless the Weyl spinor 
$\Psi_3=-\Phi_{12}$ vanishes identically. Consequently, for an appropriate source term the gyraton metric (\ref{brinkmann-original-form}) is  
of Petrov type III.

The Einstein 3-forms corresponding to the curvature 2-forms (\ref{gyraton-curv-forms}) are
\begin{align}
*G^1
&=
-\left[
\left(
\frac{1}{\rho}\frac{\partial}{\partial \rho}\rho\frac{\partial}{\partial \rho}+\frac{1}{\rho^2}\frac{\partial^2}{\partial \phi^2}\right) H
-
\frac{1}{\rho^2}J_{u\phi}
+
\frac{1}{2}\left(\frac{1}{\rho}J_\rho\right)^2
\right]
*k
\nonumber\\
&\fant{=}-
\frac{1}{2\sqrt{2}}\left[
i\left(\frac{1}{\rho}J_\rho\right)_\rho
+
\frac{1}{\rho}\left(\frac{1}{\rho}J_\rho\right)_\phi
\right]*m
-
\frac{1}{2\sqrt{2}}\left[
-
i\left(\frac{1}{\rho}J_\rho\right)_\rho
+
\frac{1}{\rho}\left(\frac{1}{\rho}J_\rho\right)_\phi
\right]*\bar{m},
\label{gyraton-einstein-forms1}\\
*G^2
&=
\frac{1}{2\sqrt{2}}\left[
-
i\left(\frac{1}{\rho}J_\rho\right)_\rho
+
\frac{1}{\rho}\left(\frac{1}{\rho}J_\rho\right)_\phi
\right]
*k,\label{gyraton-einstein-forms2}
\end{align}
and the expression for the $*G^3$ can be obtained by the complex conjugate of that of $*G^2$.
As an important consequence of the off-diagonal metric component $J$ in the ansatz, one now has an additional nonvanishing component of the Einstein tensor, namely 
$G^1_{\fant{a}2}=-\bar{G}^{2}_{\fant{a}0}$. 

For the energy-momentum tensor field to be consistent with the metric ansatz for a spinning null fluid (\ref{gyraton-metric-polar-form}), the general, traceless
energy-momentum 1-forms are necessarily have the form
\be\label{gyratonic-energy-mom-null-form}
*T^1=-\varrho *k-\mathcal{J} *m-\bar{\mathcal{J}} *\bar{m},\qquad *T^2=\bar{\mathcal{J}} *k
\ee
in terms of a complex function $\mathcal{J}\equiv j_1+ij_2$ (here the numerical subscripts 1 and 2 to $j$ are just labels for the real and the imaginary parts  of $\mathcal{J}$)  and the energy density $T_{00}=\varrho$. By definition $-\sqrt{2}j_1$ and $\sqrt{2}j_2$ then correspond to the components of the energy-momentum 
tensor relative to the orthonormal coframe componets $\theta^2$ and $\theta^3$ defined by $m=2^{-1/2}(\theta^2+i\theta^3)$, respectively.
For $\mathcal{J}=0$, the energy momentum tensor (\ref{gyratonic-energy-mom-null-form}) representing a gyratonic source relative to a complex null NP coframe reduce to that of a pure radiation with energy density $\varrho$.  The only restriction on the components of the gyratonic energy-momentum is that they have to satisfy the constraint $D*T^a=0$ for the consistency of the field equations with the first Bianchi identity.

Eventually, the Einstein's field equations using the Einstein 3-forms (\ref{gyraton-einstein-forms1})-(\ref{gyraton-einstein-forms2}) with the gyratonic energy-momentum form source terms (\ref{gyratonic-energy-mom-null-form}) reduce to the following system of differential equations:
\be\label{gyraton-field-eqn1}
\left(
\frac{1}{\rho}\frac{\partial}{\partial \rho}\rho\frac{\partial}{\partial \rho}+\frac{1}{\rho^2}\frac{\partial^2}{\partial \phi^2}\right) H
-
\frac{1}{\rho^2}J_{u\phi}
+
\frac{1}{2}\left(\frac{1}{\rho}J_\rho\right)^2
=
\kappa^2 \varrho,
\ee
and 
\be\label{gyraton-field-eqn2}
\frac{1}{2}\left(\frac{1}{\rho}J_\rho\right)_\rho
=
-\kappa^2 j_2,\qquad
\frac{1}{2\rho}\left(\frac{1}{\rho}J_\rho\right)_\phi
=
\kappa^2 j_1
\ee
for the metric functions $H$ and $J$. 

The field equations for the exterior of the source is obtained by setting $\varrho=0$, $j_1=0$ and $j_2=0$ in (\ref{gyraton-field-eqn1}) 
and (\ref{gyraton-field-eqn2}). In this case the field equations
(\ref{gyraton-field-eqn2}) imply that
$
\rho^{-1}J_\rho
$
is a function of the coordinate $u$ only and that $J$ has the general form
\be
J(u,\rho,\phi)
\equiv
\rho^2\omega(u)+\chi(u,\phi)
\ee
for some functions $\omega(u)$ and $\chi(u,\phi)$. Using this result and by defining 
\be
\tilde{H}
\equiv
H
-\frac{1}{16}\omega^2\rho^2,
\ee
 (\ref{gyraton-field-eqn1}) can be rewritten as a Poisson equation of the form
\be\label{profile-eqn-gyraton}
\left(\frac{1}{\rho}\frac{\partial}{\partial \rho}\rho\frac{\partial}{\partial \rho}+\frac{1}{\rho^2}\frac{\partial^2}{\partial \phi^2}\right) \tilde{H}
=
\frac{1}{\rho^2}\chi_{u\phi}.
\ee
For axially symmetric source with $\chi_\phi=0$, (\ref{profile-eqn-gyraton}) further reduces to a Laplace's equation in the transverse planes.
In this particular case, one has $\Psi_3=\Phi_{12}=0$ and that the metric ansatz is of Petrov Type N.
The general vacuum solutions to the field equations for the gyraton ansatz have been discussed in detail in \cite{frolov-fursaev,frolov-israel-zelnikov}
in four and higher dimensions using Euclidean coordinates for the transverse planes.

\section{Rosen form of the \texorpdfstring{$pp$}{}-wave metric}

Although the Kerr-Schild (or equivalently, the Brinkmann) form of the $pp$-wave metric is more commonly used and convenient  in many cases of interest, for example, in a discussion of the Penrose limits, the Rosen coordinates  turn out to be  essential 
(See, for example, the self-contained, and elegant review on the Penrose limits by Blau \cite{blau}). The coordinate singularities that arise in the Rosen form originally lead to some confusion about the gravitational wave metrics \cite{einstein-rosen,kennefick-speed-of-thought,kennefick-phys-today}.

\subsection{The Rosen form of \texorpdfstring{$pp$}{}-wave metric: Connection and curvature expressions relative to a coordinate coframe}

The general $pp$-wave metric can be expressed  in the Rosen form as
\be\label{general-rosen-form}
g=
-du\ot dv-dv\ot du+g_{AB}(u)dx^A\ot dx^B
\ee
where the capital Latin indices  run over the space coordinates. The metric coeffcients $g_{AB}$
are functions of the real null coordinate $u$ only and the transverse character of the metric is explicit.
For the metric of the form (\ref{general-rosen-form}),  the nonvanishing Christoffel symbols and the corresponding connection 1-forms relative to the natural coframe with the coordinates labeled as $x^a=\{u, v, x^A\}$, can be calculated from the general formula
\be
\Gamma^{a}_{bc}
=
\frac{1}{2}g^{ad}\left(\partial_{b} g_{dc}+\partial_{c} g_{bd}-\partial_{d} g_{bc}\right)
\ee
where the lower-case Latin indices $a,b,\ldots $ run over the range $0, 1, 2, 3$, whereas the capital Latin indices $A, B, \ldots $ run over the range $2, 3$. 
The nonvanishing  Levi-Civita connection 1-forms, in terms of Christoffel symbols expressed as $\Gamma^{a}_{\fant{a}b}=\Gamma^{a}_{cb}dx^c$,  turn out to be  of the form
\be
\Gamma^{1}_{\fant{a}A}=\frac{1}{2}g'_{AB}dx^B,\qquad \Gamma^{A}_{\fant{a}0}=g^{AB}\Gamma^{1}_{\fant{a}B},
\qquad  \Gamma^{A}_{\fant{a} B}=\frac{1}{2}g^{AC}g'_{CB}du,
\ee
where a prime denotes  partial derivative with respect to the null coordinate $u$. One can show that $\Omega^{0}_{\fant{A}0}=\Omega^{0}_{\fant{A}A}=\Omega^{A}_{\fant{B}B}=0$ and the only nonvanishing curvature 2-form is $\Omega^{1}_{\fant{a}A}$. The components  of the curvature 2-form $\Omega^{1}_{\fant{a}A}$ can readily be calculated from Cartan's second structure equation as 
\begin{align}
\Omega^{1}_{\fant{a}A}
&=
d\Gamma^{1}_{\fant{a}A}+\Gamma^{1}_{\fant{a}B}\wdg \Gamma^{B}_{\fant{a} A}
\nonumber\\
&=
\frac{1}{2}\left(
g''_{AB}-\frac{1}{2}g'_{AC}g^{CD}g'_{DB}
\right)du\wdg dx^B.\label{rosen-curvature-form}
\end{align}
Consequently, the nonvanishing components of the Riemann tensor relative to the coordinate coframe are of the form
\be
R_{0A0B}
=
-
\frac{1}{2}\left(
g''_{AB}
-
\frac{1}{2}g'_{AC}g^{CD}g'_{DB}
\right)
\ee
expressed entirely in terms the derivatives of the two-dimensional metric coefficients $g_{AB}$ and its inverse $g^{AB}$ \cite{exact-sol-SKMHH}.
The scalar curvature $R=R^{ab}_{\fant{ab}ab}$ vanishes identically and there is only one nonvanishing component of the Ricci tensor 
corresponding to $R_{00}=g^{ab}R_{0a0b}=g^{AB}R_{0A0B}$ and therefore Ricci 1-form can be written as
\be\label{Ricci-rosen-form}
R^1
=
i_A\Omega^{A1}
=
\frac{1}{2}\left(
g^{AB}g''_{AB}-\frac{1}{2}g^{AB}g'_{BC}g^{CD}g'_{DA}
\right)du
\ee
where $i_A$ denotes a contraction operator with respect to the natural basis frame field $\partial_A$, namely, $i_A\equiv i_{\partial_A}$ and 
$\Omega^{1A}=g^{AB}\Omega^{1}_{\fant{b}B}$. The expression for the Ricci tensor (\ref{Ricci-rosen-form}) is in accordance with the corresponding expressions relative to an orthonormal coframe.

As a side remark, note that the curvature components (\ref{rosen-curvature-form}) can be rewritten formally as a covariant exterior derivative 
of the form 
\be\label{rosen-covariant-der}
\Omega^{1A}
=
D^{(2)}\Gamma^{1A}
\ee
with the covariant exterior derivative $D^{(2)}$ defined in terms of the Levi-Civita connection 1-forms $\Gamma^{A}_{\fant{a}B}$ as 
\be
D^{(2)}
\equiv
d+
\Gamma^{A}_{\fant{A}B}.
\ee
Consequently, the Ricci 1-form can be expressed as a contraction of the covariant exterior derivative (\ref{rosen-covariant-der}) and the vacuum field equations 
can be expressed as
\be
i_AD^{(2)}\Gamma^{1A}=0
\ee   
which corresponds to the coordinate coframe version of the expression (\ref{simplified-einstein-form-for-pp}) for the Einstein 3-form.

The general coordinate transformations relating the Kerr-Schild form to the Rosen form of the $pp$-wave metric are given in \cite{exact-sol-SKMHH}.
For the particularly simple  case of the plane wave metrics, the coordinate transformation relating the two  equivalent form of the metrics
is given below after calculating its curvature relative to a null coframe in the following subsection.

\subsection{Connection and curvature forms a plane wave metric in the Rosen form}

A particular plane wave metric of the form,  having a linear polarization corresponding to a + linear polarization mode,  is given by
\be\label{rosen-form-plane-wave+pol}
g=-dt\ot dt+p^2dx\ot dx+q^2dy\ot dy+dz\ot dz
\ee
where the metric coefficients $p, q$ are assumed to be functions of the real null coordinate $u$ and the real null coordinates are  $u=\frac{1}{\sqrt{2}}(t-z)$ and $v=\frac{1}{\sqrt{2}}(t+z)$. It is convenient to adopt the following set of basis null coframe 1-forms
\be
k=du,\qquad l=dv,\qquad m=\tfrac{1}{\sqrt{2}}\left(pdx+iqdy\right).
\ee
The nonvanishing connection 1-form relative to the null coframe is
\be
\omega^{1}_{\fant{a}2}
=
\frac{1}{2}\left(
\frac{p'}{p}
-
\frac{q'}{q}
\right)m
+
\frac{1}{2}\left(
\frac{p'}{p}
+
\frac{q'}{q}
\right)\bar{m}
\ee
where the prime denotes a derivative with respect to the null coordinate $u$.
The corresponding curvature 2-form turns out to be of the form
\be
\Omega^{1}_{\fant{a}2}
=
\frac{1}{2}\left(
\frac{p''}{p}
+
\frac{q''}{q}
\right)k\wdg \bar{m}
+
\frac{1}{2}\left(
\frac{p''}{p}
-
\frac{q''}{q}
\right)k\wdg m
\ee
where one can read off the nonvanishing curvature spinors as
\be
\Phi_{22}
=
-
\frac{1}{2}\left(
\frac{p''}{p}
+
\frac{q''}{q}
\right),
\qquad
\Psi_4
=
-
\frac{1}{2}\left(
\frac{p''}{p}
-
\frac{q''}{q}
\right).
\ee
Note that Weyl spinor $\Psi_4$ turns out to be have a  real magnitude indicating that the metrics of the form (\ref{rosen-form-plane-wave+pol}) 
have  + linear polarization mode. However, for the general Rosen form (\ref{general-rosen-form}), it is not straightforward to decouple 
the linear + and $\times$ polarization  modes \cite{cropp-visser}. 

 The only nonvanishing Einstein 3-form  becomes
\be
*G^1
=
\left(
\frac{p''}{p}
+
\frac{q''}{q}
\right)*k.
\ee
Evidently,  as it is  the case for the profile function in the Brinkmann form, the metric coefficients $p(u), q(u)$ are  not fully determined by the Einstein field equations.

An example of a conformally flat, electrovacuum $pp$-wave solution presented by Brdicka in 1951 \cite{brdicka} can be expressed
by specifying the metric coefficients defined in this section as
\be
p^2=1+\sin(\omega u),\qquad q^2=1-\sin(\omega u)
\ee
where $\omega$ is a constant.
In this case one has
\be
\frac{p''}{p}=\frac{q''}{q}=-\tfrac{1}{4}\omega^2
\ee
and consequently, Weyl curvature spinors vanish identically and at the same time requiring a constant null  electromagnetic component $\Phi_{2}$.

Another interesting example of $pp$-waves in the Rosen form is defined by the metric functions of the form
\be\label{cut-paste-composite}
p=1-au\Theta(u),\qquad q=1+au\Theta(u)
\ee
where $\Theta(u)$ stands for the Heaviside step function. The corresponding metric  describes an impulsive gravitational plane wave in vacuum with impulsive curvature components $\Psi_4=a\delta(u)$.  

As a side remark regarding  the impulsive wave metric defined by (\ref{cut-paste-composite}), note that the Penrose's ``Cut and Paste" method of constructing  impulsive waves in Minkowski background is carried out first by cutting the Minkowski spacetime along a null hyperplane and then reattaching the two halves 
by a suitable warp along the cut \cite{penrose-cut-paste}.
The warp is introduced by a transformation of the coordinates $\{u, v, \zeta,\bar{\zeta}\}$ of the Minkowski spacetime defined in terms of an arbitrary
function \cite{exact-sol-griffiths-podolsky}. By making use of the Penrose's junction conditions one can show that the original and the transformed coordinates can be combined into an expression of type given in (\ref{cut-paste-composite}). 

Finally, note also that one can rotate the coordinates of the transverse planes by an angle of amount $\pi/4$ to have linear $\times$ polarization. In terms of the rotated coordinates, the metric takes the form 
\be
g=
-dt\ot dt+\frac{1}{2}\left(p^2+q^2\right)\left(dx\ot dx+dy\ot dy\right)+\left(p^2-q^2\right)dx\ot dy+dz\ot dz
\ee
without altering  the equation satisfied by the metric functions $p$ and $q$ as it has been noted above in Section \ref{polarization-sect}.

By making use of  the general Rosen form (\ref{general-rosen-form}), the plane gravitational waves having more general  polarization modes have been discussed 
 by Cropp and Visser \cite{cropp-visser}  recently.

\subsection{Coordinate transformation from the Rosen form to the Kerr-Schild form}

In this subsection the coordinate transformation  relating the  Rosen form to Kerr-Schild form is briefly discussed for a particular type of plane wave 
metrics (\ref{rosen-form-plane-wave+pol}).

First, in order to have a flat metric for transverse planes, it is convenient to introduce the new spatial coordinates as
\be
X=px,\qquad Y=qy.
\ee
Consequently, by omitting the tensor product for convenience of the notation and calculating the coordinate differentials one finds
\be
p^2dx^2
+
q^2dy^2
=
dX^2+dY^2-(p'^2x^2+q'^2y^2)du^2
-
2xpp'dudx-2yqq'dudy.
\ee
By inspecting these  expressions, one can see that it is convenient to introduce the following new real null coordinates
\be
U=u,\qquad V=v+\tfrac{1}{2}x^2pp'+\tfrac{1}{2}y^2qq'.
\ee
Similarly, by calculating the coordinate differentials for the real null coordinates, one finds
\be
2dudv
=
2dUdV
-
2xpp'dudx-2ydudy
-
\left(
X^2\frac{p''}{p}+Y^2\frac{q''}{q}
+
x^2p'^2+y^2q'^2
\right)du^2.
\ee
By combining these results the cross terms are canceled out and  one eventually ends up with
\be
g=
-dU\ot dV-dV\ot dU
+\left(
X^2\frac{p''}{p}+Y^2\frac{q''}{q}\right)dU\ot dU+dX\ot dX+dY\ot dY.
\ee
For the metric of the form given in Eq. (\ref{rosen-form-plane-wave+pol})  satisfying the vacuum Einstein field equations, the metric coefficients $p,q$ are not independent and consequently, in this case the metric can be rewritten in the more familiar Kerr-Schild form as
\be
g=
-dU\ot dV-dV\ot dU
-
\frac{q''}{q}
\left(
X^2-Y^2\right)dU\ot dU+dX\ot dX+dY\ot dY
\ee
where the profile function can be identified as
\be\label{transformed-profile-funct}
H(U, X, Y)
=
\frac{q''}{2q}
\left(
X^2-Y^2\right).
\ee
Note that the expression (\ref{transformed-profile-funct}) for the profile function is consistent with the fact that the function $q(u)$ is not  determined by the field equations.

\subsection{Rosen form of the \texorpdfstring{$pp$}{}-wave metric: Connection and curvature expressions relative to a  semi-null coframe}

It is always possible to introduce a set of orthonormal basis frame fields $\theta^{A}$ for $A=2,3$  to the space part  of the metric (\ref{general-rosen-form}), 
so that it takes the form
\be\label{rosen-form-hybrid-coframe}
g=
\eta_{ab}\theta^a\ot \theta^b
=
-\theta^{0}\ot \theta^{1} -\theta^{1}\ot \theta^{0} +\delta_{AB}\theta^A\ot \theta^B,
\ee
where  $\theta^0=du$, $\theta^1=dv$ and the capital Latin indices  $A, B$ only assume the numerical values 2 and 3 relative to the hybrid coframe defined in 
(\ref{rosen-form-hybrid-coframe}). Then the constant metric components  $\eta_{ab}$ in (\ref{rosen-form-hybrid-coframe}) can be written in the form of a matrix as
\be
\eta_{ab}
=
\begin{pmatrix}
  \begin{matrix}
  0 & -1 \\
  -1 & 0
  \end{matrix}
  & \rvline & \bigzero \\
\hline
  \bigzero & \rvline &
  \begin{matrix}
  +1 & 0 \\
  0 & +1
  \end{matrix}
\end{pmatrix}.
\ee
The semi-null coframe basis above has been previously used by Obukhov \cite{obukhov-kundt} in a  study of Kundt class of 
metrics generalized to arbitrary dimensions also taking a cosmological constant into account. 

Since the metric components with the space indices depend only on the null coordinate $u$, $g_{AB}=g_{AB}(u)$, it is possible to write
\be\label{f_AB-def}
d\theta^A\equiv -f^{A}_{\fant{A}B}\theta^{0B}
\ee
for some functions $f^{A}_{\fant{A}B}$ of the coordinate $u$. By definition, the indices of $f_{AB}$ and $f^{AB}$ are raised/lowered by two dimensional flat metric
$\delta_{AB}$ and $\delta^{AB}$ so that $f^{A}_{\fant{A}B}=f_{AB}=f^{AB}$. However,  one has $f_{AB}\neq f_{BA}$ in general.

The Cartan's first structure equations can be solved for the connection 1-forms relative to the hybrid coframe,   and one finds that 
$\omega^{0}_{\fant{0}0}=0=\omega^{0}_{\fant{a}A}$, and that the nonvanishing connection 1-forms can be written in the form
\begin{align}
\omega^{1}_{\fant{A}A}
&=
-\frac{1}{2}
\left(
f_{AB}+f_{BA}
\right)\theta^B,
\label{hybrid-conn1}\\
\omega^{A}_{\fant{A}B}
&=
+\frac{1}{2}
\left(
f^{A}_{\fant{A}B}-f_{B}^{\fant{B}A}
\right)\theta^0.\label{hybrid-conn2}
\end{align}

By feeding the expressions for the connection 1-forms (\ref{hybrid-conn1}), (\ref{hybrid-conn2}) and the exterior derivative (\ref{f_AB-def}) into the  Cartan's structure equations
\be
\Omega^{1}_{\fant{a}A}
=
d\omega^{1}_{\fant{a}A}
+
\omega^{1}_{\fant{a}B}\wdg \omega^{B}_{\fant{a}A},
\ee
one finds
\be
\Omega^{1}_{\fant{a}A}
=
r_{AB}\theta^{0}\wdg \theta^B
\ee
with
\be\label{r_AB-def}
r_{AB}
\equiv
-\frac{1}{2}
\left[
f'_{AB}+f'_{BA}
-
\left(f_{AC}+f_{CA}\right)f^{C}_{\fant{A}B}
+
\frac{1}{2}
\left(f_{AC}-f_{CA}\right)
\left(f^{C}_{\fant{B}B}+f_{B}^{\fant{A}C}\right)
\right],
\ee
where  a prime denotes a derivative with respect to coordinate $u$.

The remaining curvature 2-forms vanish identically: 
$\Omega^{0}_{\fant{0}A}=0=\Omega^{A}_{\fant{A}B}$. As a result, the only nonvanishing component of the corresponding Riemann tensor turns out to be of the form
\be
R^{1}_{\fant{a}A0B}
=
r_{AB}.
\ee
The symmetry of the Riemann tensor with respect to exchange of the first and the second pair of indices requires $r_{AB}=r_{BA}$.
The only contraction of the resulting Riemann tensor is $R^{1A}_{\fant{1A}0A}=r^{A}_{\fant{A}A}\equiv r$  and therefore, $R=0$ identically as expected. 
The  nonvanishing Ricci 1-form can be written as $R^1=i_a\Omega^{a1}=i_A\Omega^{A1}=r(u) du$ (cf. equation (\ref{simplified-einstein-form-for-pp})) with the function $r(u)$ having the explicit form 
\be\label{rosen-hybrid-ricci}
r
=
-
f'^{A}_{\fant{A}A}+\frac{1}{2}f_{AB}\left(f^{AB}+f^{BA}\right).
\ee

The expression in equation (\ref{rosen-hybrid-ricci}) is the semi-null coframe version of the general formula (\ref{Ricci-rosen-form}) for the coordinate coframe. Once a convenient set of orthonormal coframes $\{\theta^{A}\}$  is introduced for the Rosen form of the metric (\ref{general-rosen-form}),  an explicit expression for(\ref{rosen-hybrid-ricci}) can be  obtained simply by calculating the exterior derivatives of the basis coframe 1-forms.

Moreover,  the components of the Weyl 2-form relative to the semi-null coframe (\ref{rosen-form-hybrid-coframe}) can readily be found. 
By making use of the general curvature expression (\ref{curvature-general-expansion}), one finds
\be
C^{1}_{\fant{a}A}
=
\left(
r_{AB}-\frac{1}{2}\delta_{AB}r
\right)
\theta^{0B},
\ee
and the only nonvanishing component of the Weyl tensor is
\be\label{hybrid-weyl-form1}
C^{1}_{\fant{a}A0B}
=
r_{AB}-\frac{1}{2}\delta_{AB}r.
\ee

The trace function $r(u)$ is determined by the Einstein's field equations, and in particular  
for the vacuum, one has $r=0$. Consequently, the expression in (\ref{hybrid-weyl-form1}) further reduces to  
\be\label{hybrid-weyl-form2}
C^{1}_{\fant{a}A0B}
=
r_{AB}
\ee
in this case.

The expression in (\ref{hybrid-weyl-form1})  for the Weyl tensor  allows one to find the corresponding  Weyl spinor $\Psi_4$ in terms of $f_{AB}$ and its derivatives by replacing the $\theta^2$ and $\theta^3$  with the complex conjugate coframe basis fields $m$ and $\bar{m}$. 
Using the above expressions for the Weyl 2-form, one can show that relative to the complex null coframe basis 
\be
k=du,\quad l=dv,\quad m=\frac{1}{\sqrt{2}}\left(\theta^2+i\theta^3\right),\quad \bar{m}=\frac{1}{\sqrt{2}}\left(\theta^2-i\theta^3\right)
\ee
which are defined in terms of the semi-null coframe (\ref{rosen-form-hybrid-coframe}), is of the form
\be\label{null-rotated-curv-exp}
\Omega^{1}_{\fant{a}2}
=
\frac{1}{2}\left(r_{22}+r_{33}\right)k\wdg \bar{m}+\left[\frac{1}{2}\left(r_{22}-r_{33}\right)-ir_{23}\right]k\wdg m.
\ee
Eventually, the real and imaginary parts of $\Psi_4$  ($k\wdg m$ component of $\Omega^{1}_{\fant{a}2}$ above), which determine the + and $\times$ polarization modes, are expressed in terms of the real  functions $r_{AB}(u)$ defined in (\ref{r_AB-def}).  In general,  this result implies that
a given Rosen form of a pp-wave metric has a superposition of + and $\times$ polarization modes \cite{cropp-visser}, given explicitly by the expression 
in square bracket in (\ref{null-rotated-curv-exp}).

\section{Petrov type N gravitational waves in curved backgrounds}

It is also possible to construct Petrov type N gravitational waves with various background spacetimes. For these type solutions to the Einstein field equations,  the transverse wavefronts are not usually flat. Constructing such a solution for the modified gravitational models are  are more difficult  and thus  a straightforward  extension of such a solution studied in this section to modified gravitational models studied in the second part is technically more difficult.

\subsection{AdS-waves}\label{ads-waves}

It was shown by Siklos in 1985 \cite{siklos} that the only Einstein spacetimes that are conformal to the $pp$-waves are so-called AdS-wave  spacetimes.
The resulting gravitational wave metric is also referred as the Siklos metric.
The Siklos metric  is a particular example of the family of metrics introduced by Ozv\`{a}th et al. 
\cite{ozvath-robinson-rozga} and by Garci\'a and Pleba\'nski \cite{garcia-plebanski,salazar-garcia-plebanski}.

The Siklos metric is in a form that is explicitly conformal to the $pp$-wave metric  in a form analogous to that of the Brinkmann form explicitly reads
\be\label{ads-wave-ansatz}
g=\frac{2a^2}{(\zeta+\bar{\zeta})^2}
\left(
-du\ot dv-dv\ot du-2Hdu\ot du+d\zeta\ot d\bar{\zeta}+d\bar{\zeta}\ot d\zeta
\right).
\ee
For the vanishing profile function, namely $H=0$, the metric (\ref{ads-wave-ansatz}) reduces to the well-known AdS spacetime.

A natural null coframe for the metric (\ref{ads-wave-ansatz}) can be chosen as
\be\label{ads-basis-coframe}
k=\frac{\sqrt{2}a}{\zeta+\bar{\zeta}}du,\qquad l=\frac{\sqrt{2}a}{\zeta+\bar{\zeta}}(dv+Hdu), \qquad m=\frac{\sqrt{2}a}{\zeta+\bar{\zeta}}d\zeta
\ee
for which it can be rewritten in the NP form given in Eq. (\ref{NP-general-form-metric}).

The connection 1-forms and the corresponding curvature 2-forms for the coframe (\ref{ads-basis-coframe}) can be found as
\be
\omega^{0}_{\fant{a}3}
=
-\frac{1}{\sqrt{2}a}k,
\qquad
\omega^{1}_{\fant{a}2}
=
\frac{\zeta+\bar{\zeta}}{\sqrt{2}a}H_\zeta k-\frac{1}{2a}l,
\qquad
\omega^{3}_{\fant{a}3}
=
\frac{1}{\sqrt{2}a}(m-\bar{m})
\ee
and
\be
\begin{split}
\Omega^{0}_{\fant{a}3}
&=
-\frac{1}{a^2}k\wdg m,
\\
\Omega^{1}_{\fant{a}2}
&=
-
\frac{(\zeta+\bar{\zeta})^2}{2a^2}H_{\zeta\zeta} k\wdg m
-
\frac{(\zeta+\bar{\zeta})^2}{2a^2}\left[H_{\zeta\bar{\zeta}}-\frac{1}{\zeta+\bar{\zeta}}(H_{\zeta}+H_{\bar{\zeta}})\right]k\wdg \bar{m}
-
\frac{1}{a^2}l\wdg \bar{m}
\\
\Omega^{0}_{\fant{a}0}
&=
\frac{1}{a^2}k\wdg l,
\qquad
\Omega^{3}_{\fant{a}3}
=
\frac{1}{a^2}m\wdg\bar{m}
\end{split}
\ee
respectively. From the curvature expressions, it follows that the curvature spinors are of the form
\be\label{ads-curvature-spinors}
\begin{split}
\Phi_{22}
&=
\frac{(\zeta+\bar{\zeta})^2}{2a^2}\left[H_{\zeta\bar{\zeta}}-\frac{1}{\zeta+\bar{\zeta}}(H_{\zeta}+H_{\bar{\zeta}})\right]
\\
\Psi_4
&=
\frac{(\zeta+\bar{\zeta})^2}{2a^2}H_{\zeta\zeta},
\qquad
R=
-\frac{12}{a^2}
\end{split}
\ee
where the diagonal components of the curvature 2-form leads to the expression for the scalar curvature $R$.
For the vanishing profile function,  the spacetime has now an  anti-de Sitter metric with $\Omega^{ab}=-a^{-2}\theta^{a}\wdg \theta^b$.
The corresponding Einstein 3-forms are of the form
\be
\begin{split}
*G^0&=\frac{3}{a^2}*k,\\
*G^1&=-\frac{1}{a^2}(\zeta+\bar{\zeta})^2
\left[
H_{\zeta\bar{\zeta}}-\frac{1}{\zeta+\bar{\zeta}}(H_\zeta+H_{\bar\zeta})
\right]*k
+
\frac{3}{a^2}*l,
\\
*G^2&=\frac{3}{a^2}*m.\\
\end{split}
\ee
Consequently, the field equations are satisfied with a cosmological constant $\Lambda=-{3}/{a^2}$ by construction.
Then, the  Einstein-Maxwell  equations with an negative cosmological constant lead to the profile function that satisfies the differential equation
\be\label{ads-profile-funct-eqn}
H_{\zeta\bar{\zeta}}-\frac{1}{\zeta+\bar{\zeta}}(H_{\zeta}+H_{\bar{\zeta}})=\kappa^2 f_\zeta f_{\bar{\zeta}}
\ee
in the notation of the previous section.
For $f=0$, the general homogeneous solution to equation (\ref{ads-profile-funct-eqn}) is of the form \cite{siklos}
\be
H(u,\zeta, \bar{\zeta})
=
\frac{1}{2}(\zeta+\bar{\zeta})(h_{\zeta}+\bar{h}_{\bar{\zeta}})-(h+\bar{h})
\ee
where $h=h(u, \zeta)$ is an arbitrary function analytic in $\zeta.$

\subsection{Impulsive waves with an arbitrary cosmological constant}

Podolsk\'y and Griffiths  \cite{griffiths-podolsky-arbitary-cosmo} showed that the impulsive $pp$-wave metrics constitute an exceptional case for  the Siklos's conclusion that   the only Einstein metrics conformal to the $pp$-wave metrics occur with a negative cosmological constant. They obtained impulsive wave metrics
by generalizing the Penrose's ``cut and paste" method to the case with an  arbitrary cosmological constant.
By making use of the Brinkmann form by using a Dirac delta distribution function introduced by Podolsk\'y and  Griffiths, nonexpanding impulsive gravitational waves  in a maximally  symmetric background   can be studied in a unified manner.

As in the previous case with an adS background, the metric can be taken to be of the form
\be
g=
\frac{1}{P^2}\left(
-du\ot dv-dv\ot du-2H(u,\zeta,\bar{\zeta})du\ot du+d\zeta\ot d\bar{\zeta}+d\bar{\zeta}\ot d\zeta
\right)
\ee
conformal to the $pp$-wave metric (\ref{pp-electro-ansatz}). The conformal factor $P$ in this case is given explicitly as
\be
P=1+\frac{\Lambda}{6}(-uv+\zeta\bar{\zeta})
\ee
with the constant $\Lambda$ can assume arbitrary values. The profile function is assumed to be of the distributional form
involving a delta function
\be\label{impulsive-profile-func}
H(u,\zeta, \bar{\zeta})
=
\delta(u) h(\zeta,\bar{\zeta})
\ee
where the function $h(\zeta,\bar{\zeta})$ is analytic in $\zeta$.

Possible  backgrounds can be obtained by assuming  a desired numerical value for the constant $\Lambda$.
For $\Lambda=0$  one recovers  the flat two dimensional transverse planes for the $pp$-wave metric (\ref{pp-electro-ansatz}) whereas for
the cases $\Lambda>0$ and $\Lambda<0$, one obtains de Sitter and anti-de Sitter backgrounds respectively. For $\Lambda>0$, the wavefront is a sphere $S^2$
whereas for $\Lambda<0$, it is a hyperbolic space $H^2$.
These wavefronts defined by  $u=0$ can described by the two dimensional metric of the explicit form
\be\label{wave-front-metric}
\frac{d\zeta\ot d\bar{\zeta}+d\bar{\zeta}\ot d\zeta}{1+\frac{\Lambda}{6}\zeta\bar{\zeta}}.
\ee
As in the case of the $pp$-waves with the Minkowski background, the Einstein field equations  can be written in terms of the Laplacian operator on the two dimensional  metric (\ref{wave-front-metric}). On the wavefront defined by the two-dimensional metric (\ref{wave-front-metric}), the Laplace operator $\Delta$ is  explicitly  given by
\be\label{laplacian-max-symmetric}
\Delta=2\left(1+\frac{\Lambda}{6}\zeta\bar{\zeta}\right)^2\partial_\zeta\partial_{\bar{\zeta}}.
\ee
 Relative to the basis coframe defined by
\be
k=P^{-1}du,\qquad l=P^{-1}(dv+Hdu),\qquad m=P^{-1}d\zeta
\ee
the connection 1-forms are of the form
\be
\begin{split}
\omega^{0}_{\fant{a}3}
&=
\frac{\Lambda}{6}(-\zeta k+um),
\\
\omega^{1}_{\fant{a}2}
&=
\frac{\Lambda}{6}(-\bar{\zeta} l+v\bar{m})+\left(1+\frac{\Lambda}{6}\zeta\bar{\zeta}\right)H_\zeta k,
\end{split}
\qquad\qquad
\begin{split}
\omega^{0}_{\fant{a}0}
&=
-\frac{\Lambda}{6}(ul-vk),
\\
\omega^{3}_{\fant{a}3}
&=
\frac{\Lambda}{6}(\bar{\zeta}m-\zeta m).
\end{split}
\ee
The corresponding curvature 2-forms take the form
\begin{align}
\Omega^{0}_{\fant{a}3}
&=
\frac{\Lambda}{3}k\wdg m,
\\
\Omega^{1}_{\fant{a}2}
&=
\left(1+\frac{\Lambda}{6}\zeta\bar{\zeta}\right)\left[
-\frac{\Lambda}{6}H+\frac{\Lambda}{6}
\left(
\zeta H_{\zeta}+\bar{\zeta} H_{\bar{\zeta}}
\right)
-\left(1+\frac{\Lambda}{6}\zeta\bar{\zeta}\right)H_{\zeta\bar{\zeta}}
\right]k\wdg \bar{m}
\nonumber\\&-
\left(1+\frac{\Lambda}{6}\zeta\bar{\zeta}\right)H_{\zeta\zeta}k\wdg m
+
\frac{\Lambda}{3}l\wdg \bar{m},
\\
\Omega^{0}_{\fant{a}0}
-
\Omega^{3}_{\fant{a}3}
&=
-\frac{\Lambda}{6}(k\wdg l+m\wdg \bar{m})
\end{align}
Note that one makes essential use of the distributional form of the profile function given in (\ref{impulsive-profile-func}) in order to obtain the connection and the curvature forms above. The expressions for the  curvature 2-forms can be used to calculate the Einstein 3-forms as
\be
\begin{split}
*G^0
&=\Lambda*k,\qquad  *G^2=\Lambda*m,
\\
*G^1
&=
2\left(1+\frac{\Lambda}{6}\zeta\bar{\zeta}\right)\left[
-\frac{\Lambda}{6}H+\frac{\Lambda}{6}
\left(
\zeta H_{\zeta}+\bar{\zeta} H_{\bar{\zeta}}
\right)
-
\left(1+\frac{\Lambda}{6}\zeta\bar{\zeta}\right)H_{\zeta\bar{\zeta}}
\right]*k+\Lambda *l.
\end{split}
\ee
The nonvanishing curvature spinors are given by
\be
\begin{split}
R&=-\Lambda,\qquad  \Psi_4=\left(1+\frac{\Lambda}{6}\zeta\bar{\zeta}\right)^2 H_{\zeta\zeta},
\\
\Phi_{22}
&=
\left(1+\frac{\Lambda}{6}\zeta\bar{\zeta}\right)
\left[
\left(1+\frac{\Lambda}{6}\zeta\bar{\zeta}\right)
H_{\zeta\bar{\zeta}}
-
\frac{\Lambda}{6}
\left(
\zeta H_{\zeta}+\bar{\zeta} H_{\bar{\zeta}}
\right)
+
\frac{\Lambda}{6}H
\right].
\end{split}
\ee

In terms of the Laplacian operator (\ref{laplacian-max-symmetric}), and the redefined profile function
\be
H\equiv \frac{1}{\sqrt{2}}\left(1+\frac{\Lambda}{6}\zeta\bar{\zeta}\right)\tilde{H},
\ee
 the nonvanishing Ricci spinor can be written in the compact form
\be
\Phi_{22}
=
\frac{1}{2\sqrt{2}}\left(1+\frac{\Lambda}{6}\zeta\bar{\zeta}\right)\left(\Delta+\frac{2\Lambda}{3}\right)\tilde{H}.
\ee
Assuming that $\tilde{H}(u,\zeta,\bar{\zeta})=\tilde{h}(\zeta,\bar{\zeta})\delta(u)$, the general solution to the vacuum field equation $\Phi_{22}=0$ can be written in the form \cite{griffiths-podolsky-arbitary-cosmo}
\be
\tilde{h}
=
f_{\zeta}+\bar{f}_{\bar{\zeta}}-\frac{\Lambda}{3}\frac{\bar{\zeta}f+\zeta\bar{f}}{\left(1+\frac{\Lambda}{6}\zeta\bar{\zeta}\right)}
\ee
and consequently, one can also show that the nonvanishing curvature component has the explicit form 
\be
\Psi_4
=
\tfrac{1}{\sqrt{2}}\left(1+\frac{\Lambda}{6}\zeta\bar{\zeta}\right)^2f_{\zeta\zeta\zeta}\delta(u)
\ee
in terms of a function $f$ analytic in the coordinate $\zeta$. 

\subsection{Impulsive waves in  direct product spacetimes}

Nonexpanding impulsive gravitational waves can also be studied in backgrounds of the product form $M_1\times M_2$ with the metric
of the form \cite{ortaggio1,ortaggio2}
\be\label{impulsive-in-product-background}
g=
-\frac{1}{\Omega^2}\left(
du\ot dv+dv\ot du+2H(u,\zeta,\bar{\zeta})du\ot du\right)
+
\frac{1}{\Sigma^2}\left(
d\zeta\ot d\bar{\zeta}+d\bar{\zeta}\ot d\zeta
\right)
\ee
where the conformal factors for the two-dimensional parts are explicitly given by
\be
\Omega(u,v)
\equiv
1-\frac{\epsilon_1}{a^2}uv,\qquad
\Sigma(\zeta,\bar{\zeta})
\equiv
1+\frac{\epsilon_2}{b^2}\zeta\bar{\zeta}.
\ee
$\epsilon_1$ and $\epsilon_2$ assume the numerical values $0,\mp1$. The constants $a,b$ have the dimension of $L$. The profile function is assumed to be of the distributional form: $H(u, \zeta, \bar{\zeta})=\delta(u)h(\zeta, \bar{\zeta})$. As in the case for the previous section, an impulsive profile function is essential for the construction of the $pp$-wave-type gravitational waves on the spacetimes of the form $M_1\times M_2$ as well.  The $pp$-waves with impulsive profile function with flat background is simply obtained by setting $\epsilon_1=\epsilon_2=0$.

In line with the choice of coframe in the previous cases, the convenient  set of basis null coframe 1-forms is of the form
\be
k=\Omega^{-1}du, \qquad l=\Omega^{-1}(dv+Hdu),\qquad m=\Sigma^{-1}d\zeta.
\ee
The nonvanishing connection 1-forms relative to the null basis then take the form
\be
\omega^{1}_{\fant{a}2}=\Sigma H_\zeta k,
\qquad
\omega^{0}_{\fant{a}0}=-\frac{\epsilon_1}{2a^2}(uk-vl),\qquad \omega^{3}_{\fant{a}3}=\frac{\epsilon_2}{2b^2}(\bar{\zeta}m-\zeta\bar{m}).
\ee
Consequently, the corresponding nonvanishing curvature 2-forms are explicitly given by
\be
\begin{split}
&\Omega^{1}_{\fant{a}2}=-{(\Sigma^2 H_{\zeta})}_{\zeta} k\wdg  m-\Sigma H_{\zeta\bar{\zeta}} k\wdg\bar{m},
\\
&\Omega^{0}_{\fant{a}0}-\Omega^{3}_{\fant{a}3}=-\frac{\epsilon_1}{a^2}k\wdg l+\frac{\epsilon_2}{b^2}m\wdg \bar{m}.
\end{split}
\ee
With the help of the general formula (\ref{null-coframe-curvature-scalars}), one can readily find that the NP curvature scalars  as
\be
\begin{split}
\Phi_{22}
&=
\Sigma^2 H_{\zeta\bar{\zeta}},\qquad \Phi_{11}=\frac{1}{4}\left(-\frac{\epsilon_1}{a^2}+\frac{\epsilon_2}{b^2}\right),
\\
\Psi_2
&=
-\frac{1}{6}\left(\frac{\epsilon_1}{a^2}+\frac{\epsilon_2}{b^2}\right), \qquad \Psi_4={(\Sigma^2 H_{\zeta})}_{\zeta},
\\
R
&=
2\left(\frac{\epsilon_1}{a^2}+\frac{\epsilon_2}{b^2}\right).
\end{split}
\ee
Note that the curvature spinors $\Phi_{11}$, $R$ and $\Psi_2$ are all constants expressed given in terms of the parameters characterizing the two 
dimensional manifolds and that  each of the two-dimensional spacetimes considered in $M_1\times M_2$ can be parameterized by a constant, denoted by $\epsilon_1$ and $\epsilon_2$ above. Some two-dimensional spacetimes admitting a physical interpretation are tabulated in Table 1.

Evidently, the ansatz (\ref{impulsive-in-product-background}) describes an  impulsive gravitational wave that propagates in a Petrov type D background spacetime provided that the parameters are chosen to have a nonvanishing  NP scalar $\Psi_2$. Moreover, for some  appropriate  values of the parameters, the background spacetime admits a constant non-null electromagnetic component $\Phi_{11}$ and a cosmological constant by construction as well.

The explicit expressions for NP scalars also suggest that the constant non-null electromagnetic component can formally be decoupled from the cosmological constant term by defining new parameters
\be
\Lambda_\mp\equiv\frac{1}{2}\left(\frac{\epsilon_1}{a^2}\mp\frac{\epsilon_2}{b^2}\right).
\ee

\begin{table}[H]
\caption{Table of  product spacetimes \texorpdfstring{$M_1\times M_2 $}{} \cite{ortaggio2}.}
\begin{center}
\begin{tabular}{|c|c|c|c|}
  \hline
  \emph{Spacetime} &\quad $M_1\times M_2$\quad                 & $\epsilon_1$  & $\epsilon_2$\\
  \hline\hline
Minkowski          &  $\mathbb{R}^{(1,1)}\times \mathbb{R}^{2}$& 0&0 \\
\hline
Nariai             &  $ dS^2\times S^2$                        &$+1$&$+1$ \\
\hline
anti-Nariai        &  $AdS^2\times H^2$                        &$-1$&$-1$\\
\hline
Bertotti-Robinson  &  $AdS^2\times S^2$                        &$-1$ & $+1$\\
\hline 
\multirow{2}{*}{Pleba\`{n}ski-Hacyan}&$\mathbb{R}^{(1,1)}\times S^2$ & 0 &$+1$
       \\ \cline{2-4}
                                     &$AdS^2\times\mathbb{R}^{2}$ &$-1$&0\\
\hline
\end{tabular}
\end{center}
\end{table}

 In terms of $\Lambda_\mp$, the Einstein 3-forms are explicitly given by the convenient form as
\be
\begin{split}
*G^0&=-\Lambda_+*k-\Lambda_-*k,
\\
*G^1&=-\Delta H*k-\Lambda_+*l-\Lambda_-*l,
\\
*G^2&=-\Lambda_+*m+\Lambda_-*m,
\end{split}
\ee
where $\Lambda_+$ is like a cosmological constant term whereas $\Lambda_-$ replaces   the non-null electromagnetic field $\Phi_{11}$ corresponding to
the diagonal components of the Ricci tensor $R^{a}_{\fant{a}b}$ that are trace-free.
The Laplacian defined on the wavefronts $u=0$ now has the explicit form $\Delta\equiv2\Sigma^2\partial_\zeta\partial_{\bar{\zeta}}$.

\subsection{Gravitational waves on the Melvin universe}

The Einstein-Maxwell theory has two well-known metrics admitting a constant magnetic field.  One is  the Bertotti-Robinson 
metric, which is of the product form $AdS^2\times S^2$, mentioned briefly in the preceding subsection and the other is the 
Melvin universe which has the topology of $\mathbb{R}^4$ which is a suitable candidate considered as a background metric in different physical contexts of interest.

Melvin Universe \cite{bonnor1954,melvin1964} is a cylindrically symmetric, static, electrovacuum spacetime described by  the metric
\be\label{melvin-metric-form1}
g=\Sigma^{2}(-dt\ot dt+dz\ot dz+d\rho\ot d\rho)+\frac{\rho^2}{\Sigma^{2}}d\phi\ot d\phi
\ee
where the function $\Sigma$ depending only on  the radial coordinate is of the form
\be\label{sigma-definition}
\Sigma
\equiv
1+\tfrac{1}{4}B^2\rho^2,
\ee
and the constant parameter $B$  is related to the electromagnetic field strength along the symmetry axis.
The metric ansatz (\ref{melvin-metric-form1}) requires a non-null electromagnetic field of the form
\be\label{melvin-self-dual-faraday}
\mathcal{F}
=
\frac{B^2}{2\Sigma^2}(k\wdg l-m\wdg \bar{m})
\ee
relative to the suitable null coframe basis $\{k, l, m, \bar{m}\}$ to be defined in terms of a suitable coordinate system below.

The Melvin universe has a number of interesting geometrical properties.
For example, it is conjectured by Garfinkle and Glass \cite{garfinkle-glass} that the Melvin universe is the only geodesically complete static spacetime.
Yet another interesting property, as  shown by Thorne \cite{stability-of-melvin-throne}, is that the Melvin universe is stable against radial perturbations.

The generalization of Melvin solution and the gravitational waves on these solutions with nonlinear  electromagnetic theories, such as Born-Infeld electrodynamics, minimally coupled to gravity have been studied by Gibbons and Herdeiro \cite{gibbons-herdeiro}.
Radu and Slagter \cite{radu-slagter} presented a Melvin-type solution in an Einstein-Maxwell-Dilaton gravity with a Liouville-type potential term for the dilaton.
Garfinkle and Melvin \cite{garfinkle-melvin-rotating-melvin} obtained a metric for a rotating magnetic universe by making use of Melvin spacetime metric as a seed metric in Ehlers transformation. By boosting the Schwarzchild-Melvin solution \cite{ernst} which describes a Schwarzchild black hole immersed in a constant magnetic field, an extension of the Aichelburg-Sexl gravitational wave solution  has been given by Ortaggio \cite{sch-melvin-boosted}.
A constant magnetic field solution to the low energy-limit of an effective string theory model has been presented by Tseytlin \cite{tseytlin-melvin}. 
(See, also \cite{suyama})

It is possible to introduce $pp$-wave type gravitational waves traveling along the $z$-axis on the Melvin universe background 
\cite{sch-melvin-boosted,garfinkle-melvin} by introducing the real null coordinates $u,v$ defined by
\be
u=\frac{1}{\sqrt{2}}(t-z),\qquad v=\frac{1}{\sqrt{2}}(t+z)
\ee
to the metric (\ref{melvin-metric-form1}). In terms of a profile function $H=H(u,\rho,\phi)$, and the real null coordinates $u,v$ the metric explicitly reads
\be\label{melvin-pp-waves}
g=\Sigma^{2}(-du\ot dv-dv\ot du-2Hdu\ot du)+\Sigma^{2}d\rho\ot d\rho+\frac{\rho^2}{\Sigma^{2}}d\phi\ot d\phi
\ee
The metric on the two dimensional wavefronts defined by $u=0$  is of the form
\be\label{2d-wavefront-for-melvin}
g^{(2)}
=
\Sigma^{2}d\rho\ot d\rho+\frac{\rho^2}{\Sigma^{2}}d\phi\ot d\phi
\ee
which has the topology of $\mathbb{R}^2$ and therefore the metric (\ref{melvin-pp-waves}) describes nonexpanding and non-twisting waves on the wavefronts with the metric (\ref{2d-wavefront-for-melvin}). The transverse character of the metric (\ref{melvin-pp-waves}) implies that the profile function $H$ satisfies
the Laplacian on the two dimensional wavefronts with the metric (\ref{2d-wavefront-for-melvin}) \cite{garfinkle-melvin}.
Consequently, as in the previous cases, the $u$ dependence of the profile function  will remain undetermined leading to a Laplace's equation in the transverse plane, as will explicitly be shown below.

The Laplacian corresponding to the two-dimensional metric (\ref{2d-wavefront-for-melvin}) has an explicit expression akin to (but not exactly identical to) that of the Laplacian in plane polar coordinates. To elucidate the difference, it is worth to derive it.  

The Laplacian can explicitly be found  by calculating the expression $d\star dh$ for a test function $h=h(\rho, \phi)$. $\star$ denotes the Hodge dual in the transverse plane with respect to the volume element $\star1=\rho d\rho\wdg  d\phi$ that is identical to that of $\mathbb{R}^2$ expressed in plane polar coordinates.
To this end, it is convenient to introduce orthonormal basis 1-forms
\be
e^1=\Sigma d\rho, \qquad e^2=\frac{\rho}{\Sigma}d\phi
\ee
spanning the transverse plane.
In terms of these basis 1-forms, one has $\star1=e^1\wdg e^2$ and thus $\star e^1=e^2$ and $\star e^2=-e^1$. The exterior derivatives of the basis 
1-forms are
\be
de^1=0,\qquad de^2=\frac{1}{\rho}\left(\frac{\rho}{\Sigma}\right)_\rho\star1.
\ee
These formula are sufficient to calculate the Laplacian and one has
\begin{align}
d\star dh
&=
d\left(\frac{1}{\Sigma}h_\rho \star e^1+\frac{\Sigma}{\rho}h_\phi\star e^2\right)
\nonumber\\
&=
d\left(\frac{h_\rho}{\Sigma}\right)\wdg \star e^1+\frac{h_\rho}{\Sigma} d\star e^1+d\left(\frac{\Sigma}{\rho}h_\phi\right)\wdg\star e^2
+
\left(\frac{\Sigma}{\rho}h_\phi\right)d\star e^2.
\end{align}
The expression on the right hand side then reduces to
\be
d\star dh
=
\frac{1}{\rho}\left[\frac{\rho}{\Sigma}\left(\frac{h_\rho}{\Sigma}\right)_\rho
+
\frac{h_\rho}{\Sigma}\left(\frac{\rho}{\Sigma}\right)_\rho
\right]\star1
+
\frac{\Sigma^2}{\rho^2}h_{\phi\phi}
\star1
\ee

Consequently, after a careful but straightforward calculation, the Laplacian for the metric (\ref{2d-wavefront-for-melvin}) can succinctly be expressed  in  the form \be\label{2d-laplacian-melvin}
\Delta
=
\frac{1}{\rho}\frac{\partial}{\partial \rho}\frac{\rho}{\Sigma^2}\frac{\partial}{\partial \rho}+\frac{\Sigma^2}{\rho^2}\frac{\partial^2}{\partial\phi^2}.
\ee
The expression (\ref{2d-laplacian-melvin}) for the Laplacian on the two-dimensional space with metric (\ref{2d-wavefront-for-melvin}) reduces to the familiar expression of Laplacian on the transverse plane in the plane polar coordinates for $\Sigma=1$, or equivalently, $B=0$. One can also compare the rest of the formula for the Melvin universe to  those given in Section \ref{polar-coords}. The expression for the Laplacian (\ref{2d-laplacian-melvin}) is identical to that of the one  given by Kadlecov\`a \cite{kadlecova-phd-thesis} previously.

After the brief digression on the explicit calculation of the Laplacian, one can now calculate the connection and the curvature for the
gravitational waves on the Melvin universe.
A natural set of null coframe basis that one can adopt is of the form
\be\label{melvin-null-coframe1}
k=\Sigma du, \qquad l=\Sigma (dv+Hdu), \qquad
m
=
\frac{1}{\sqrt{2}}
\left(
\Sigma d\rho+\frac{i\rho}{\Sigma}d\phi
\right).
\ee
and this choice of the null coframe basis closely follows  that of the basis null frame fields adopted by Ortaggio \cite{sch-melvin-boosted} up to some
changes in the numerical factor multiplying the profile function.

For convenience note that the set of null basis frame fields associated to the coframe (\ref{melvin-null-coframe1}) is
\be\label{melvin-null-frame1}
D
=
-\frac{1}{\Sigma} \partial_v, \qquad 
\Delta=-\frac{1}{\Sigma} (\partial_u-H\partial_v), 
\qquad
\delta
=
\frac{1}{\sqrt{2}}
\left(
\frac{1}{\Sigma}\partial_\rho+\frac{i\Sigma}{\rho}\partial_\phi
\right).
\ee
where they are denoted, as before, by $D=k^\mu\partial_\mu$, $\Delta=l^\mu\partial_\mu$  and $\delta=m^\mu\partial_\mu$
(here the differential operator $\Delta$ is not to be confused with the Laplacian operator defined in Eq. (\ref{2d-laplacian-melvin}) on the transverse  space).

By solving the Cartan's structure equations, one can show that
the non-vanishing connection 1-forms Relative to the null coframe (\ref{melvin-null-coframe1})  are conveniently expressed in the form
\begin{align}
\omega^{0}_{\fant{0}3}
&=
\frac{1}{2\sqrt{2}}\frac{B^2\rho}{\Sigma^2}k,
\qquad
\omega^{1}_{\fant{0}2}
=
\bar{\delta} Hk
+
\frac{1}{2\sqrt{2}}\frac{B^2\rho}{\Sigma^2}l,
\\
\omega^{3}_{\fant{3}3}
&=
-\frac{1}{\sqrt{2}}\frac{1}{\Sigma^2\rho}\left(1-\tfrac{1}{4}B^2\rho^2\right)(m-\bar{m}),
\end{align}
where $\bar{\delta}$ is the complex conjugate of the differential operator appearing in the  definition  of the basis frame fields (\ref{melvin-null-frame1}).

The above expressions for the connection 1-forms then lead to the corresponding curvature 2-forms that are of the form
\begin{align}
\Omega^{0}_{\fant{0}3}
&=
-\frac{B^2}{2\Sigma^{4}}\left(1-\tfrac{1}{4}B^2\rho^2\right)k\wdg m,
\\
\Omega^{1}_{\fant{0}2}
&=
\left[
-\bar{\delta}\bar{\delta}H
+
\sqrt{2}\left(\frac{1}{\Sigma}\right)_\rho \bar{\delta }H
+
\frac{1}{\sqrt{2}\rho}\left(\frac{\rho}{\Sigma}\right)_\rho\bar{\delta}H
\right]k\wdg{m}
-
\frac{B^2}{2\Sigma^{4}}\left(1-\tfrac{1}{4}B^2\rho^2\right)l\wdg \bar{m}
\nonumber\\
&+
\left[
-{\delta}\bar{\delta}H
+
\frac{1}{\sqrt{2}}\left(\frac{1}{\Sigma}\right)_\rho \left(\delta+\bar{\delta}\right)H
-
\frac{1}{\sqrt{2}\rho}\left(\frac{\rho}{\Sigma}\right)_\rho\bar{\delta}H
\right]k\wdg\bar{m},
\\
(\Omega^{0}_{\fant{0}0}
-
\Omega^{3}_{\fant{0}3})
&=
\frac{B^4\rho^2}{2\Sigma^{4}}k\wdg l
+
\frac{B^2}{2\Sigma^{4}}\left(2-\tfrac{1}{4}B^2\rho^2\right)m\wdg \bar{m}.
\end{align}

By making use of the above expressions in connection with the general expressions (\ref{null-coframe-curvature-scalars}), the nonvanishing NP curvature scalars can be identified  as
\begin{align}
\Phi_{11}
&=
\frac{B^2}{2\Sigma^4},
\qquad
\Psi_{2}
=
-\frac{B^2}{2\Sigma^4}\left(1-\tfrac{1}{4}B^2\rho^2\right),
\\
\Phi_{22}
&=
\frac{1}{2\Sigma^2\rho^2}
\left[
\rho\frac{\partial}{\partial\rho}\left(\rho\frac{\partial H}{\partial\rho}\right)
+
\Sigma^4\frac{\partial^2 H}{\partial\phi^2}
\right],
\label{melvin-phi-22}
\\
\Psi_{4}
&=
\bar{\delta}\bar{\delta}H
-
\left[
\sqrt{2}\left(\frac{1}{\Sigma}\right)_\rho 
+
\frac{1}{\sqrt{2}\rho}\left(\frac{\rho}{\Sigma}\right)_\rho
\right]
\bar{\delta}H.
\end{align}

The corresponding Einstein 3-forms can explicitly be written as
\be
\begin{split}
*G^0
&=
-\frac{B^2}{\Sigma^4}*k,
\\
*G^1
&=
-\frac{1}{\Sigma^2\rho^2}
\left[
\rho\frac{\partial}{\partial\rho}\left(\rho\frac{\partial H}{\partial\rho}\right)
+
\Sigma^4\frac{\partial^2 H}{\partial\phi^2}
\right]*k-\frac{B^2}{\Sigma^4}*l,
\\
*G^2
&=
\frac{B^2}{\Sigma^4}*m.
\end{split}
\ee
Therefore, by using the simplified the expression for $\Phi_{22}$ given in Eq. (\ref{melvin-phi-22}), the electrovacuum field equations with a static magnetic field is $\Phi_{22}=0$ and it explicitly leads to the following equation on the transverse space
\be\label{melvin-reduced-field-eqn}
\left(
\rho\frac{\partial}{\partial \rho}\rho\frac{\partial}{\partial \rho}
+
\Sigma^4\frac{\partial^2}{\partial\phi^2}
\right) H=0.
\ee
It is interesting to note that the differential operator involved in (\ref{melvin-reduced-field-eqn}) is not  the Laplacian defined in 
(\ref{2d-laplacian-melvin}). However, (\ref{melvin-reduced-field-eqn}) reduces to the corresponding expression of the $pp$-wave expressed in terms of the plane polar coordinates.

Note also that the Maxwell's equations $d\mathcal{F}=0$ is identically satisfied for the self-dual 2-form  ansatz given in (\ref{melvin-self-dual-faraday}) with the basis coframe defined as in Eq. (\ref{melvin-null-coframe1}). On the other hand, although one can readily  introduce an additional electromagnetic null component $\Phi_2$ that satisfies $d\mathcal{F}=0$ along with the non-null  component $\Phi_1$. However, such a null component is not compatible with the metric ansatz (\ref{melvin-metric-form1}) because one then has an additional  nonvanishing Ricci tensor component $\Phi_{12}$ of the form $\Phi_{12}=\kappa^2 \Phi_1\bar{\Phi}_2$ in this case.

In the case $\partial_\phi H=0$, the differential equation (\ref{melvin-reduced-field-eqn}) can readily be solved for $H=H(u,\rho)$ to obtain a solution which is a polynomial in $\rho$. However, the general form of the profile function $H$ cannot be expressed in terms of elementary functions which can be used to find a general  form of $\Psi_4$. Consequently, general form of the NP scalars lead to the fact that the metric (\ref{melvin-metric-form1}) is of Petrov type D, static electrovacuum  metric and that the metric (\ref{melvin-pp-waves}) describes $pp$-wave-type gravitational wave   on a Petrov type D background spacetime.

For the purpose of comparison of the above formula to some other NP curvature scalars that appear in the literature, it is convenient to consider another coframe  defined as
\be\label{melvin-null-coframe2}
\tilde{k}=\Sigma^2 du, \qquad
\tilde{l}=dv+Hdu, \qquad
\tilde{m}
=
\frac{1}{\sqrt{2}}
\left(
\Sigma d\rho+\frac{i\rho}{\Sigma}d\phi
\right).
\ee
The two coframes (\ref{melvin-null-coframe1}) and (\ref{melvin-null-coframe2}) are simply related by
\be\label{null-coframe-tr-melvin}
\tilde{k}=\Sigma k,\qquad \tilde{l}=\Sigma^{-1}l,\qquad \tilde{m}=m
\ee
where the real coefficient $\Sigma$ is defined as in (\ref{sigma-definition}). The corresponding connection 1-forms are then related by
\be
\begin{split}
&\tilde{\omega}^{0}_{\fant{a}3}=\Sigma\omega^{0}_{\fant{a}3},
\qquad
\tilde{\omega}^{1}_{\fant{q}2}=\Sigma^{-1}\omega^{1}_{\fant{a}2},
\\
&\tilde{\omega}^{0}_{\fant{a}0}-\tilde{\omega}^{3}_{\fant{a}3}=\omega^{0}_{\fant{a}0}-\omega^{3}_{\fant{a}3}+\Sigma^{-1}d\Sigma,
\end{split}
\ee
and making use of these transformation formulas, one can also show that the curvature 2-forms also transform as
\be
\begin{split}
&\tilde{\Omega}^{0}_{\fant{a}3}=\Sigma\Omega^{0}_{\fant{a}3},
\qquad
\tilde{\Omega}^{1}_{\fant{q}2}=\Sigma^{-1}\Omega^{1}_{\fant{a}2},
\\
&\tilde{\Omega}^{0}_{\fant{a}0}-\tilde{\Omega}^{3}_{\fant{a}3}=\Omega^{0}_{\fant{a}0}-\Omega^{3}_{\fant{a}3}.
\end{split}
\ee
Consequently, the NP curvature scalars belonging to the null coframe (\ref{melvin-null-coframe2}) are related to those of (\ref{melvin-null-coframe2})
by an overall factor of $\Sigma$ and also replacing the profile function expressions  $\Sigma H $ by $H$ in the resulting expressions;
\be
\tilde{\Phi}_{22}=\Sigma^{-2}\Phi_{22},\qquad \tilde{\Psi}_4=\Sigma^{-2}\Psi_4.
\ee

By making use of this recipe and the associated formula for the null coframe transformation (\ref{null-coframe-tr-melvin}), it is possible to find that the expression for curvature spinor $\tilde{\Phi}_{22}$ for the null coframe (\ref{melvin-null-coframe2}) takes the form
\be
\tilde{\Phi}_{22}
=
\frac{1}{2\Sigma^4\rho^2}
\left[
\rho\frac{\partial}{\partial\rho}\rho\frac{\partial}{\partial\rho}
+
\Sigma^4\frac{\partial^2}{\partial \phi^2}
\right]H.
\ee
One can rearrange terms in the equation satisfied by the profile function to obtain
\be
\rho\frac{\partial}{\partial\rho}\left(\rho\frac{\partial H}{\partial\rho}\right)
+
\Sigma^4\frac{\partial^2 H}{\partial \phi^2}
=0.
\ee
Following the original treatment of Garfinkle and Melvin \cite{garfinkle-melvin}, one can introduce a new radial variable
\be
 x\equiv \frac{1}{4}B^2\rho^2
\ee
and the equation for the profile function can be separated by assuming that the profile function is of the form 
\be
H(u,\rho,\phi)=a(u)h(x)\cos (m\phi)
\ee
with $m$ being an integer. Consequently, the radial equation for $h(x)$ takes the form
\be\label{melvin-reduced-profile-eqn}
x\frac{d}{dx}\left(x\frac{d h}{dx}\right)
-
\frac{1}{4}m^2(1+x)^4h
=0.
\ee
This result is in accordance with the general result that the contribution of the Kerr-Schild term to the curvature tensor is proportional to a term of the form $\Delta H$ relative to the null coframe (\ref{melvin-null-coframe2}), under general assumptions related to the generalized Kerr-Schild transformations 
\cite{garfinkle-melvin,garfinkle-vachaspati}. Subsequently, the expression originally found by Garfinkle and Melvin  is adopted in a more detailed study of gravitational waves on the Melvin universe by Ortaggio \cite{sch-melvin-boosted} by using the coframe (\ref{melvin-null-coframe1}).

For $m=0$ in (\ref{melvin-reduced-profile-eqn}), corresponding to an axially symmetric field configuration, it is easy to find  an exact solution for $h(x)$.  
One finds that for the axially symmetric solution, the expression for $h(x)$ is of the form
$h(x)
=
\ln x
$.
The equation for the profile function  (\ref{melvin-reduced-profile-eqn}) is in the form originally given by Garfinkle and Melvin  
and the solutions to this equation are discussed extensively in \cite{garfinkle-melvin}.

\subsection{Some further remarks}

The family of $pp$-wave metrics are exact solutions in a number modified gravitational theories, for example, the $pp$-wave solutions was studied in the supergravity theories \cite{aichelburg-beler-dereli1,aichelburg-beler-dereli2,aichelburg-beler-dereli3,aichelburg-beler-dereli4,cariglia-gibbons-guven-pope} and in the gravitational models relevant to the low energy limit of the  string theories \cite{bergshoeff1,bergshoeff2}.
The plane wave solutions to Yang-Mills type non-Abelian gauge fields was first studied by Coleman \cite{coleman}. Shortly after, G\"uven \cite{guven-E-YM} presented the extension of Coleman's solutions to curved spacetime. Dereli and G\"uven   \cite{dereli-guven-susy-ym} later generalized the Coleman's solutions to the non-Abelian  Yang-Mills gauge fields with supersymmetry. Trautmann \cite{trautman} also presented  the plane wave solutions to non-Abelian gauge fields and to the quadratic curvature model discussed in Sect. \ref{qc-section} below.

Another remarkable property of the gravitational plane waves,  due to Penrose,  is that any spacetime has a plane wave as a limit \cite{penrose,blau}. The Penrose limit is later  extended in \cite{guven-PLB200}  to  all five of the string theories by G\"uven. It is also shown by Penrose \cite{penrose-cauchy-hypersurface} that there is no global Cauchy hypersurface for the plane wave metrics.

It has been shown by D\"uzta\c{s} and Semiz \cite{duztas-semiz} that the decoupling  of a massive  vector field requires a covariantly constant null vector field in the background spacetime and therefore this interesting result evidently implies that the background spacetime is to be a $pp$-wave spacetime.

The impulsive gravitational wave solutions to some popular alternative gravitational models has been studied previously in \cite{barrabes-hogan} by Barrab\`{e}s and Hogan.  The gravitational wave solutions of some modified gravity theories, such as $f(R)$ and the models involving higher curvature terms in their Lagrangian,  have previously been studied \cite{linearized-wave-sol1,linearized-wave-sol2,linearized-wave-sol3} in the linearized approximation.
These solutions to the linearized equations are relevant to the discussion below because it is well-known that the $pp$-wave metrics  constitute their own linearizations.

It is well-known that the curvature invariants of a $pp$-wave metric vanishes. (See \cite{schmidt} for a nice proof using a fixed point of the vector field generating a homothety but is not an isometry). It has been shown in \cite{pravda} that the  spacetimes with all the scalar invariants constructed from the
Riemann tensor and its covariant derivatives vanish  if the spacetime is of Petrov type III, N or O. In this case,  the multiple  null eigenvector of the Weyl tensor is geodesic with all the optical scalars vanishing. Later, these defining properties are used to construct generalized $pp$-waves in higher dimensions 
\cite{CMPPPZ,OPP}. In four dimensions, the curvature invariants corresponding to a type N metric involving a cosmological constant are studied  in 
\cite{bicak-pravda}. The spacetimes with vanishing curvature invariants in higher dimensions have been studied in \cite{CMPP}.

Because the curvature invariants of $pp$-wave metric vanish, the Karlhede algorithm \cite{karlhede,karlhede2}, involving the covariant derivatives of the Riemann curvature tensor, is used to obtain an invariant classification of the $pp$-wave spacetimes in \cite{karlhede-class1} and, recently, in \cite{karlhede-class2} using the exterior algebra of differential forms.

\section{\texorpdfstring{$pp$}{}-waves in modified gravitational models}

From the point of  view of a given modified gravitational model, it is an interesting  question to investigate whether  the simple family of the $pp$-wave type metrics can be lifted to a set of gravitational wave solutions.

\subsection{\texorpdfstring{$pp$}{}-waves in the Brans-Dicke  theory}

The Brans-Dicke theory of gravity \cite{BD-original} is one of the most popular scalar-tensor theories of gravity. Although it was proposed to incorporate to Machian ideas of inertia into the general relativity theory by introducing a geometry-matter coupling via a dynamical scalar field,  it has now become popular in the context of  the $f(R)$-type modified gravity models. Some peculiar properties of the radiative metrics in BD theory have been studied in \cite{robinson-winicour,mcintosh} and the radiative metrics in the linear approximation in BD theory has previously been studied in \cite{wagoner} by Wagoner.
More recently, the $pp$-wave solutions for the BD theory with Maxwell field studied by Robinson \cite{robinson}  who
observed that the BD-Maxwell theory admits solutions with BD scalar with propagating scalar and non-vanishing Maxwell field in the
Minkowski background. Interestingly, a certain  part of the result pertaining to BD vacuum case rediscovered recently by Robinson are the $pp$-wave solutions presented some time ago in \cite{tupper}. Similar  solutions  to the scalar tensor theories involving a potential term has also been reported in \cite{zanelli}. The cylindrically symmetric gravitational wave solutions generalizing those given  by Einstein and Rosen \cite{einstein-rosen} has also been discussed in \cite{lokman-delice} recently. As a side remark related to the paper by Einstein and Rosen, the reader is referred  to the References \cite{kennefick-speed-of-thought,kennefick-phys-today}  for the interesting historical accounts on the gravitational waves.

In order to be able to use the null coframe language in connection with the exterior algebra developed in the first part, it is convenient to write the BD field equations relative to a null or orthonormal coframe. This can be achieved for example by using a first order formalism where the connection
and the coframe 1-forms are assumed to be the independent gravitational variables. The field equations  for the pseudo-Riemannian metric (equivalently, the equations for coframe 1-forms) are then obtained by constraining the independent connection 1-form to be a Levi-Civita connection as a subcase. Such a formulation has also been worked out, for example, in \cite{dereli-tucker-PLB} for the formulation of BD theory including  the fermion fields.
The details of the application of the first order formalism to the BD Lagrangian is provided below for convenience.

Expressed in terms of the differential forms,  the total Lagrangian 4-form for the original BD theory interacting with matter fields reads
\be\label{BD-lag}
L_{tot.}
=
L_{BD}[\phi, \theta^a, \omega^{a}_{\fant{a}b}]+\frac{8\pi}{c^4}L_{m}[g,\psi]
\ee
where the gravitational part in the so-called Jordan frame is
\be
L_{BD}
=
\frac{\phi}{2} \Omega_{ab}\wdg *\theta^{ab}-\frac{\omega}{2\phi}d\phi\wdg *d\phi.
\ee
The matter fields with the Lagrangian  $L_{m}[g, \psi]$ are assumed couple to the metric minimally and are also assumed to be independent of the BD scalar field. The gravitational coupling constant is replaced by a dynamical scalar field $\phi^{-1}$ with a corresponding kinetic terms for the scalar field.
$\omega$ is the free BD parameter and in general the corresponding general relativistic expression is recovered in the $\omega\mapsto\infty$ limit. However, such a correspondence is  not always warranted \cite{bhadra-nandi,baykal-delice-bd-grg}, as the case with matter energy-momentum tensor having a vanishing trace furnishes a well-known counter example.

In the general framework of first order formalism for gravity, the independent gravitational variables are the set of basis coframe 1-forms $\{\theta^a\}$ and the connection 1-forms $\{\omega^{a}_{\fant{a}b}\}$. The local Lorentz invariance of a gravitational Lagrangian forbids the gravitational action to have explicit dependence on $\{\omega^{a}_{\fant{a}b}\}$ and the exterior derivatives  $\{d\theta^a\}$ and $\{d\omega^{a}_{\fant{a}b}\}$. However, instead of the explicit dependence on derivatives  $\{d\theta^a\}$ and $\{d\omega^{a}_{\fant{a}b}\}$, a gravitational Lagrangian can have dependence on $\Theta^a$ and $\Omega^{a}_{\fant{a}b}$, respectively. Moreover, the minimal coupling prescription for the matter fields also implies that $d\theta^a$ and $d\omega^{a}_{\fant{a}b}$ occur only in the gravitational sector in a total Lagrangian with matter fields. On the other hand,
the BD scalar field  $\phi$  couples nonminimally to the metric simply because it multiplies the scalar curvature. As a consequence  of the nonminimal coupling, the BD scalar  field is a dynamical field even in the absence of the kinetic term for it.

The vanishing torsion constraint for the independent connection 1-form can be implemented into the variational procedure by introducing Lagrange multiplier 4-form term
\be\label{zero-torsion-constraint-lag}
L_C
=
\lambda_a\wdg (d\theta^a+\omega^{a}_{\fant{a}b}\wdg \theta^b)
\ee
to the original Lagrangian form $L_{BD}$, where the Lagrange multiplier 2-form $\lambda_a$ is a vector-valued 2-form enforcing the constraint $\Theta^a=0$.
The Lagrangian for the extended gravitational part then has the explicit form
\be
L_{ext.}[\phi, \theta^a, \omega^{a}_{\fant{a}b}, \lambda_a]
=
L_{BD}[\phi, \theta^a, \omega^{a}_{\fant{a}b}]+L_C[\theta^a,  \omega^{a}_{\fant{a}b}, \lambda^a].
\ee
The total variational derivative of $L_{ext.}$ with respect to the independent variables can be found as
\begin{align}
\delta L_{ext.}
&=
\delta\phi
\left(
\frac{1}{2}R*1
-
\frac{\omega}{2\phi}d*d\phi
+
\frac{\omega}{2\phi^2}d\phi\wdg *d\phi
\right)
+
\delta \theta_a\wdg
\left(
\frac{\phi}{2}\Omega_{bc}\wdg *e^{abc}
+
D\lambda^a
+
\frac{\omega}{\phi}*T^a[\phi]
\right)
\nonumber\\
&+
\delta\omega_{ab}\wdg
\tfrac{1}{2}
\left[
D\phi*\theta^{ab}
-
(\theta^a\wdg \lambda^b-\theta^b\wdg \lambda^a)
\right]
+
\delta\lambda_a\wdg \Theta^a\label{total-var-der}
\end{align}
up to an  omitted exact form.
The energy-momentum forms of the scalar field $*T^a[\phi]=T^a_{\fant{a}b}[\phi]*\theta^b$ stand for
\be
*T^a[\phi]
\equiv
\tfrac{1}{2}((i_ad\phi)*d\phi+d\phi\wdg i_a*d\phi).
\ee

The field equations for the connection then read
\be\label{bd-conn-eqn}
D(\phi *\theta^{ab})
=
\theta^a\wdg \lambda^b-\theta^b\wdg \lambda^a
\ee
and
these equations can be considered as equation for the Lagrange multiplier 2-forms $\lambda^a$. (\ref{bd-conn-eqn}) can  uniquely be solved for the Lagrange multiplier 2-form by calculating its contractions and taking the constraint $\Theta^a=0$ into account. Explicitly, by calculating two successive contractions and subsequently combining them,  one finds
\be\label{lag-mult-expression}
\lambda^a
=
*(d\phi\wdg \theta^a)
\ee
as the unique solution.
Consequently, using the expression (\ref{lag-mult-expression}) for the Lagrange multiplier form in the metric field equations induced by the coframe variational derivative $\delta L_{ext.}/\delta \theta^a\equiv *E^a$ in (\ref{total-var-der})  read
\be\label{BD-eqns1}
*E^a
\equiv
-\phi *G^a+D*(d\phi\wdg \theta^a)+\frac{\omega}{\phi}*T^a[\phi]+\frac{8\pi}{c^4}*T^a[F]=0
\ee
where $*E^a=E^a_{\fant{a}b}*\theta^b$ is vector-valued 1-form and $*T^a[F]$ stands for the energy-momentum forms for the Faraday 2-form field $F$ derived from the variational derivative of the Maxwell Lagrangian 4-form with respect to basis coframe 1-forms and it is defined  by  Eq. (\ref{maxwell-en-mom-3-form}) above.

As a consequence of the diffeomorphism invariance of the BD Lagrangian, it follows from the corresponding Noether  identity that $D*E^a=0$ \cite{hehl-mccrea-mielke-neemann}.
Consequently, from  the relation $D*T^a[\psi]=0$ one can derive the geodesic postulate for point-like  test particles.
Explicitly, with the help of the identities
\begin{align}
D\left(\phi*G^a\right)
&=
d\phi\wdg *G^a,
\\
D^2*(d\phi\wdg \theta^a)
&=
d\phi \wdg*R^a,
\\
D\left(\frac{\omega}{\phi}*T^a[\phi]\right)
&=
(i^ad\phi)\left(\frac{\omega}{2\phi}d*d\phi-\frac{\omega}{2\phi^2}d\phi\wdg *d\phi\right),
\end{align}
one eventually arrives at
\be
D*E^a
=
\tfrac{1}{2}(i^ad\phi)\left(\frac{\omega}{\phi}d*d\phi-\frac{\omega}{\phi^2}d\phi\wdg *d\phi+R*1\right)
\ee
as expected. The right hand side vanishes identically provided that the field equation for the BD scalar is satisfied since the terms on the right hand side are proportional to the field equations for the BD scalar given below.

In addition, the field equations for the BD scalar that follows from the variational derivative $\delta L_{BD}/\delta \phi$ is given by
\be\label{scalar eqn1}
\omega d*d\phi-\frac{\omega}{\phi}d\phi\wdg *d\phi+\phi R*1=0.
\ee
The BD scalar couples to the matter energy-momentum through the last term in (\ref{scalar eqn1}). In fact, by
combining it with the trace of the metric equations,  the equation for the BD scalar simplifies to
\be\label{reduced-scalar-eqn}
d*d\phi
=
\frac{8\pi}{c^4}\frac{1}{2\omega+3}*T[\psi]
\ee
where $T[\psi]\equiv T^{a}_{\fant{a}a}[\psi]$ is the trace  of the matter energy-momentum  tensor.
As pointed out above, the reduced scalar field equation (\ref{reduced-scalar-eqn}) follows  from Bianchi identity for the BD field equations together with the  trace. Since the only matter field present in the discussion is Maxwell field  $F$, $T[F]=0$ identically and the BD scalar satisfies a homogeneous equation corresponding to (\ref{reduced-scalar-eqn}).

The BD-Maxwell field equations  can be written by specializing the indices of the field equations (\ref{BD-eqns1}) to a complex null coframe. For $a=0, 1, 2$ relative to a null coframe  can explicitly be written in the form
\be\label{BD-eqn-general-null-form}
\begin{split}
*E^0
&=
-\phi*G^0+D*(d\phi\wdg k)+\frac{\omega}{\phi}*T^0[\phi]+\frac{8\pi }{c^4}*T^0[F]
=0,
\\
*E^1
&=
-\phi*G^1+D*(d\phi\wdg l)+\frac{\omega}{\phi}*T^1[\phi]+\frac{8\pi }{c^4}*T^1[F]
=0,
\\
*E^2
&=
-\phi*G^2+D*(d\phi\wdg m)+\frac{\omega}{\phi}*T^2[\phi]+\frac{8\pi }{c^4}*T^2[F]
=0,
\end{split}
\ee
respectively. Note that the above expression for $*E^0$ and $*E^1$ are real whereas the expressions for the components $*E^2$ and $*E^3$ are complex conjugates.
It is worth to note that, the BD field equations (\ref{BD-eqn-general-null-form}) can be written out   explicitly in terms of NP quantities with some straightforward work and further definitions identifying the components of the differential forms in terms of original NP curvature spinors and spin coefficients. On the other hand, for the modified gravitational models it is not, in general, possible  to use the field equations of the form $E_{ab}=\kappa^2 T_{ab}$
to simply determine the anti-selfdual part of the curvature expression as in the case of GR in the NP formalism.

As a consequence of the simplicity of the $pp$-wave metrical ansatz, the Einstein field equations are quite restrictive in admitting a matter
source. The $pp$-wave metric ansatz above admit only the scalar field ansatz of the form
$\phi=\phi(u)$ and thus $d\phi=\phi_u du$. For such a  scalar field there is only one non-vanishing component of the energy-momentum form of the form
\be
*T^1[\phi]
=
-\phi_u^2*k.
\ee
Moreover, with the assumption $\phi=\phi(u)$, the only  non-vanishing term among $D*(d\phi\wdg \theta^a)$ is for $a=1$ for which
\be
D*(d\phi\wdg l)
=
d*(d\phi\wdg l)
+
{\omega}^{1}_{\fant{a}2}\wdg *(d\phi\wdg m)
+
\bar{\omega}^{1}_{\fant{a}2}\wdg *(d\phi\wdg \bar{m}).
\ee
The second and the third terms are complex conjugate of one another making the left hand side a real 3-form.
Moreover, one has $\omega^{1}_{\fant{a}2}\wdg *(d\phi\wdg {m})=0$ and consequently
\be
D*(d\phi\wdg l)
=
-\phi_{uu}*k.
\ee
Moreover, the terms involving derivatives of the scalar field can be combined to have
\begin{align}
D*(d\phi\wdg l)
+
\frac{\omega}{\phi}T^1[\phi]
&=
d*(d\phi\wdg l)+\frac{\omega}{\phi}d\phi\wdg *(d\phi\wdg l)
\nonumber\\
&=
\frac{\phi^{-\omega}}{(1+\omega)}
d*\left[d(\phi^{(1+\omega)})\wdg l\right].\label{combination-of-scalar-terms}
\end{align}
By a direct calculation, it is possible to show that the expression on the right hand side has the non-vanishing component
\be
D*(d\phi\wdg l)
+
\frac{\omega}{\phi}T^1[\phi]
=
-\left(\phi_{uu}+\omega\frac{\phi_u^2}{\phi}\right)*k
\ee
for $\phi=\phi(u)$.
Eventually, the only nontrivial equation $*E^1=0$ reduces to
\be
-\phi d*dl
+
d*(d\phi\wdg l)
+
\frac{\omega}{\phi}d\phi\wdg *(d\phi\wdg l)
+
2|\Phi_2|^2*k=0.
\ee
Consequently, $*E^1=0$ can compactly be rewritten as
\be\label{simplified-BD-eqn}
-
\phi d*dl
+
\frac{\phi^{-\omega}}{(1+\omega)}d*\left[d(\phi^{(1+\omega)})\wdg l\right]
+
2|\Phi_2|^2*k=0.
\ee

It is interesting to  note that  the BD field equations do not fully determine the profile function as in the GR case. For a profile function with a reasonable
dependence on the real null coordinate $u$, the  $\omega\mapsto\infty$ limit yields the GR equations canceling out the second term, provided that one assumes $\phi\mapsto \phi_0=\kappa^{-2}$ in this limit.

For the BD theory, the following $pp$-wave  solutions can be constructed:
\begin{itemize}

\item[(1)] For the vacuum case with $\Phi_2=0$, there are two subcases depending on the numerical value of the BD parameter $\omega$, one with $\omega=-1$ and the other with  $\omega+1\neq0$. In this case, the first and the second terms in (\ref{simplified-BD-eqn}) are set equal to zero separately: $d*dl=-2H_{\zeta\bar{\zeta}}*k=0$ and
\be
d*\left[d(\phi^{(1+\omega)})\wdg l\right]=0.
\ee
Now calculate  $d*\left[df(u)\wdg l\right]$ for an arbitrary function of $f(u)$ and the basis coframe $l$. One finds
\be
d*\left[df(u)\wdg l\right]
=
-f''*k-if'(dm\wdg \bar{m}-m\wdg d\bar{m})
\ee
where a prime denotes an ordinary derivative with respect to the coordinate $u$. The second term on the right hand side vanishes by the definition of the coframe for the $pp$-wave metric: $dm=dd\zeta\equiv0$.  The first term, on the other hand, vanishes if and only if $f''=0$. In other words, $d*[df(u)\wdg l]=0$ is satisfied iff $f(u)\sim u$ up to a constant. Thus, with the assumption $\omega\neq1$, one finds
\be
\phi(u)=\phi_0 u^{1/(1+\omega)}.
\ee

The case $\omega=-1$ has to be treated separately starting from (\ref{combination-of-scalar-terms}).
For $\omega=-1$, Eq. (\ref{combination-of-scalar-terms}) explicitly becomes
\be
D*(d\phi\wdg l)
-
\frac{1}{\phi}*T^1[\phi]
=
\phi d*[(d\ln \phi)\wdg l]
\ee
where $d\ln \phi\equiv \phi^{-1}d\phi$ and thus, in this case the right hand side vanishes identically iff $\phi(u)=\phi_0 e^u$.
These solutions had been  reported in \cite{tupper} and then  they are rediscovered in \cite{robinson} recently including the electromagnetic field into the discussion. The flat spacetime requires a vanishing Riemann tensor (or equivalently the vanishing curvature 2-form) according to  the well-known Riemann theorem and that the flat spacetime is a trivial solution of $G_{ab}=0$. The BD field equations $E_{ab}=0$ with the $pp$-wave metric ansatz, on the other hand, also admit   the flat spacetime solution together with a propagating scalar field in the flat background.

\item[(2)]
For the electrovacuum case in the BD theory, it is possible to construct the following general expression
\be\label{case-general}
H(u, \zeta)
=
h(u, \zeta)
+
\bar{h}(u, \bar{\zeta})
-
|\zeta|^2\frac{(\phi^{1+\omega})_{uu}}{2(1+\omega)\phi^{1+\omega}}
-
\frac{8\pi}{c^4\phi}f\bar{f}=0
\ee
for profile function. This particular solution was, somewhat surprisingly, reported in \cite{robinson} recently. The electrovacuum solutions of BD  in common with  the vacuum solutions of  GR now follow if
\be\label{case-1a}
|\zeta|^2\frac{(\phi^{\omega+1})_{uu}}{(\omega+1)\phi^{(1+\omega)}}+\frac{16\pi}{c^4}f\bar{f}=0
\ee
is satisfied separately from the vanishing of the Einstein tensor.
Note that, as in the previous case, the scalar field equation  (\ref{case-1a}) is satisfied for the flat spacetime as well.
This particular case corresponds to the propagating Maxwell and scalar fields in the flat background and the fields are referred to as
non-gravitating waves in \cite{robinson} by Robinson.

\end{itemize}

In the manner of solving BD field equations with $pp$-wave metric ansatz as  in the cases (1) and (2) above,  the scalar field equations satisfy the same field equations irrespective of the  assumptions either $H_{\zeta\bar{\zeta}}=0$ or $H=0$ because the BD field equations in this ansatz can be decoupled into two separate equations, one for the metric, and one for the BD scalar field together with the energy momentum component coming from the  Maxwell field. Consequently,  the BD field equations admit flat spacetime solutions with propagating BD scalar.

Evidently, the source of non-gravitating propagating scalar field can be traced back to the constraint term, namely the presence of the term $D*(d\phi\wdg \theta^a)$ which arises from the nonminimal coupling of the BD scalar field to the curvature \cite{zanelli}.  In the flat spacetime,  the term $D*(d\phi\wdg \theta^a)$ becomes
\be
d*(d\phi\wdg dx^a)
=
(P^{a}_{\fant{a}b}\phi)*dx^b
\ee
where $*$ now stands for the Hodge dual for Minkowski spacetime whereas the second order differential operator
$P_{ab}$ is explicitly given by
\be
P_{ab}
=
\partial_{a}\partial_{b}-\eta_{ab}\Box
\ee
which  is a projection operator in Minkowski spacetime with $\Box\equiv\eta^{ab}\partial_a\partial_b$. The $P_{ab}$ is also a transverse differential operator:  $\partial^a P_{ab}\equiv0$. Its trace is given by $P^{a}_{\fant{a}a}=-3\Box$.

By ``switching off" the gravitational interaction in the  theory, the field equations allow one to have a propagating scalar solution in the  Minkowski background. Explicitly, by introducing  a potential term $V(\phi)*1$ into the original BD Lagrangian, the field equations in this case reduce to
\be
P_{ab}\phi+\frac{\omega}{\phi}T_{ab}[\phi]+\eta_{ab}V(\phi)=0.
\ee
By requiring the existence of nontrivial solutions to these equations,  one determines the form of the self-interaction potential term and these solutions are studied in \cite{zanelli} recently.

\subsection{\texorpdfstring{$pp$}{}-waves in a metric \texorpdfstring{$f(R)$}{} gravity}

The simplest modification of the General theory of relativity encompassing sufficient generality involves the modification of the Einstein-Hilbert Lagrangian  to a general function of the scalar curvature of the form $f(R)$. Although the field equations for $f(R)$ models have been worked out long time ago in \cite{buchdahl-f(R)}, these fourth order models are studied intensively  by  the current motivations arising mainly from the recent cosmological observations. See, for example,  Refs. \cite{faraoni-sotiriou1,faraoni-sotiriou2,nojiri-odintsov-f(R)-rev}, for a thorough review on various aspects of $f(R)$ gravity models.

It is well-known that a generic $f(R)$ gravitational  model  with the Lagrangian
\be\label{f(R)-lag}
L
=
\tfrac{1}{2}
f(R)*1
\ee
have a dynamically equivalent scalar-tensor model  \cite{faraoni-sotiriou2, st-equivalent}. By introducing an auxiliary Lagrangian
\be\label{aux-lag}
L_{aux.}
=
\left\{f(\chi)+f'(\chi)(R-\chi)\right\}*1
\ee
in terms an auxiliary field $\chi$ with the prime denoting a derivative with respect to $\chi$. Assuming that $f''\neq 0$ and by using the field equations $\delta L_{aux.}/\delta\chi=0 $ that follow from (\ref{aux-lag}) to eliminate  the auxiliary scalar field $\chi$, one ends up with the equivalent Lagrangian of the form
 \be\label{f(R)-equivalent}
 L_{ST}
 =
 \frac{\phi}{2}\Omega_{ab}\wdg *\theta^{ab}-\tfrac{1}{2}V(\phi)*1
 \ee
similar to the original BD Lagrangian (\ref{BD-lag}) with the BD parameter $\omega=0$ while having  an additional potential term for the nonminimally coupled BD-type scalar field. However, the equivalent BD-type scalar-tensor model has no kinetic term for the scalar field $\phi$.
 The potential term in (\ref{f(R)-equivalent}) is obtained by the Legendre transform of the function $f(R)$:
\be\label{legendre}
V(\phi)=R(\phi)f'(R(\phi))-f(R(\phi))
\ee
 and consequently, the potential term $V(\phi)$  with $\phi\equiv f'(R)=df/dR$ in the resulting scalar-tensor equivalent Lagrangian is defined by
the Legendre transformation. (\ref{legendre}) also implies
$
{dV}/{d\phi}=R
$
and consequently
\be
f(R(\phi))
=
\phi\frac{dV}{d\phi}-V(\phi).
\ee

The contribution of the  $\tfrac{1}{2}V(\phi)*1$ to the metric field equations that follow from (\ref{f(R)-equivalent}) is of the form of a variable cosmological term $\tfrac{1}{2}V(\phi)*\theta^a$ However, such a term is incompatible with the $pp$-wave metric ansatz (\ref{pp-wave-ansatz}).
Thus, it is more convenient to discuss $pp$-wave solutions by making use of the Lagrangian (\ref{f(R)-lag}) as has previously been discussed, for example, in Ref. \cite{mohseni} recently by Mohseni.

The derivation of the field equations for the $f(R)$ models with minimally coupled matter fields $\psi$ can be obtained along the lines of
the BD equations which are derived   in some detail above (see also Ref. \cite{baykal-epjp}) and  it is straightforward to show that the metric field equations that follow from the coframe variation of the  Lagrangian 4-form (\ref{f(R)-lag}) are
\be\label{f(R)-eqns}
-f' *G^a+D*(df'\wdg \theta^a)+\tfrac{1}{2}(f-Rf')*\theta^a+\kappa^2 *T^a[\psi]=0
\ee
in the form similar to the BD field equations (\ref{BD-eqns1}). Here, $T^a[\psi]$ stands for energy-momentum 1-form for a matter field $\psi$. Because the $pp$-wave metric ansatz is incompatible with a cosmological constant, the form of a generic function $f(R)$ has to be restricted for the $pp$-wave ansatz to solve the corresponding field equations (\ref{f(R)-eqns}).

The fourth order terms can explicitly be written out in the form
\be
D*(df'\wdg \theta^a)
=
f''D*(dR\wdg \theta^a)+f'''dR\wdg i^a*dR
\ee
 and that this term vanishes identically for the $pp$-wave ansatz for which  $R=0$ identically. Consequently, for the ansatz (\ref{pp-wave-ansatz})  the metric field equations boil down  to the form
 \be\label{reduced-f(R)-eq}
-f'(0) *G^a+\tfrac{1}{2}f(0)*\theta^a+\kappa^2 *T^a[\psi]=0.
 \ee
 The cosmological-like  term for the scalar-tensor equivalent Lagrangian persists in Eq. (\ref{reduced-f(R)-eq}) as well and  the only way for the $pp$-wave ansatz to satisfy these equations is now transformed to the condition that  $f(0)=0$. In this particular case, the
 $f(R)$  theory field equations reduces to Einstein field equations with a new gravitational coupling constant $\kappa^2/f'(0)$.  Hence, the electrovacuum solution to the Einstein's field equations  with the effective coupling constant $\kappa^2/f'(0)$ are also solutions to $f(R)$ models with  the function $f(R)$
satisfying the condition $f(0)=0$ and $f'(0)>0$. Explicit forms for some of $f(R)$ models relevant to the cosmological applications satisfying this condition have been reported in \cite{mohseni}, in a study of  Aichelburg-Sexl type solutions in various modified gravity models.
It is also interesting to note that introducing a cosmological term $\Lambda*\theta^a$  to the field equations (\ref{f(R)-eqns}), the condition the compatibility condition $f(0)=0$ then becomes $f(0)+2\Lambda=0$. As a consequence one can conclude that  in general the $f(R)$ models admit a particular $pp$-wave solution with a cosmological constant term.

The linearized field equations of the $f(R)$ model has recently been studied in \cite{linearized-wave-sol1}  by making use of the scalar-tensor equivalent models of such theories and showed  explicitly that there is a massive scalar mode  of gravitational radiation in addition to the transverse modes.

\subsection{\texorpdfstring{$pp$}{}-waves in a nonminimal \texorpdfstring{$f(R)$}{} gravity}

Another popular $f(R)$ model that allow one to discuss $pp$-waves in the manner in line  with the discussion above is the model that has been recently introduced in \cite{nojiri-odintsov-non-min-fR,allemandi,bertolami} and it involves  two analytical functions $f_1(R)$ and $f_2(R)$.
The gravitational waves in the context of non-minimal matter coupling studied in this section has also previously been studied in \cite{} as well.

 The Lagrangian 4-form of the model can be written in the  form
\be\label{non-minimal-f(R)-lag}
L_{n.m.}
=
\frac{1}{2}
f_1(R)*1+[1+\lambda f_2(R)]L_m
\ee
with a new parameter $\lambda$ giving the strength of the nonminimal coupling of matter to the modified gravitational Lagrangian function $f_2(R)$.
As before, the matter Lagrangian 4-form $L_m\equiv \mathcal{L}_m*1$ is assumed to depend on the metric tensor but not on the
connection 1-forms. One can show that the metric field equations $*E^a=0$ that follow from $\delta L_{n.m.}/\delta \theta_a\equiv *E^a$  can be written compactly as
\begin{align}
&-(f_1'+2\lambda f_2'\mathcal{L}_m) *G^a
+
D*\left[d\left(f_1'+2\lambda f_2'\mathcal{L}_m\right)\wdg \theta^a\right]
\nonumber\\
&
\qquad+
\tfrac{1}{2}[(f_1-Rf_1')-2\lambda R f_2'\mathcal{L}_m ]*\theta^a
+
(1+\lambda f_2 )*T^a[\psi]=0\label{nonminimal-mod-field-eqns}
\end{align}
 in the same way as the $f(R)$  field equations (\ref{f(R)-eqns}) by using the constrained first order formalism \cite{baykal-delice-var}.
The derivation of the field equations   (\ref{nonminimal-mod-field-eqns}) proceeds first by extending the Lagrangian 4-form (\ref{non-minimal-f(R)-lag})
by the constraint term (\ref{zero-torsion-constraint-lag}) and then solving the connection equations for the Lagrange multiplier  and then subsequently
using it to obtain the total variational derivative with respect to the coframe as the metric field equations (\ref{nonminimal-mod-field-eqns})  as in the BD case. The matter energy-momentum 3-form $*T^a[\psi]$ is defined as the variational derivative $\delta L_{m}/\delta \theta_a$ as in the GR case.

 An important feature of the model that follow from the Lagrangian (\ref{non-minimal-f(R)-lag}) can be explained briefly in the present notation as follows. The Lagrangian 4-form (\ref{non-minimal-f(R)-lag})  is  apparently invariant under an arbitrary coordinate transformation and thus leads to the Noether identity $D*E^a=0$. (See Ref. \cite{obukhov-puetzfeld} for general conservation laws derived from Lagrange-Noether methods for the models with nonminimal couplings).  One can show, by  a direct computation of the covariant exterior derivative of the field equations (\ref{non-minimal-f(R)-lag}), that
\be\label{modified-cov-div}
D*T^a[\psi]=\frac{\lambda f'_2}{1+\lambda f_2 } dR\wdg \left(i^aL_m-*T^a[\psi]\right)
\ee
as a result of the nonminimal coupling of matter fields.
Note that for $\lambda=0$  the above expression reduce to the previous case and implies the usual covariant expression $D*T^a[\psi]=0$ for the matter energy-momentum forms as one should expect on consistency grounds.

The modified covariant expression (\ref{modified-cov-div}) implies that the
massive test particles do not follow geodesic curves that invalidates the principle of equivalence. It is argued \cite{bertolami} that the term on the right hand side can be interpreted as an extra force arising from the particular nonminimal matter coupling defined by (\ref{non-minimal-f(R)-lag}).
Accordingly, as will be exemplified below,  such a nonminimal coupling is naturally bound to modify the matter field equations as well.

Let us consider, for example, the electromagnetic field as the matter Lagrangian present in the above model.  Let us assume further that $dF=0$  which can also be imposed to the field equations by introducing an appropriate Lagrange multiplier term. Explicitly, for the Maxwell Lagrangian 4-form of the form
\be
L_m=L_m[g, F]=-\tfrac{1}{2}F\wdg*F
\ee
the electromagnetic field equation given by the variational derivative $\delta L_{n.m}/\delta F=0$ is modified to the form \cite{sert1,sert2,sert3,sert4}
\be
d[(1+\lambda f_2)*F]=0.
\ee
Note that the modified electromagnetic field equation can also be rewritten in the alternate form
\be\label{modified-reduced-em-field}
d*F
+
\frac{\lambda f_2'}{1+\lambda f_2} dR\wdg *F=0.
\ee
It is possible to show that Eq. (\ref{modified-reduced-em-field}) can also be derived, in a somewhat indirect manner, by making use of the general formula (\ref{modified-cov-div}). For the null electromagnetic field ansatz (\ref{pp-electro-ansatz}), the modified field equation
(\ref{modified-reduced-em-field}) simplifies to the familiar source-free Maxwell's equation $d*F=0$.

Now returning to discussion of the gravitational waves, by inserting the $pp$-wave metric ansatz into the field equations (\ref{nonminimal-mod-field-eqns}) and assuming that the only matter Lagrangian present is the electromagnetic Lagrangian and noting  the fact that for the null fields $L_m=F\wdg *F=0$, Eq. (\ref{nonminimal-mod-field-eqns}) reduces to
\be
-f_1'(0) *G^1
+
\tfrac{1}{2}f_1(0)*l
+
(1+\lambda f_2(0) )*T^1[F]=0.
\ee
As in the previous case, with the further assumption $f_1(0)=0$, which is now required for the satisfaction of both the electromagnetic  and the metric field equations, one reobtains  Einstein field equations in the form
\be
-
*G^1
+
\left(\frac{1+\lambda f_2(0) }{f_1'(0) }\right)*T^1[F]=0
\ee
with the gravitational coupling constant $\kappa^2$ is now replaced by the constant factor in parenthesis on the right hand side.
Consequently, one can apply the formula given in  the previous  section to write down the $pp$-wave solutions of the form given in (\ref{electrovac-gr-sol}) to the gravitational model with the Lagrangian (\ref{non-minimal-f(R)-lag}).

\subsection{\texorpdfstring{$pp$}{}-waves in a gravitational model with a non-minimal Maxwell coupling}

The nonminimal coupling of general matter Lagrangian studied in the preceding section discussed the particular nonminimal coupling of type $f(R)F\wdg *F$. On the other hand, one can formulate  a more general nonminimally coupled Einstein-Maxwell system by considering  all mathematically admissible coupling terms involving curvature and the square of the Faraday tensors. In particular, there are many mathematically admissible interaction terms of the general form $RF^2$ that one can consider \cite{gurses-halil}.  Such coupling terms can explicitly be written in  the following forms:  $F\wdg F_{ab}*\Omega^{ab}$, $F\wdg F_{ab}\Omega^{ab}$, $F^a\wdg {R}_{a}\wdg *F$ or $F^a\wdg {R}_{a}\wdg F$ \cite{dereli-sert-pp-wave,goenner-nonminimal-coupling} where $R_{a}$ is the Ricci 1-form $R_{a}=R_{ab}\theta^b$ that can be defined in terms of the contraction $R_a\equiv i_b\Omega^{b}_{\fant{a}a}$.

The $pp$-wave solution to the nonminimal coupling involving the term of the particular  form $F\wdg F_{ab}*\Omega^{ab}$ has been recently studied in \cite{dereli-sert-pp-wave} based on the Lagrangian of the form
\be\label{dereli-sert-lag}
L
=
\frac{1}{2\kappa^2}\Omega_{ab}\wdg*\theta^{ab}
-
\frac{1}{2}F\wdg *F
+
\frac{\gamma}{2}F\wdg F_{ab}*\Omega^{ab}.
\ee
where $\gamma$ is a coupling constant and $F_{ab}$ denotes   the components of the Faraday 2-form $F=\tfrac{1}{2}F_{ab}\theta^{a}\wdg \theta^{b}$ relative to an orthonormal coframe. Note that the nonminimal coupling term involving the Riemann tensor can explicitly  be written out in the form
\be\label{prasanna-type-rf^2-coupling}
F\wdg F_{ab}*\Omega^{ab}
=
\tfrac{1}{2}F_{ab}R^{ab}_{\fant{cd}cd}F^{cd}*1.
\ee
The expression (\ref{prasanna-type-rf^2-coupling}) has the same form relative to a coordinate basis as well and
that the nonminimal $RF^2$ coupling of this particular  type was first considered by Prasanna \cite{prasanna} long time ago.

By using the constrained first order formalism  it is possible to obtain the metric field equations that follow from (\ref{dereli-sert-lag}) which can be written in the form \cite{dereli-sert-pp-wave}
\be\label{dereli-sert-mereic-eqn}
-\frac{1}{\kappa^2}*G_a+*T_a[F]+D\lambda_a+\gamma F_{ac}(i_bF)\wdg * \Omega^{bc}+\gamma *T_a[F,\Omega]=0
\ee
by using the  auxiliary 3-form definition
\be
*T_a[F,\Omega]
\equiv
-\tfrac{1}{4}F_{bc}\left(i_a F\wdg*\Omega^{bc}+i_a\Omega^{bc}\wdg *F-F\wdg i_a *\Omega^{bc}-\Omega^{bc}\wdg i_a*F\right)
\ee
and likewise, the 3-form $*T_a[F]$ denotes the electromagnetic energy-momentum 3-form defined in Eq. (\ref{maxwell-en-mom-3-form}) above.
The vector-valued Lagrange multiplier 2-form $\lambda^a$ is obtained by solving the connection equations and it has the explicit expression given by
\be
\lambda^a
=
\gamma i_bD (F^{ba}*F)+\frac{\gamma}{4}\theta^{a}\wdg i_bi_c D(F^{bc}*F).
\ee
In addition, the modified field equations for the Faraday 2-form  then takes the form $dF=0$ and
\be
d*(F-\gamma F_{ab}\Omega^{ab})=0
\ee
which involves third order the partial derivatives of the metric variable in general.

For the electrovacuum $pp$-wave metric ansatz (\ref{pp-wave-ansatz}) and (\ref{pp-electro-ansatz}) above, the nonminimal interaction term vanishes identically and consequently the field equations for the Faraday 2-form  reduce to Maxwell's equations $dF=d*F=0$. Moreover, another consequence of the vanishing of the nonminimal coupling term is that $*T_a[F,\Omega]\equiv0$. The term of the form $\gamma F_{ac}(i_bF)\wdg * \Omega^{bc}$ vanishes for the electrovacuum $pp$-waves
identically as well.

For the $pp$-wave metric ansatz (\ref{pp-wave-ansatz}), the only nontrivial contribution of the nonminimal coupling terms arises from the Lagrange multiplier 2-forms. Explicitly, $\lambda^1$ component can be expressed in the form
\be\label{nonminimal-einstein-maxwell-lag-multiplier-form}
\lambda^1
=
\gamma\left[f_{\zeta\zeta}f_{\bar{\zeta}}*(k\wdg m)+f_{\bar{\zeta}\bar{\zeta}}f_{\zeta}*(k\wdg \bar{m})\right]
\ee
whereas one has $\lambda^0=\lambda^2=0$.
Using the result (\ref{nonminimal-einstein-maxwell-lag-multiplier-form}) in the expression for  the  covariant exterior derivative of the Lagrange multiplier 2-form, one finds
\be
 D\lambda^1
 =
2\gamma f_{\zeta\zeta}f_{\bar{\zeta}\bar{\zeta}}*k.
\ee
Now taking all these results into account, the metric equation (\ref{dereli-sert-mereic-eqn}) for $a=1$ eventually leads to
the following second order partial differential equation
\be\label{dereli-sert-eqn}
H_{\zeta\bar{\zeta}}
=
\kappa^2 f_{\zeta}\bar{f}_{\bar{\zeta}}
-
\gamma\kappa^2f_{\zeta\zeta}\bar{f}_{\bar{\zeta}\bar{\zeta}}.
\ee
The general solution to Eq. (\ref{dereli-sert-eqn}) can be written in the form
\be
H(u, \zeta, \bar{\zeta})
=
h(u, \zeta)+\bar{h}(u, \bar{\zeta})
+
\kappa^2 f\bar{f}
-
\gamma\kappa^2f_{\zeta}\bar{f}_{\bar{\zeta}}
\ee
in terms of the complex functions $h(u, \zeta)$ and $f(u, \zeta)$ having arbitrary $u$-dependence that are analytic in the variable $\zeta$ as in the previous cases.

A family of solution to the field equation  (\ref{dereli-sert-eqn}) has been reported recently  in \cite{dereli-sert-pp-wave}  associated with  a partially massless spin-2 photon and a partially massive spin-2 graviton that were introduced  in \cite{deser-waldron} by Deser and Waldron previously. For the  particular $pp$-wave solution reported in (\ref{dereli-sert-eqn}), the electromagnetic function $f$ introduced into the electrovacuum ansatz has explicitly the form
\be
f(u,\zeta)
=
f_1(u)\zeta+f_2(u)\zeta^2
\ee
corresponding to a superposition of two null electromagnetic fields.

\subsection{\texorpdfstring{$pp$}{}-waves in a general quadratic curvature gravity}\label{qc-section}

Considered as a low energy limit of some theory of quantum gravity, the General Relativity is expected to receive correction terms
involving higher powers of curvature tensor to the Einstein-Hilbert  action. In particular, the quadratic curvature terms in the effective action
are essential  for the power-counting renormalizability \cite{stelle}, although they are not free from the problem of the ghosts. The quadratic curvature terms also appear in the low-energy effective action in string theory \cite{deser-redlich}. More recently, the modified gravity models involving the quadratic curvature terms following from the Lagrangian $f(R, R_{ab}R^{ab}, R^{abcd}R_{abcd})$ in a cosmological context are discussed in \cite{carroll-et-al}. The complicated modified gravitational models of this type are usually discussed in the to investigate the effect of terms that gain significance in the case where the spacetime has a small curvature. Most general ghost-free gravitational Lagrangian having nonlocal terms but  having a better UV behavior, and also leading to general relativity in IR limit has been introduced  in \cite{biswas-mazumdar}. The quantum aspects of such a ghost-free and singularity-free theory is alsodiscussed in \cite{mazumdar}.

The general quadratic curvature gravity  leads to a set of fourth order field equations and naturally, there are not as many exact solutions as in the general theory of relativity. The gravitational wave solutions to the quadratic curvature (QC) gravity has been studied in 
\cite{adamowicz1,obukhov-pereira-rubilar,adamowicz2,adamowicz3} and recently in \cite{garcia-macias-putzfeld-socorro,obukhov-plane-waves-in-mag,pasic-barakovic} in the more general setting  of metric-affine gravity.
The QC gravity has a long history \cite{schmidt-history-of-fourth-order-grav} initiated shortly  after the introduction of GR. One of the earliest example of the quadratic curvature gravity is the  conformally invariant gravity following from  the square of the Weyl's conformal tensor introduced by Bach 
\cite{bach} in 1921.

The gravitational wave solutions to quadratic curvature gravity with matter source have also been discussed in the literature. $pp$-wave solutions with a Dirac delta
source, have been discussed in \cite{campanelli-lousto}. More recently, a perturbative approach has been introduced to obtain some particular solution
to Ricci-squared gravity \cite{neto-odliyo} by de Rey Neto et al.

The particular QC model discussed in this section has been introduced as a Yang-Mills type action  in \cite{sky1,sky2,sky3,sky4} and following the terminology introduced in \cite{mielke1,mielke2}, the Lagrangian will be denoted by $L_{SKY}$ below with the acronym ``SKY" standing for Stephenson-Kilmister-Yang. The scale invariant gravitational  Lagrangian 4-form, depending on both the metric (owing to the presence of the Hodge dual) and the connection expressed in terms of curvature 2-form, explicitly reads
\be\label{sky-lag}
L_{SKY}
=
\tfrac{1}{2}\Omega_{ab}\wdg*\Omega^{ab}.
\ee
The vacuum  field equations for the pseudo-Riemannian metric that follow from (\ref{sky-lag}) by using the constrained coframe variational derivative  in the first order formalism can be written in the form
\be\label{sky-eqn}
*E^a
=
D\lambda^a+*T^a[\Omega]=0
\ee
with the Lagrange multiplier 2-form having the explicit expression
\be\label{sky-eqn2}
\lambda^a
=
2i_bD*\Omega^{ba}+\tfrac{1}{2}\theta^a\wdg i_bi_c D*\Omega^{bc}.
\ee
Consequently, (\ref{sky-lag}) leads to the equations that are fourth order in the partial derivatives of the metric components as indicated by the   Lagrange multiplier term in (\ref{sky-eqn}). The terms that are quadratic in curvature components, namely $*T^a[\Omega]$ term in (\ref{sky-eqn}), have the explicit form
\be\label{sky-qc-term-in-field-eqn}
*T_a[\Omega]
\equiv
-\tfrac{1}{2}\left(i_a\Omega^{bc}\wdg*\Omega_{bc}-\Omega_{bc}\wdg i_a*\Omega^{bc}\right)
\ee
which arise from the commutation of the variational derivative with the Hodge dual operator. The energy-momentum-like 3-form term (\ref{sky-qc-term-in-field-eqn}) can be obtained in a way similar to a matter energy-momentum 3-form calculated by a coframe variation.

As in the previous cases, the $pp$-wave metric ansatz (\ref{pp-electro-ansatz}) simplifies the metric equations (\ref{sky-eqn}) considerably. It is straightforward to show that all the components of $*T_a[\Omega]$ vanishes identically in this case. The only nonvanishing contribution of the Lagrange multiplier term 
then turns out to be an expression in terms of the covariant exterior derivative $D*\Omega^{1}_{\fant{a}2}$ and its complex conjugate $D*\Omega^{1}_{\fant{a}3}$.
By using the general expression
\be
D*\Omega^{ab}
=
d*\Omega^{ab}+\omega^{a}_{\fant{a}c}\wdg *\Omega^{cb}+\omega^{b}_{\fant{a}c}\wdg *\Omega^{ac}
\ee
for the covariant exterior derivative, it is straightforward to show that
\be
D*\Omega^{1}_{\fant{a}2}
=
-2H_{\zeta\zeta\bar{\zeta}}*k
\ee
and consequently, using the fact that $dk=0$ identically  one can also show that the Lagrange multiplier can be written in the form
\be
\lambda^1
=
-
4*d(H_{\zeta\bar{\zeta}} k).
\ee
Using this result in (\ref{sky-eqn}), and also noting that $D\lambda^1=d\lambda^1$,  one eventually  ends up with
\be
*E^1
=
-4d*d(H_{\zeta\bar{\zeta}}k)=0.
\ee
By using the fact that $\Phi_{22}=2H_{\zeta \bar{\zeta}}$, this can be rewritten as an equation for the 1-form $\Phi_{22}k$ as
\be\label{QC-krechtman-simp-eqn}
*E^1
=
-2d*d(\Phi_{22}k)=0
\ee
which is to be compared to the corresponding GR expression (\ref{simplified-einstein-form-for-pp}).
The general solution to the homogeneous equation (\ref{QC-krechtman-simp-eqn})  can be written 
$\frac{1}{2}\Phi_{22}=\varphi(u, \zeta)+\bar{\varphi}(u, \bar{\zeta})$ in terms of an arbitrary function $\varphi(u, \zeta)$
analytic in $\zeta$. Subsequently, the homogeneous fourth order QC field equations can be written as an inhomogeneous equation in the form 
$H_{\zeta\bar{\zeta}}=\varphi(u, \zeta)+\bar{\varphi}(u, \bar{\zeta})$. 

It is interesting to note that the field equation (\ref{sky-eqn}) can also be rewritten in the following compact form as
\be\label{qc-reduced-eqn}
*E^1
=
2d*d*d*dl=-4H_{\zeta\zeta\bar{\zeta}\bar{\zeta}}*k=0
\ee
for the real profile function $H$ and the basis coframe field $l$ by using the relation (\ref{simplified-einstein-form-for-pp}).

These interrelations suggest, among other things,  that the linearized form of the metric field equations for the quadratic curvature theory \cite{deser-tekin-qc-energy1,deser-tekin-qc-energy2,deser-tekin-qc-energy3,baykal-qc-energy} has formally the structure similar to those of GR and that the resulting 
differential operator is a consequence of the curvature tensor. 

Another  observation related to the expression (\ref{qc-reduced-eqn}) is that it is probably the simplest subcase of a remarkable theorem due to G\"urses et al. \cite{gurses-hervik-sisman-tekin}. The theorem allows one to find the exact solution to a wide class of modified gravitational models governed by the Lagrangian depending on a function of the Riemann tensor $f(R^{ab}_{\fant{ab}cd})$. (see, \cite{amati,horowitz-steif} and also Theorem 1.3 in \cite{hervik-pravda-pravdova}). For the Kundt class of Petrov type N metrics with all the scalar invariants being constant,  the theorem  states that any  symmetric,  second-rank tensor constructed from the Riemann tensor and its covariant derivatives  can be expressed  as a  linear combination of  the metric components, the  traceless Ricci tensor  components and the higher covariant derivatives of the traceless Ricci tensor  components. 
Consequently, one is bound to end up with (\ref{qc-reduced-eqn}) for any gravitational model based on any quadratic curvature invariant 
for the pp-wave metric ansatz (\ref{pp-wave-ansatz}).

Now consider the quadratic curvature gravity model obtained by adding the Einstein-Maxwell Lagrangian
\be\label{qc+einstein-eqn}
L_{E-M}
=
\frac{1}{2\kappa^{2}}\Omega_{ab}\wdg*\theta^{ab}-\frac{1}{2}F\wdg*F
\ee
to the quadratic curvature Lagrangian (\ref{sky-lag}).  One can show that the field equation following from the total Lagrangian can be written in the form
\be\label{3-form-total-qc-eqn}
d*d(\ell^2+*d*d)l=2\kappa^2f_{\zeta}\bar{f}_{\bar{\zeta}}*k
\ee
for the $pp$-wave metric ansatz. In Eq. (\ref{qc+einstein-eqn}), $\ell^2$ stands for an appropriate coupling constant for the quadratic curvature terms in the total Lagrangian. The fourth order equation (\ref{3-form-total-qc-eqn}) for 3-forms can be written as a second order equation in the form
\be\label{reduced-3-form-total-qc-eqn}
d*d\sigma=2\kappa^2f_{\zeta}\bar{f}_{\bar{\zeta}}*k
\ee
in terms of the 1-form $\sigma$ as an equation analogous to Eq. (\ref{3-form-eqn-gr}) by defining the auxiliary 1-form $\sigma $ as
\be\label{auxiliary-sigma-def}
\sigma
\equiv
(*d*d+\ell^2)l.
\ee
Now for a given electromagnetic potential $f(\zeta,\bar{\zeta})$, Eq. (\ref{reduced-3-form-total-qc-eqn}) is now to be solved for 1-form $\sigma$ and subsequently the resulting expression for it is to be used to solve Eq.  (\ref{auxiliary-sigma-def}) for the 1-form $l=dv+Hdu$, or equivalently the profile function $H$ to obtain a solution.

For Maxwell field coupled to pure QC gravity with Lagrangian (\ref{sky-lag}) with an appropriate  coupling constant $\gamma$, 
one can show that the general solution to (\ref{reduced-3-form-total-qc-eqn}) can be written in the form 
\be
H=H_h+H_p,
\ee
as a sum of a homogeneous solution $H_h(u, \zeta, \bar{\zeta})$ and a particular solution $H_p(u, \zeta, \bar{\zeta})$ with the explicit form
\be
H_p(u,\zeta,\bar{\zeta})
=
-\frac{\gamma}{2}\int^{\zeta}f(u, \zeta')d\zeta'\int^{\bar{\zeta}}\bar{f}(u, \bar{\zeta}')d\bar{\zeta}'.
\ee

In order to discuss the linearized QC field equations around the flat Minkowski background, it is convenient  to consider a more general QC Lagrangian instead of Lagrangian (\ref{sky-lag}). For a discussion in parallel  to  the SKY Lagrangian above,  the expressions for the corresponding  field equations relative to an orthonormal/null coframe is required. In this manner the explicit calculations with the linearized equations serve to justify the universality property of the pp-wave metric  somewhat indirectly and laboriously only  for an unnecessarily  restricted (but practically convenient) subset of generic gravitational models.

The most general QC Lagrangian  and the field equations can be expressed in terms of differential forms as follows.
In  four spacetime dimensions, the general quadratic curvature gravity Lagrangian can be written in a preliminary form as
\be\label{gen-qc-lag}
L=
\gamma \Omega_{ab}\wdg *\Omega^{ab}+\alpha R_a\wdg *R^a+\beta R^2*1
\ee
involving the quadratic curvature terms built out of the contractions of the curvature 2-form, namely the Ricci-squared and  scalar curvature-squared terms
with respective coupling constants $\alpha,\beta$ and $\gamma$ being another coupling constant.

At this point, it is convenient to recall the Lovelock's theorem \cite{lovelock,lovelock2}, stating that the most general gravitational actions that generalize Einstein-Hilbert  action leading to second order field equations in metric components involve the Einstein-Hilbert action complemented with a cosmological constant term  and terms with curvature polynomials,  the dimensionally-continued Euler-Poincar\'{e} (EP) forms (these forms are also commonly known as the Gauss-Bonnet terms and the famous Chinese  geometer Chern was the first to use dimensionally-continued Euler-Poincar\'{e} forms to generalize
the Gauss-Bonnet theorem to higher dimensions \cite{chern}). In four spacetime dimensions, the EP term involving quadratic curvature expression does not contribute to  the gravitational field equations. Thus, one can exploit the EP forms that are quadratic in curvature components  to remove the redundancy in the general Lagrangian (\ref{gen-qc-lag}) without loss of generality. Explicitly,  the dimensionally-continued Euler-Poincare term
\be\label{EP-form1}
L_{EP}
=
\tfrac{1}{4}\Omega_{ab}\wdg \Omega_{cd}* \theta^{abcd}
\ee
can be expressed in  the following  linear combinations of quadratic terms:
\begin{align}
L_{EP}
&=
\tfrac{1}{2}\Omega_{ab}\wdg* \Omega^{ab}-R_a\wdg *R^a+\tfrac{1}{4}R^2*1
\\
&=
\tfrac{1}{2}C_{ab}\wdg* C^{ab}-\frac{1}{2}\left(R_a\wdg *R^a-\tfrac{1}{3}R^2*1\right)
\end{align}
by a straightforward computation in four dimensions.  After some tedious computation in exterior algebra starting from the expression in Eq. (\ref{EP-form1}), it is possible to show that the $L_{EP}$ term can be written  as an exact form as
\be\label{corrected-EP-form}
L_{EP}
=
d\left(
\omega_{ab}\wdg \Omega^{cd}-\tfrac{1}{3}\omega_{ae}\wdg \omega^{e}_{\fant{a}b}\wdg \omega^{cd}
\right)\epsilon^{ab}_{\fant{ab}cd}
\ee
in a form remarkably similar to the gravitational Chern-Simons term arising from another topological term, the Pontryagin form, namely the 4-form $\Omega_{ab}\wdg \Omega^{ba}$ \cite{mielke1, mielke2}, cf. Eq. (\ref{CS-3-form}) below.
Consequently, because $L_{EP}$ is an exact form that will not contribute to the field equations in four dimensions, and one can add  $L_{EP}$ to the general quadratic curvature action (\ref{gen-qc-lag}) to eliminate one of the terms in (\ref{gen-qc-lag}) in favor of the remaining two. Following this custom, one can  write the most general quadratic curvature Lagrangian in the form
\be\label{qc-reduced-action}
L=
\alpha R^2*1+\beta R_a\wdg *R^a
\ee
in \emph{four spacetime dimensions}. The $pp$-wave solutions to the model that follows from (\ref{qc-reduced-action}) has previously been studied in, for example, \cite{buchdahl1,buchdahl2} previously. With  the coupling constant satisfying the relation $\alpha=-3\beta$, the most general Lagrangian (\ref{qc-reduced-action}) corresponds to  the conformally-invariant quadratic curvature model introduced by Bach mentioned above and the corresponding field equations is expressed in terms of  so-called Bach tensor \cite{cotton}.

In spacetime dimensions $n\geq4$, the Petrov III and N solutions to general the quadratic curvature gravity model defined by the Lagrangian (\ref{qc-reduced-action}) supplemented with the term (\ref{corrected-EP-form}) have been recently studied in \cite{malek-pravda} using the classification of Weyl curvature tensor \cite{HD-petrov-class} that extends the four dimensional classification of Weyl tensor in four dimensions.

In the present notation, the general form of the field equations that follow from (\ref{qc-reduced-action}) have been reported in \cite{baykal-grg-qc} and they explicitly read
\be\label{qc-general-eqs}
*E^a
=
2D*D\left\{\beta R^a+\left(2\alpha+\tfrac{1}{2}\beta\right) R\theta^a\right\}+*T^a[\alpha,\beta]=0
\ee
where the quadratic term $*T_a[\alpha,\beta]$ has the explicit form
\be\label{qc-general-eqs2}
*T_a[\alpha,\beta]
\equiv
\Omega_{bc}\wdg i_a *X^{bc}-\tfrac{1}{2}i_a(\Omega_{bc}\wdg *X^{bc})
\ee
and the auxiliary tensor-valued 2-form $X^{ab}$ standing for
\be\label{auxiliary-qc-2form}
X^{ab}
\equiv
\theta^a\wdg (\beta R^b+\alpha R\theta^b)-\theta^b\wdg(\beta R^a+\alpha R\theta^{a}).
\ee
The trace of the metric field equations can be found by the wedge product $\theta_a\wdg *E^a= E^{a}_{\fant{a}a}*1\equiv E*1$. One finds
\be
E*1
=
-\left[4\alpha (n-1)+\beta n\right]d*dR+\tfrac{1}{2}(n-4)\Omega_{ab}\wdg *X^{ab}.
\ee
In four spacetime dimensions with the coupling constants satisfying $\alpha+3\beta=0$, the trace vanishes identically and one can shown that the QC Lagrangian  
corresponds to the $C_{ab}\wdg *C^{ab}$ up to a boundary term.
These results  are in harmony with the results previously reported, for example in  \cite{buchdahl1}, by using the tensorial methods. 

In connection with the particle content,  the novel effect arising from the quadratic curvature model is that there is now additional  transverse massive  scalar and the massive spin-2 (ghost) modes. See, for example, Refs. \cite{linearized-wave-sol3,rey-neto} for further details on this issue.

\subsection{QC gravity linearized  around a flat background}

With the field equations formulated in terms of the differential  forms and the exterior algebra at hand, it is now straightforward to show by inspection that the most general quadratic curvature vacuum equations (\ref{qc-general-eqs}) reduce to $d*dR^1=0$ leading to the result given in Eq. (\ref{qc-reduced-eqn}). 

The quadratic curvature field equations in the linearized form can also be put in a compact form as in (\ref{simplified-einstein-form-for-pp}) for the 
Einstein's field equations. As it has been noted above, it is well-known that the $pp$-waves constitute their own generalization. 
In particular, it is possible to show explicitly that the general QC field equations (\ref{qc-general-eqs}) linearized around Minkowski background  reproduces the fourth order linear partial differential equation (\ref{qc-reduced-eqn}) for the profile function \cite{baykal-dereli-linearization}.
Explicitly, by considering a metric $g^L$ defined in terms of the metric perturbation coefficients $h_{ab}$ as
\be\label{lin-metric}
g^L
=
\eta+2h_{ab}e^a\ot e^b
=
h_a \ot e^a+e^a\ot h_a
\ee  
around the flat metric $\eta=\eta_{ab} e^a\ot e^b$, one can show that the $pp$ wave metric (\ref{pp-wave-ansatz}) linearizes the general quadratic curvature field equations. (A label $L$ refers to a tensorial quantity linear in $h_{ab}=h_{ba}$). In the second equality on the right hand side (\ref{lin-metric}), $h_a$ is the perturbation 1-form defined as $h_a=h_{ab}e^b$ in terms of the basis coframe 1-forms of the Minkowski spacetime $\{e^a\equiv dx^a\}$ and 
$x^a=\{u, v, \zeta, \bar{\zeta}\}$ for convenience.
For the $pp$-wave metric ansatz $g=g^L$ by assumption and that the only nonvanishing perturbation 1-form can be expressed in terms of the profile function 
$H=H(u,\zeta,\bar{\zeta})$ as $h^1=Hdu$ relative to the null coframe of the Minkowski metric.

At the linearized level, the field equations (\ref{qc-general-eqs}) can conveniently be examined by introducing the  1-form 
$E_a^L=E_{ab}^Le^b$ by $(*E^a)^L=\star E^a_L$ where $\star$ denotes the  Hodge dual operator in the Minkowski spacetime. 
The linearized vacuum equations $\star E^a_L=0$ then read
\be\label{lin-qc-eqns}
d\star d [\beta G_L^a+(2\alpha+\beta)R_L dx^a]=0
\ee
where $G^L_a$ is linearized Einstein 1-form $G^L_a=G^L_{ab}e^b$ and $R^L$ is the linearized curvature scalar and the term $*T[\alpha, \beta]$ vanishes identically at the linearized level. Further simplification of the linearized equations (\ref{lin-qc-eqns}) follows because  $R^L=0$ and $R^a_L=G^a_L$ for the $pp$-wave metric.
Consequently, the linearized vacuum equations take the form 
\be
\beta d\star dG^a_L=0.
\ee

It is well-known that general quadratic curvature gravity defined by (\ref{qc-reduced-action}) admits also maximally symmetric vacua as well as the solutions of the form a product form $M_1\times M_2$ of two two dimensional manifolds $M1$ and $M_2$. The type N gravitational wave solutions of  quadratic curvature gravity on these backgrounds  have in general nonvanishing $*T[\alpha, \beta]$   (see, for example, \cite{baykal-gr-waves}).

The linearized Einstein 3-forms $\star G^a_L\equiv (*G^a)_L$ can be 
expressed in terms of the perturbation 1-forms as
\be
\star G^a_L=d\left[e^b\wdg\star \left( dh_b\wdg e^a\right)\right]
\ee
where they reduce to the form
\be
\star G^a_L=d\star dh^a
\ee
in general in the gauge (transverse-traceless gauge) defined by $\partial_a h^{a}_{\fant{a}b}=0$ and $h=0$, or equivalently assuming that  $i_adh^a=0$. One also has $R^L=0$  and that the vacuum equations (\ref{lin-qc-eqns}) eventually reduce to form
\be\label{reduced-lin-qc}
d\star d \star d\star dh^a=0
\ee 
which yields the result (\ref{qc-reduced-eqn}) obtained above for the SKY Lagrangian provided that  one specializes to the null coframe in (\ref{reduced-lin-qc})
with the perturbation 1-form $h^{1}=H(u,\zeta,\bar{\zeta}) du$ which defines the profile function for the $pp$-wave metric (\ref{pp-wave-ansatz}).

The general quadratic curvature gravity are linearized around a curved background, for example, in \cite{gurses-sisman-tekin2012}
explicitly showing that a particular Kerr-Schild form of AdS wave metric linearizes the field equations. 

\subsection{The vacua of general QC gravity}

The field equations (\ref{sky-eqn}) and (\ref{qc-general-eqs}) hold in general in  $n\geq3$ spacetime dimensions. They evidently admit an $n$ dimensional flat spacetime as a vacuum solution. In addition, they also admit maximally symmetric spacetime vacuum solutions as well. 
Although the solutions considered in the current paper are all confined to four spacetime dimensions, the formulation of the field equations allows a brief discussion of the quadratic curvature gravity assuming  a general spacetime dimensions $n\geq3$. 

Let us consider a maximally symmetric spacetime its  the 2-form curvature of the explicit form
2-form of the form
\be\label{max-sym-curv-form}
\Omega^{ab}
=
k\theta^{ab}
\ee
for some constant $k$. Such a spacetime satisfies Einstein's $\Lambda$-vacuum field equations of the form 
\be
*G^a+\Lambda*\theta^a=0
\ee
with a cosmological constant $\Lambda$ given by
\be
k
=
\frac{2\Lambda}{(n-1)(n-2)}.
\ee

The field equations corresponding to the general lagrangian (\ref{gen-qc-lag}) can be put into the form similar to the one given in (\ref{qc-general-eqs}) for the reduced QC Lagrangian simply by combining (\ref{sky-eqn}), (\ref{sky-eqn2}) and (\ref{sky-qc-term-in-field-eqn}).
For the curvature 2-forms of the form (\ref{max-sym-curv-form}), the auxiliary 2-forms the fourth order terms vanishes identically as a consequence of vanishing torsion for the Levi-Civita connection. In this case, one can readily show that the corresponding auxiliary 2-forms become
\be
X^{ab}
=
2\gamma \Omega^{ab}+\beta (\theta^a\wdg R^\beta-\theta^b\wdg R^a)+2\alpha R\theta^{ab}.
\ee
For the curvature 2-forms of the form (\ref{max-sym-curv-form}) the above  expression for the auxiliary 2-form reduces to
\be\label{max-sym-auxiliary-2form}
X^{ab}
=
2\left[\gamma+(n-1)(\alpha n+\beta)\right]k\theta^{ab}.
\ee
Consequently, (\ref{max-sym-auxiliary-2form}) can be inserted into the general expression (\ref{qc-general-eqs2}) to show that the general quadratic curvature field equations that follow from (\ref{gen-qc-lag}) reduce to 
\be\label{qc-max-sym-vacuum-eqns}
*E^a
=
(n-4)(n-1)\left[\gamma+2(n-1)(\alpha n+\beta)\right]k^2*\theta^a=0
\ee
for vacuum.

It is interesting to note that (\ref{qc-max-sym-vacuum-eqns}) are satisfied identically in four spacetime dimensions for $n=4$ for all values of the QC coupling constants $\alpha, \beta, \gamma$. In this regard, four spacetime dimensions can be considered as an exceptional case. 

The brief discussion above on maximally symmetric vacuum solutions of general QC gravity vacuum equations suggests that one can also consider
gravitational waves around these background vacua admitted by the quadratic curvature gravity.  Such solutions 
in generic gravity theories, which are based on arbitrary functions of Riemann tensor, have previously been  studied in 
\cite{gurses-hervik-sisman-tekin,gurses-sisman-tekin2012,gurses-sisman-tekin} in AdS background. $pp$-waves  and AdS wave-type solutions in the context of modified gravitational theories lacking Lorentz invariance have recently  been discussed in \cite{gurses-senturk} by G\"urses and \c{S}ent\"urk.

\subsection{AdS-wave solutions in QC gravity}

Now, we have the result that QC gravity admits maximally symmetric vacuum solutions, it is possible to construct 
an AdS-wave solution by incorporating the  the results presented in Sect. \ref{ads-waves} into the QC field equations  (\ref{qc-general-eqs}) above confining the discussion back to four spacetime dimensions. The results presented in this section can be extended to other type N gravitational waves in curved  backgrounds discussed in the first part (See, for example, \cite{baykal-gr-waves}) for the QC gravity. 

For the AdS-wave ansatz (\ref{ads-wave-ansatz}), the auxiliary 2-form $X^{ab}$ are given by
\begin{align}
X^{01}
&=
-(4\alpha+\beta)\frac{3}{a^2}k\wdg l,
\\
X^{02}
&=
-(4\alpha+\beta)\frac{3}{a^2}k\wdg m,
\\
X^{12}
&=
-(4\alpha+\beta)\frac{3}{a^2}l\wdg m-\beta\Phi_{22}k\wdg m,
\\
X^{23}
&=
-(4\alpha+\beta)\frac{3}{a^2}m\wdg \bar{m},
\end{align}
where $\Phi_{22}$ has an explicit expression  given in (\ref{ads-curvature-spinors}). One can show that for the above auxiliary 2-forms lead to result 
$*T^a[\alpha, \beta]=0$ for $a=0, 2, 3$ for the AdS-wave metric. The only nonvanishing component reads
\be
*T^1[\alpha,\beta]
=
\frac{2\beta}{a^2}\Phi_{22}*k.
\ee

Furthermore, $*E^0=0=*E^2$ are satisfied identically  and the remaining QC equation eventually leads to the following equation 
\be
*E^1
=
2\beta D*\left(d R^1+\omega^{1}_{\fant{a}2}\wdg  R^2+ \bar{\omega}^{1}_{\fant{a}2}\wdg  R^3\right)+\frac{2\beta}{a^2}\Phi_{22}*k=0
\ee
for the profile function. After some algebra, one can show that this equation further reduces to a linear equation in terms of Ricci spinor
\be\label{qc-ads-profile-eqn}
\left(*d*d+\frac{1}{2a^2}\right)\Phi_{22}k=0
\ee
which has a formal resemblance to the equation satisfied by the profile function in the curved background with now 
the profile function is replaced by the traceless Ricci tensor component $\Phi_{22}$ consistent with the case of the pp-waves discussion in general QC gravity. 
However, note that the Hodge dual in this case is defined relative to the null coframe defined in (\ref{ads-wave-ansatz}) with an AdS background. 

Recently, it has been shown \cite{gurses-sisman-tekin-universal-kundt} that Kerr-Schild-Kundt class of metrics are universal metrics 
\cite{deser-polarization,bergshoeff1,gibbons,amati,coley-pope,hervik-pravda-pravdova} in the sense that they solve generic gravity equations for vacuum with a cosmological constant. The equation (\ref{qc-ads-profile-eqn}) for the profile function is in accordance with the mathematical statement of an all-encompassing theorem  obtained in \cite{gurses-sisman-tekin-universal-kundt}. In the narrow context of pure QC gravity discussion here, the universality property implies that the QC gravity model governed by the SKY Lagrangian 4-form  (\ref{sky-lag}) also leads to the equation (\ref{qc-ads-profile-eqn}) for the profile function for the metric ansatz (\ref{ads-wave-ansatz}). Note that the SKY Lagrangian (\ref{sky-lag}) is not equivalent to the general QC  Lagrangian (\ref{qc-reduced-action}).

\subsection{\texorpdfstring{$pp$}{}-waves in a tensor-tensor gravity with  torsion}

The quadratic curvature gravity model studied in this section  involves a symmetric second rank tensor $\Phi=\Phi_{ab}\theta^a\ot \theta^b$ \cite{dereli-tucker-pp-waves-energy-in-tensor-tensor-model}. The Lagrangian of such a tensor-tensor model  involves an interaction  term of the form $\Phi^{ab}\Omega_{ac}\wdg*\Omega^{c}_{\fant{a}b}$ and the particularly interesting feature of this model is that the field equations yield   the Bell tensor \cite{bell1,bell2,bell3,bell4}.

Following closely the definition given in \cite{dereli-tucker-pp-waves-energy-in-tensor-tensor-model}, the fourth-rank Bell tensor  $B$ can be written in the form
\be
B\equiv T_{abc}\ot \theta^a\ot \theta^b\ot \theta^c
\ee
where three-indexed 1-form $T_{abc}$ can be defined in terms of the following expressions  of the curvature 2-forms
\be\label{bell-def}
*T_{abc}
\equiv
\tfrac{1}{2}
\left(
i_a\Omega_{bd}\wdg *\Omega^{d}_{\fant{a}c}
-
\Omega_{bd}\wdg i_a*\Omega^{d}_{\fant{a}c}
\right)
\ee
in the current notation. For the curvature 2-forms corresponding to a Levi-Civita  connection, the Bell tensor defined by (\ref{bell-def}) has some mathematical properties in common with the energy-momentum 3-form  $*T^a[F]$ of the Maxwell field defined in Eq. (\ref{maxwell-en-mom-3-form}) above.
Remarkably, for the null coframe defined by Eq. (\ref{pp-coframe-def}), the non-vanishing component of the Bell tensor is
\be\label{nonzero-bell-comp}
*T^{1}_{\fant{a}00}
=
-2\left(H_{\zeta\bar{\zeta}}H_{\zeta\bar{\zeta}}+H_{\zeta\zeta}H_{\bar{\zeta}\bar{\zeta}}\right)*k.
\ee

The Lagrangian 4-form introduced by Dereli and Tucker  \cite{dereli-tucker-pp-waves-energy-in-tensor-tensor-model}  can  explicitly be written in the form
\be\label{tensor-tensor-lag}
L=
\frac{1}{2\kappa^2}\Omega_{ab}\wdg *\theta^{ab}
+
\tfrac{1}{2}\Phi_{ab}\Omega^{a}_{\fant{a}c}\wdg *\Omega^{cb}
+
\tfrac{1}{2}D\Phi_{ab}\wdg *D\Phi^{ab}
-
\tfrac{1}{2}F\wdg *F
\ee
with the help of a symmetrical second rank tensor $\Phi_{ab}$ coupling to a particular quadratic curvature term. The issue of energy and momentum carried by a gravitational wave was also discussed in \cite{obukhov-pereira-rubilar} in the context of teleparallel gravity  by making use of the Bell tensor.

Assuming that the connection and the coframe 1-forms are independent gravitational variables, the field equations that follow from the total variational derivative of the Lagrangian 4-form (\ref{tensor-tensor-lag})  yields the gravitational field equations \cite{dereli-tucker-pp-waves-energy-in-tensor-tensor-model}
\be\label{tensor-tensor-coframe-eqn}
-\frac{1}{\kappa^2}*G^a+*T^{a}[F]+\lambda*T^{a}[D\Phi]+\Phi^{bc}*T^{a}_{\fant{a}bc}=0
\ee
for the coframe 1-forms where the energy-momentum 3-forms $*T^a[D\Phi]$ is defined as
\be
*T^a[D\Phi]
=
\tfrac{1}{2}
\left(
i_a D\Phi_{bc}\wdg* D\Phi^{bc}
+
D\Phi_{bc}\wdg i_a* D\Phi^{bc}
\right).
\ee
The equations of motion for the independent connection are
\be\label{tensor-tensor-connection-eqn}
-\frac{1}{2\kappa^2}\Theta_c\wdg *\theta^{abc}
+\tfrac{1}{2}D\left(\Phi^{a}_{\fant{a}c}*\Omega^{cb}-\Phi^{b}_{\fant{a}c}*\Omega^{ca}\right)
+
\lambda\left(\Phi^{a}_{\fant{a}c}D*\Phi^{cb}-\Phi^{b}_{\fant{a}c}*D\Phi^{ca}\right)=0.
\ee

The above gravitational field equations are defined in a Riemann-Cartan spacetime in general. Although the Ricci tensor in this case is defined as in the pseudo-Riemannian case, and likewise the Einstein 3-form is constructed from the curvature 2-form  of a general connection with a non-vanishing torsion as in the pseudo-Riemannian case. However, the Einstein tensor is not symmetrical in general in the non-Riemannian case with a non-vanishing torsion. Accordingly, the covariant exterior derivative is defined  with a more  general connection with a non-vanishing torsion as well.

These equations are to be supplemented with the equations
\be\label{phi-tensor-eqns}
\lambda D*D\Phi^{ab}-\tfrac{1}{2}\Omega^{a}_{\fant{a}c}\wdg*\Omega^{cb}=0
\ee
for the tensor field $\Phi_{ab}$ and the Maxwell's equations $d*F=dF=0$.
In contrast to the pseudo-Riemannian case, the field equations are obtained by constraining the independent connection  to be a Levi-Civita connection, are more complicated then the field equations above in the non-Riemann geometry with torsion.

Keeping  in mind that  the only contribution of Bell tensor to the coframe equations is of the form given in Eq. (\ref{nonzero-bell-comp}),
it is natural to assume that a compatible ansatz for the tensor $\Phi$ is of the form $\Phi=\Phi_{ab} \theta^a\ot \theta^b=\Phi_{11} l\ot l$ with the only non-vanishing components $\Phi_{11}$ and the choice $\Phi_{11}=const.$ renders $\Phi$ a covariantly constant tensor. (To avoid confusion, the non-vanishing constant component is denoted by $\Phi_c$ below) As a consequence of the judicious  choice for $\Phi_{ab}$, all $*T^a[D\Phi]$ vanish identically as well.   Furthermore, one also obtains that $\Theta^a=0$ from the independent connection equation and that Eq. (\ref{phi-tensor-eqns}) are satisfied identically. Eventually, it is consistent to write the coframe equations (\ref{tensor-tensor-coframe-eqn}) in terms of the pseudo-Riemannian quantities. Fot the $pp$-wave ansatz (\ref{pp-electro-ansatz}), this then yields a nonlinear partial differential equation of the form
\be\label{tensor-tensor-profile-function-pde}
-H_{\zeta\bar{\zeta}}
+
\kappa^2\Phi_c \left(H_{\zeta\bar{\zeta}}H_{\zeta\bar{\zeta}}
+
H_{\zeta\zeta}H_{\bar{\zeta}\bar{\zeta}}\right)
+
\kappa^2f_\zeta\bar{f}_{\bar{\zeta}}=0.
\ee
A particularly simple solution to Eq. (\ref{tensor-tensor-profile-function-pde}) can be constructed with a homogeneous profile function of the
form
\be\label{tensor-tensor-sol-ansatz}
H(u,\zeta,\bar{\zeta})
=
h_1(u) \zeta^2+h_1(u)\bar{\zeta}^2+h_2(u)\zeta\bar{\zeta}
\ee
with $f(u,\zeta)=\alpha(u)\zeta$ corresponding to the electromagnetic part of the ansatz. $h_1(u), h_2(u)$ and $\alpha(u)$ are  real functions of the variable $u$ satisfying
\be
h_2=\kappa^2 \Phi_c(h_2^2+h_1^2)+\kappa^2\alpha^2.
\ee
As one can observe from (\ref{nonzero-bell-comp}) that,  for a positive $\Phi_c$,  the  solution  of the  form  (\ref{tensor-tensor-sol-ansatz}) then leads to a positive definite expression for $T_{000}$  \cite{dereli-tucker-pp-waves-energy-in-tensor-tensor-model} admitting finite values only.

\subsection{\texorpdfstring{$pp$}{}-waves in the Chern-Simons modified GR}

The Chern-Simons modified gravity is a parity violating extension of GR introduced by Jackiw and Pi \cite{jackiw-pi}. It is also motivated by
the string theory \cite{alexander-yunes-phys-rep}. In this modified gravity model, the three dimensional  CS-topological current  is embedded into four spacetime dimensions. It has subsequently found diverse applications, for example,  in the context of the inflationary models and in the study of primordial gravitational waves as well as many other topics in cosmology.

The Lagrangian 4-form for the CS modified GR employs the Pontryagin topological term in addition to the familiar Einstein-Hilbert term.
In particular, the CS term is favored by string theory predicting the Pontryagin topological correction term in  the low energy limit.
A derivation of the CS modified GR field equations from a truncation of a low-energy effective heterotic string theory models involving the Kalb-Ramond  field, and a dilaton field was recently given in \cite{adak-dereli}, and a  similar derivation was also presented in \cite{smith-erickeck-et-al} at around the same time.

The field equations for the CS modified GR model follow from the Lagrangian 4-form
\be\label{CS-lagrangian}
L_{CS}
=
\tfrac{1}{2}\Omega_{ab}\wdg *\theta^{ab}
-
\tfrac{1}{8}\theta(x)\Omega_{ab}\wdg\Omega^{ba}
\ee
where the first term is the familiar Einstein-Hilbert Lagrangian 4-form and
the second term, which is known as the Pontryagin term, is an exact form
\be\label{Pontryagin-term}
\Omega_{ab}\wdg\Omega^{ba}=dK
\ee
with the Chern-Simons 3-form $K$ defined as
\be\label{CS-3-form}
K=
\omega_{ab}\wdg \Omega^{ba}-\tfrac{1}{3}\omega^{a}_{\fant{a}b}\wdg \omega^{b}_{\fant{a}c}\wdg \omega^{c}_{\fant{a}a}
\ee
and therefore the term in (\ref{Pontryagin-term}) does not contribute to the field equations for a constant $\theta$.
Written in this form, the Pontryajin term depends  on the connection 1-form and it contributes to the coframe equations
only through the Lagrange multiplier term $L_C$ of the form (\ref{zero-torsion-constraint-lag}) introduced to impose the vanishing torsion constraint for the independent connection in the first order formalism.

The total variational derivative of the extended action $L_{ext.}= L_{CS}+L_C$ with respect to the variables can be found in a straightforward manner  as
\begin{align}
\delta L_{ext.}
&=
\delta \theta_a\wdg \left(-*G^a+D\lambda^a\right)
-
\delta \theta \tfrac{1}{8}\Omega_{ab}\wdg \Omega^{ba}
+
\delta \lambda_a\wdg \Theta^a
\nonumber\\
&\fant{=}+
\delta{\omega}_{ab}\wdg
\left\{
\tfrac{1}{2}D*\theta^{ab}
-
\tfrac{1}{4}D(\theta\Omega^{ab})
-
\tfrac{1}{2}\left(\theta^a\wdg \lambda^b-\theta^b\wdg\lambda^a\right)\
\right\}
\end{align}
up to an omitted exact form.  Because the contribution of the topological terms to the coframe equations results  from the Lagrange multiplier term, as before, one first solves the  independent connection equations $\delta L_{ext.}/\delta\omega_{ab}=0$ for the Lagrange multiplier 2-form.

The connection equations which  explicitly read
\be
D*\theta^{ab}
-
\tfrac{1}{2}D(\theta\Omega^{ab})
-
\left(\theta^a\wdg \lambda^b-\theta^b\wdg\lambda^a\right)=0
\ee
can be solved for the Lagrange multiplier 2-form to have
\be
\lambda^a
=
-i_b(\Omega^{ba}\wdg d\theta)-\tfrac{1}{4}\theta^a\wdg  i_bi_c(\Omega^{bc}\wdg d\theta)
\ee
as the unique solution. At this point, it is convenient to define auxiliary vector-valued 2-form $P^a$ as
\be
P^a\equiv
i_b(\Omega^{ba}\wdg d\theta)
\ee
and its contraction $P\equiv i_a P^a$.  In terms of the vector-valued auxiliary form $P^a$ and its contraction, the Lagrange multiplier 2-form can be written conveniently as
\be
\lambda^a
=
-(P^a-\tfrac{1}{4}\theta^a\wdg P)
\ee
formally resembling the expression for the Schouten 1-form defined in 2+1 dimensions in the context of topologically massive gravity \cite{tmg}.
In 2+1 dimensions, the Cotton 2-form is derived from the Schouten 1-form \cite{cotton}, $L^a=R^a-\tfrac{1}{4}R\theta^a$. In general, the Cotton tensor is defined in any spacetime dimension $D\geq3$, but the definition depends explicitly on the spacetime dimensions.

In the present geometrical framework, it is convenient to define the vector-valued 3-form $C^a=\tfrac{1}{6}C^{a}_{\fant{a}bcd}\theta^{bcd}$ by
\be
C^a\equiv D\lambda^a
=
D(P^a-\tfrac{1}{4}\theta^a\wdg P)
\ee
as well.
Note that $C^a$ defined in this way can be expressed in terms of  a symmetric traceless tensor  $C_{ab}$ by using the relation
\be\label{c-form-c-tensor-relation}
C^{ab}
=
-\tfrac{1}{2}*(\theta^a \wdg C^b+\theta^b\wdg C^a)
\ee
where the tensor $C^{ab}$ represents the orthonormal components of the tensor which is often called C-tensor in the literature. See, for example,  the review article by Alexander and Yunes \cite{alexander-yunes-phys-rep} and the references therein for an extensive discussion. The relation (\ref{c-form-c-tensor-relation}) can be used to derive a more familiar expression for the C-tensor relative to a coordinate expression. In the presentation below, the vector-valued 3-form $C^a$ will be called  ``C-form".

Eventually, after taking the matter energy-momentum forms $*T_a[\psi]\equiv \delta L_{m}/\delta \theta^a$ coming from the matter Lagrangian $L_{m}[g,\psi]$ into account, the coframe equations for the CS modified gravity  then take the form
\be\label{vacuum-cs-eqn}
*G^a+C^a=\kappa^2*T^a[\psi]
\ee
with  $\kappa^2$ denoting the coupling constant in GR. AS in the previous cases, only the vacuum and the electrovacuum solutions to these field equations will be discussed in this subsection.

The variational derivative of the extended Lagrangian with respect to CS scalar field, which can be considered as a Lagrange multiplier 0-form for the  model,
leads to the constraint
\be\label{p-constraint}
\Omega_{ab}\wdg \Omega^{ba}=0.
\ee
Eqs. (\ref{vacuum-cs-eqn}) and (\ref{p-constraint}) constitute the field equations for the CS modified gravity model expressed in terms of the exterior forms relative to a null  coframe.

It is worth to note at this point that the use of NP formalism in a study of CS modified gravity is also favored by the constraint (\ref{p-constraint}) because (\ref{p-constraint}) can be rewritten in terms of Weyl 2-forms in the form
\be\label{p-constraint-form2}
C_{ab}\wdg C^{ba}
=
0
\ee
by making use of the expansion (\ref{curvature-general-expansion}) and the first Bianchi identity satisfied by the curvature 2-form.
Consequently, because the expression  (\ref{p-constraint-form2}) involves only the  Weyl spinor scalars $\Psi_k$, the Pontryagin constraint can be considered as  a constraint on the Petrov type. It is possible to show  that (\ref{p-constraint-form2}) explicitly reads
\be
\tfrac{1}{8}C_{ab}\wdg C^{ba}
=
i\left\{
3(\Psi^2_2-\bar{\Psi}^2_2)-4(\Psi_1\Psi_3-\bar{\Psi}_1\bar{\Psi}_3)+\Psi_0\Psi_4-\bar{\Psi}_0\bar{\Psi}_4
\right\}*1
\ee
by using (\ref{null-coframe-curvature-scalars}) after some straightforward algebra.
The constraint (\ref{p-constraint}), or equivalently (\ref{p-constraint-form2}), imposing the vanishing of the Pontryagin term, is essential to have the diffeomorphism invariance of the model. Note that the Pontryagin constraint (\ref{p-constraint}) is satisfied identically for the ansatz (\ref{pp-electro-ansatz}) because the only non-vanishing Weyl spinor scalar is $\Psi_4$. Hence, it suffices  to consider the CS modified equations (\ref{vacuum-cs-eqn})  for a type N metric in general.

As a consequence of the constraint (\ref{p-constraint}), the C-form is covariantly constant and therefore the matter coupling to the CS modified gravity requires a covariantly constant matter energy-momentum tensor as in the GR. Explicitly, by making use of the first Bianchi identity satisfied by the curvature 2-form it is possible to show that
\be
DC^a
=
-\tfrac{1}{4}(i^ad\theta)\Omega_{bc}\wdg \Omega^{cb}.
\ee

The C-form is a traceless vector-valued 3-form by definition and it is covariantly constant provided that the Pontryagin constraint is satisfied. It is important to note that these properties are in common  with the Cotton 2-form in 2+1 dimensions. However, C-form has some other properties that are not in common with those of the Cotton 2-form defined in three dimensions.

In order to construct the solutions to the CS modified gravity,  one has to make additional assumptions for the CS scalar field $\theta$  along with the metric ansatz (\ref{pp-electro-ansatz}). It is convenient to start with a general CS scalar such that $\theta=\theta(u, v, \zeta, \bar{\zeta})$ and then subsequently restrict it to a convenient form as one proceeds.

By making use of curvature expressions (\ref{pp-curavture-form}) for the $pp$-wave metric ansatz (\ref{pp-wave-ansatz}), one finds the following expressions
\be\label{P-tensor-expressions}
\begin{split}
P^1
&=
-
2\theta_v H_{\zeta\bar{\zeta}}k\wdg l
+
\left(
\theta_{\bar{\zeta}} H_{\zeta\zeta}-\theta_{{\zeta}} H_{\zeta\bar{\zeta}}\right)
k\wdg m
+
\left(
\theta_{{\zeta}} H_{\bar{\zeta}\bar{\zeta}}-\theta_{\bar{\zeta}} H_{\zeta\bar{\zeta}}\right)
k\wdg \bar{m},
\\
P^2
&=
-
\theta_v H_{\bar{\zeta}\bar{\zeta}}k\wdg \bar{m}
-
\theta_v H_{\zeta\bar{\zeta}}k\wdg {m},
\end{split}
\ee
for the non-vanishing components of the auxiliary form $P^a$ and also note that $P^3$ can be obtained by the complex conjugation relation $P^3=\bar{P}^2$. Moreover, using the expressions (\ref{P-tensor-expressions}), one can find the contraction of the 2-form $P^a$ as
\be
P
=
4\theta_v H_{\zeta\bar{\zeta}}k.
\ee

By combining the above results, the non-vanishing  Lagrange multiplier 2-forms then can be expressed in  the form
\be
\begin{split}
\lambda^1
&=
-
\theta_v H_{\zeta\bar{\zeta}}k\wdg l
+
\left(
\theta_{\bar{\zeta}} H_{\zeta\zeta}-\theta_{{\zeta}} H_{\zeta\bar{\zeta}}\right)
k\wdg m
+
\left(
\theta_{{\zeta}} H_{\bar{\zeta}\bar{\zeta}}-\theta_{\bar{\zeta}} H_{\zeta\bar{\zeta}}\right)
k\wdg \bar{m},
\\
\lambda^2
&=
+
\theta_vH_{\zeta\bar{\zeta}}k\wdg m
-
\theta_v H_{\bar{\zeta}\bar{\zeta}}k\wdg \bar{m}
-
\theta_v H_{\zeta\bar{\zeta}}k\wdg {m}.
\end{split}
\ee
where $\lambda^3=\bar{\lambda}^2$.
Finally, by using the fact that the C-tensor is given by the covariant exterior derivative as
$
C^a
=
D\lambda^a,
$
one ends up with the following non-vanishing components of the C-form:
\begin{align}
C^1
&=
-i\left(
\theta_{\bar{\zeta}\bar{\zeta}} H_{\zeta\zeta}
-
\theta_{\zeta\zeta}H_{\bar{\zeta}\bar{\zeta}}
-
2\theta_{{\zeta}}H_{\zeta\bar{\zeta}\bar{\zeta}}
+
2\theta_{\bar{\zeta}}H_{\zeta{\zeta}\bar{\zeta}}
\right)
*k
\nonumber\\
&\qquad-
i\left(
\theta_v H_{\zeta\zeta}
\right)_{\bar{\zeta}}*m
+
i\left(
\theta_v H_{\bar{\zeta}\bar{\zeta}}
\right)_{{\zeta}}*\bar{m},
\label{C-form-1-exp}\\
C^2
&=
-i\left(
\theta_v H_{\bar{\zeta}\bar{\zeta}}
\right)_{{\zeta}}*k
-
i\theta_{vv}H_{\bar{\zeta}\bar{\zeta}}*\bar{m},
\label{C-form-2-exp}
\end{align}
where $C^3=\bar{C}^2$ by definition.

 An immediate observation about the general  equations of motion for the CS modified gravity equations is that, one of the vacuum equations, namely
$\theta_{vv}H_{{\zeta}{\zeta}}=0$, decouples from the rest of the field equations. Moreover, because this equation involves the only non-vanishing Weyl component $\Psi_4=H_{{\zeta}{\zeta}}$, one has to assume that $\theta_{vv}=0$ in order to maintain the Petrov type of the metric.

A  classification scheme of the solutions to the CS modified GR introduced in \cite{grumiller-yunes}, and it is convenient to discuss the  $pp$-waves solution in this regard as well. It is possible to construct $\mathcal{P}$ Class and $\mathcal{CS}$ Class solutions to the CS modified GR as follows.
\begin{itemize}

\item[(1)]$\mathcal{P}$ Class solutions (GR solutions lifted to the CS modified GR): The  solutions in this class satisfy $*G^a=\kappa^2 *T^a[F]$ and also by demanding $C^a=0$ separately.  For the  $\mathcal{P}$ Class solutions for the $pp$-wave ansatz, the field equations reduce to the following third order coupled partial differential equations
    \begin{align}
&H_{\zeta\bar{\zeta}}=\kappa^2 f_\zeta \bar{f}_{\bar{\zeta}},
\\
&
\theta_{\bar{\zeta}\bar{\zeta}} H_{\zeta\zeta}
-
\theta_{\zeta\zeta}H_{\bar{\zeta}\bar{\zeta}}
-
2\theta_{{\zeta}}H_{\zeta\bar{\zeta}\bar{\zeta}}
+
2\theta_{\bar{\zeta}}H_{\zeta{\zeta}\bar{\zeta}}=0,
\\
&
(\theta_{v} H_{{\zeta}{\zeta}})_{\bar{\zeta}}
=0, \qquad \theta_{vv}H_{\bar{\zeta}\bar{\zeta}} =0.
\end{align}
First note that the last of the above equations imply that $\theta_{vv}=0$ for a non-vanishing $\Psi_4$. For the vacuum case, these equations simplify considerably and take  the form
\begin{align}
&H_{\zeta\bar{\zeta}}=0,\label{cs-reduced-to-gr}
\\
&\theta_{\bar{\zeta}\bar{\zeta}} H_{\zeta\zeta}
-
\theta_{\zeta\zeta}H_{\bar{\zeta}\bar{\zeta}}
=0,
\label{gr-class-eqn2}\\
&
\theta_{v\bar{\zeta}} H_{{\zeta}{\zeta}}
=0. \label{gr-class-eqn3}
\end{align}
Furthermore, Eqs. in (\ref{gr-class-eqn3}) imply that $\theta_{v\zeta}=\theta_{v\bar{\zeta}}=0$.
As a consequence these relations involving only the partial derivative of the CS scalar field, one can conclude that the CS scalar must be of the form
\be\label{general-cs-scalar-sol}
\theta(u,v,\zeta,\bar{\zeta})
=
vA(u)+B(u,\zeta,\bar{\zeta})
\ee
with two undetermined functions $A=A(u)$ and $B=B(u, \zeta, \bar{\zeta})$. After determining  the general solution  Eq. (\ref{cs-reduced-to-gr}), the function $B$  then can be determined by the equations
\be\label{B-eqn}
B_{\bar{\zeta}\bar{\zeta}}H_{\zeta\zeta}-B_{\zeta\zeta}H_{\bar{\zeta}\bar{\zeta}}=0
\ee
by using (\ref{gr-class-eqn2}). The CS scalar field of the form given in (\ref{general-cs-scalar-sol}) with the function $B(u,\zeta, \bar{\zeta})$ satisfying (\ref{B-eqn}) then leads to the most general vacuum $pp$-wave solution that the CS modified GR model has in common with the GR solutions.

For the vacuum solutions of the Einstein field equations of the form $H(u, \zeta, \bar{\zeta})=h(u, \zeta)+\bar{h}(u, \bar{\zeta})$ where the function $h$ is an analytical function of $\zeta$ with arbitrary $u$-dependence, Eq. (\ref{B-eqn}) is solved by the function $B$ having the same form as $H$ and thus, one has
\be
B(u, \zeta, \bar{\zeta})
=
h(u, \zeta)+\bar{h}(u, \bar{\zeta}).
\ee

It is also possible to construct  another $\mathcal{P}$-class  solution by considering the Aichelburg-Sexl solution \cite{grumiller-yunes}.
For  the Aichelburg-Sexl solution  with $h(u,\zeta)\sim \delta(u)\ln\zeta$ which requires a null particle source term in the Einstein field equations, Eq. (\ref{B-eqn}) now becomes
\be
\zeta^2B_{\zeta\zeta}-\bar{\zeta}^2B_{\bar{\zeta}\bar{\zeta}}=0
\ee
and the resulting equation  implies that $B$ is an arbitrary function of  the real variable $|\zeta|^4$ leaving the $u$ dependence undetermined.

\item[(2)]$\mathcal{CS}$ Class  solutions (non-GR solutions):
This class of solutions can be found by solving  the general equations (\ref{vacuum-cs-eqn}). They are, in general, third order partial differential equations in the metric components and for the $pp$-wave ansatz, they can explicitly be written as
\begin{align}
&
\left(\theta_{v} H_{{\zeta}{\zeta}}\right)_{\bar{\zeta}}
=0,
\label{cs-class-eqns1}\\
&2H_{\zeta\bar{\zeta}}
-
i\left(
\theta_{\bar{\zeta}\bar{\zeta}} H_{\zeta\zeta}
-
\theta_{\zeta\zeta}H_{\bar{\zeta}\bar{\zeta}}
-
2\theta_{{\zeta}}H_{\zeta\bar{\zeta}\bar{\zeta}}
+
2\theta_{\bar{\zeta}}H_{\zeta{\zeta}\bar{\zeta}}
\right)
=2\kappa^2 f_\zeta\bar{f}_{\bar{\zeta}},\label{cs-class-eqn2}
\end{align}
In addition, it is also assumed in this case that the CS scalar satisfy the equation $\theta_{vv}=0$ as before.
With the further  simplifying assumption $\theta_v=0$, the field equations Eq. (\ref{cs-class-eqns1}) are satisfied  identically and one is left with
 Eq. (\ref{cs-class-eqn2}) with the expression in Eq. (\ref{general-cs-scalar-sol}) reducing to $\theta=\theta(u, \zeta, \bar{\zeta})$.
In this case, the functions $\theta=\theta(u, \zeta, \bar{\zeta})$ and $H=H(u, \zeta, \bar{\zeta})$ are to be determined from the  Eq. (\ref{cs-class-eqn2}) alone.

The vacuum solution presented in \cite{grumiller-yunes} by Grumiller and Yunes is constructed under the additional assumption that  $\theta_{\zeta\zeta}=0$.
The CS scalar then simplifies to the form linear in the complex coordinates $\zeta$ and $\bar{\zeta}$ which can be written as
\be
\theta
=
a(u)\zeta+\bar{a}(u)\bar{\zeta}+b(u)
\ee
with $a$ a complex functions of the real null coordinate $u$ whereas  $b=b(u)$ is a real function. Consequently, Eq. (\ref{cs-class-eqn2}) reduces to
\be\label{cs-class-reduced-eqn2}
H_{\zeta\bar{\zeta}}
+
i\left(
\theta_{{\zeta}}H_{\zeta\bar{\zeta}\bar{\zeta}}
-
\theta_{\bar{\zeta}}H_{\zeta{\zeta}\bar{\zeta}}
\right)
=0.
\ee
In order to put this equation into the form of a  Poisson equation,  it is convenient to introduce \cite{grumiller-yunes} the following field redefinition
\be\label{field-redef}
H_{\zeta\bar{\zeta}}\equiv q(u, \zeta, \bar{\zeta}).
\ee
In terms of  the new function $q$, Eq. (\ref{cs-class-reduced-eqn2}) can now be rewritten as a first order partial differential equation of the form
\be\label{non-electro-f-CS-class-sol}
q-i(aq_{\bar{\zeta}}-\bar{a}q_\zeta)=0.
\ee
One can verify that this equation has the general solution of the form
\be\label{special-solution-to-reduced-eqn}
q(u, \zeta, \bar{\zeta})
=
e^{(\zeta+\bar{\zeta})/i(\bar{a}-a)}\phi(\bar{a}\zeta+{a}\bar{\zeta}+b)
\ee
with $\phi$ being an arbitrary function of the argument $\bar{a}\zeta+{a}\bar{\zeta}+b$. Subsequently, this solution can be inserted back into the Eq.
(\ref{field-redef}) to construct a $\mathcal{CS}$-class solution by solving the resulting Poisson equation \cite{grumiller-yunes}, for example, by the method of Green's functions by introducing some appropriate boundary  conditions on the transverse planes spanned by the complex coordinates for different values of the coordinate $u$.

\end{itemize}

\section{Concluding comments}

Before the LIGO Scientific Collaboration's groundbreaking announcement confirming the first direct observation of gravitational waves emitted by a black hole binary  merger, we had  only an indirect evidence \cite{hulse-taylor1,hulse-taylor2} for the existence of the gravitational waves. (There are other ongoing efforts in various projects \cite{GRW-observe2,GRW-observe3,GRW-observe4} to observe gravitational waves as well) In the near future, the advances on the observational front will be a powerful tool for testing the viable theoretical  models of gravity \cite{eardley-lee-lightman,eardley-lee-lightman2} and some popular modified gravitational models will certainly be ruled out by the observations. Furthermore, considering the puzzling observational data evidencing a current accelerated expansion phase of the Universe contrary to the former expectations, the direct detection  of gravitational waves will probably have an impact on the theoretical efforts as well.

Contrary to the remarks in Ref. \cite{tupper}, implying that the use of the Newman-Penrose null tetrad formalism is somewhat cumbersome in deriving the $pp$-wave type solutions to the BD theory, the NP formalism provides probably the most convenient and efficient mathematical framework in any topic involving  the gravitational radiation and it is also certainly well-suited to the discussions of these issues in the context of  variety of modified gravity models
(See the remarks on the practical efficiency of the null tetrad formalism  by R. P. Kerr made in his historical account of the metric he founded \cite{kerr}). In particular, with some further appropriate development of the technical presentation \cite{baykal-gr-waves}, the discussion of the exact solutions above can be extended to  more complicated family of metrics, at the same time, also taking  the algebraic character of such solutions into account.

It is well-known  that the family of $pp$-wave metrics belongs to a more general family of metrics, known as the Kundt waves \cite{kundt1,kundt2,ehlers-kundt}. The family of Kundt waves are also described by a null geodesic with vanishing optical scalars, however the assumption that the null vector is  covariantly constant is dropped. Consequently, for the gravitational waves metrics in this family, the  transverse  planes are not flat. In addition, the metrics in the Kundt family can  have Petrov types II, D, III and N \cite{exact-sol-griffiths-podolsky}. It is shown by Bi\v{c}\'ak and Podolsk\'y \cite{bicak-podolsky1,bicak-podolsky2} that the family of Petrov type N gravitational waves with a nonvanishing expansion with a cosmological constant belongs to Kundt family of metrics whereas  the expanding solutions belong to the Robinson-Trautman family of metrics.  It also possible to construct twisting  type N solutions \cite{hauser1, hauser2,zhang-finley1,zhang-finley2}. However, the twisting type N solutions lead to a field equations that are quite involved compared the nontwisting solutions and are not discussed above. The discussion of such algebraically special solutions of the  modified gravity models will be work for the future research.

\section{Appendix: Prime companion null coframe for the \texorpdfstring{$pp$}{}-wave metric ansatz in (\ref{pp-wave-ansatz})}

As an alternative to the set of basis coframe 1-forms  (\ref{pp-coframe-def}), it is also possible to adopt the null coframe
\be\label{prime-companion-coframe}
k=Hdu +dv, \qquad l=du,\qquad  m=d\bar{\zeta}
\ee
for the $pp$-wave ansatz (\ref{pp-wave-ansatz}).
The set of coframes (\ref{prime-companion-coframe}) is related to the set coframes (\ref{pp-coframe-def}) by the interchanges $k\leftrightarrow l$ and $m\leftrightarrow \bar{m}$. Under these interchange of the basis 1-forms, the Cartan's structure  equations (\ref{first-SE}) and (\ref{SE-2}) are mapped onto themselves. This symmetry, originally called the prime symmetry,  is  a computationally useful symmetry in the NP formalism.  The  tensorial objects or scalars related by prime symmetry are said to be prime companions.
In terms of the null coframe indices, the prime symmetry corresponds to symmetry of the structure equations under the interchanges $0\leftrightarrow 1$ and $2\leftrightarrow 3$.

The connection and the curvature forms  belonging to the null coframe (\ref{prime-companion-coframe}), which are the prime companions to those of (\ref{pp-coframe-def}), are given by
\begin{align}
&\omega^{0}_{\fant{a}3}=H_{\bar{\zeta}}l,
\\
&\Omega^{0}_{\fant{a}3}=d\omega^{0}_{\fant{a}3}=-H_{\bar{\zeta}\bar{\zeta}}l\wdg \bar{m}-H_{\zeta\bar{\zeta}}l\wdg m\label{alter-curv-2-f0rm}
\end{align}
respectively. It follows from (\ref{alter-curv-2-f0rm}) that $*G^0=-2H_{\zeta\bar{\zeta}}*l$ and the corresponding curvature scalars are $\Phi_{00}=H_{\zeta\bar{\zeta}}$, and $\Psi_0=H_{\bar{\zeta}\bar{\zeta}}$.

\section*{Acknowledgments}
It is a pleasure to thank O. T. Turgut for his suggestion to write a review paper which initiated the current work and for all the subsequent encouragement while writing and  revising it. I thank \"O. Delice for pointing out the papers on the gravitational wave solutions in Brans-Dicke theory and M. 
\"Ozer for his kind invitation to the special issue of Turk. J. Phys. on General Relativity and Related Topics. 
I would also like to thank the anonymous TJP referee for his valuable suggestions on the original version that helped to improve the presentation.
For the previous revisions, I would like to thank A. N. Aliev, A. Mazumdar,  V. Pravda and S. D. Odintsov for informative correspondences.

\end{document}